\documentclass[%
 amsmath,
 aps,
 twocolumn,
 prd,
 floatfix,
 superscriptaddress,
 preprintnumbers,
 nofootinbib
]{revtex4}

\usepackage{graphicx}
\usepackage{dcolumn}
\usepackage{bm}
\usepackage{hyperref}
\usepackage{subfigure}
\usepackage{color}
\usepackage{url}

\begin{document}

\preprint{IPPP/17/13}
\preprint{Nikhef-2017-037}

\title{Double-charming Higgs identification using machine-learning assisted jet shapes}

\author{Alexander Lenz}
\email{alexander.lenz@durham.ac.uk}
\author{Michael Spannowsky}%
\email{michael.spannowsky@durham.ac.uk}
\affiliation{%
IPPP, Department of Physics, Durham University, Durham DH1 3LE, UK
}%
\author{Gilberto Tetlalmatzi-Xolocotzi}%
\email{gtx@nikhef.nl}
\affiliation{%
Nikhef, Science Park 105,
1098 XG Amsterdam
}%

\date{\today}

\begin{abstract}
We study the possibility of identifying a boosted resonance that decays into a charm pair against 
different sources of background using QCD event shapes, which are promoted to jet shapes. 
Using a set of jet shapes as input to a boosted decision tree, we find that observables utilizing the simultaneous presence of two charm quarks can access complementary information compared to approaches relying on two independent charm tags.
Focusing on Higgs associated production with subsequent $H\to c \bar{c}$ decay and on a CP-odd scalar $A$ with
$m_A \leq 10$ GeV we obtain the limits
$\mathcal{B}r(H\rightarrow c\bar{c})\leq 6.09\%$ and $\mathcal{B}r(H\rightarrow A(\rightarrow c\bar{c}) Z)\leq 0.01\%$ at $95\%$ C. L.. 
\end{abstract}

\maketitle

\section{\label{sec:intro}Introduction}

After the discovery of the Higgs boson \cite{Aad:2012tfa,Chatrchyan:2012xdj} a precise determination
of its couplings is now of fundamental importance.
The couplings of the Higgs boson to the $W$ and $Z$ bosons are already known to be in good agreement with
the standard model (SM) expectation, as can be inferred from the measurements of the Higgs decay and production rates by ATLAS and 
CMS \cite{Khachatryan:2014jba,Aad:2015gba,Khachatryan:2016vau}.\\

In the SM the Yukawa interaction describes the coupling of the Higgs boson to a fermion $f$ with a strength
given by the Yukawa-coupling $y_f^{\rm SM}$. Deviations from the SM expectation can be parametrised by 
$\kappa_f = y_f / y_f^{\rm SM}$, which can be deduced from a measurement of the signal strength $\mu_f$ 
defined as $ \mu_f = \sigma_H \mathrm{Br}_{f \bar{f}} / ( \sigma_H^{\rm SM}  \mathrm{Br}_{f \bar{f}}^{\rm SM})$.
Here $\sigma_H$ is the Higgs boson production cross section and $\mathrm{Br}_{f \bar{f}}$ is the 
branching ratio of the decay process $ H \to f \bar{f}$. 
Currently the couplings between the Higgs boson and the third generation fermions
are consistent with the SM expectations, one gets:
$\mu_t = 2.2  \pm 0.6$ \cite{Ghosh:2015gpa} (see \cite{Perez:2015aoa} for slightly older values), 
$\mu_b = 0.90 \pm 0.18^{+0.21}_{-0.19}$ \cite{ATLAS:2017bic} 
and $\mu_\tau  = 0.98 \pm 0.18$ \cite{Sirunyan:2017khh}.
However, much less is known about the couplings of the Higgs boson to fermions of the first two families:
the current bounds found by ATLAS \cite{Aad:2014xva} and CMS \cite{Khachatryan:2014aep}   are 
 $\mu_e \leq 4 \times 10^5$ and $\mu_\mu \leq 7$.
During LHC's High-Luminosity run $\mu_\mu \simeq 1$ might be achievable \cite{CMS:2013xfa}, while the electron
coupling to the Higgs is far below the experimental sensitivity. Here a future $e^+ e^-$ collider could
get close to the SM value \cite{Altmannshofer:2015qra,dEnterria:2016sca}.\\

In this paper we focus on the coupling of the charm quark to the Higgs boson.
Besides a measurement of the exclusive branching ratio $H \to J / \psi \gamma$ \cite{Aad:2015sda}, 
yielding $\kappa_c \leq 220$ \cite{Bodwin:2013gca,Koenig:2015pha,Perez:2015lra, Chisholm:2016fzg} 
inclusive $H \to c \bar{c}$ decays were studied e.g. in 
\cite{Perez:2015aoa,Perez:2015lra,Delaunay:2013pja,Brivio:2015fxa,Carpenter:2016mwd}.\\

A global fit to Higgs signal strengths gives the strongest bound of $\kappa_c \leq 6.2$ \cite{Perez:2015aoa}.
Modifications of the charm Yukawa coupling can occur in different new physics models
\cite{Giudice:2008uua,Botella:2016krk,Harnik:2012pb,Bauer:2015kzy,Altmannshofer:2016zrn,Bishara:2015cha},
it can even be zero \cite{Ghosh:2015gpa}.
Our aim is to develop a strategy that allows to set a direct upper limit on the charm Yukawa coupling.

The improvement
in our bounds on $\mu_c$, derived from inclusive analyses, depends strongly on the c-tagging efficiency 
at the LHC. While dedicated charm tagging algorithms
are relatively new \cite{Atlas:2015}, flavour tagging has been used in the identification of jets derived
from the hadronization of $b$ quarks for more than 20 years, and were employed at the Tevatron for the
discovery of the top quark \cite{Abe:1995hr, Abachi:1994td}.
Two features of the b-mesons are exploited to achieve a good b-tagging performance: 1) the dominance of semileptonic rates when a b-hadron decays and 2) the long life-time of $b$-hadrons. For the latter one can search for 
displaced secondary vertices (decay vertex) of $b$-hadrons with respect to the primary vertex
(interaction point) in a given event. This distance, known as impact parameter, is normally larger
for $b$-hadrons in comparison with that associated with
states obtained from the hadronization of light quarks ($u, d, s$) and gluons. 
A similar approach can be followed for $c$-jets. However, as tagging procedures for $b$-jets and $c$-jets are quite similar, their mutual mis-identification rates are consequently quite large. \\

In general, bottom- or charm-taggers are designed to find jets initiated by individual $b$ or $c$ quarks, allowing for a generic use of these algorithms in a wide range of applications. However, in searches for light or boosted resonances that decay into a charm or bottom pair, such algorithms might not be ideal, as they neglect correlations between the decay products. For example, if the decaying resonance is a colour singlet particle, its decay products are colour connected and soft gluon emissions of either decay product have a preference to be emitted into the cone between the quark pair \cite{Marchesini:1983bm}. Thus to increase the sensitivity in searches for new physics or Higgs boson measurements it can be beneficial to design dedicated 2-prong reconstruction algorithms that allow to utilise more information about the decaying resonances. Observables that are particularly sensitive to the radiation profile of the event are so-called event shape observables \cite{Banfi:2004nk, Banfi:2010xy}, which 
have been proposed as hypothesis-tester in the study of Higgs boson properties \cite{Englert:2012ct, Bernaciak:2012nh}. By promoting those well-studied observables to jet shape observables, applied to a fat jet, they can be used as input to a machine-learning algorithms to separate signal from large QCD backgrounds.\\

In this letter we present a procedure to identify jets initiated by $c\bar{c}$ pairs from Higgs boson decays based on the application of different event shapes and the transverse momenta of leptons ($e^{\pm}$ and $\mu^{\pm}$).
It is expected that high-$p_T$ jets arising from highly-boosted Higgs bosons have a different energy flow in comparison to jets arising from pure QCD backgrounds. We would like to emphasize that in the double tagging 
strategy presented in this work, we study the energy
distribution of the full jet associated with the boosted Higgs bosons decaying into the c and the $\bar{c}$ quark,
without separating the corresponding subjets after hadronization. 
Our analysis is based on fully showered and hadronised Monte Carlo events and the results obtained can be considered as an upper bound to a more complete study when detector effects are also included.\\

The structure of the paper is as follows: In Sec.~\ref{sec:GenandSel} we describe the event generation and the selection 
criteria.
Then in Sec. ~\ref{sec:Atlas_analysis} we present the performance of our approach for the selection of the SM Higgs boson
$H$ against different
sources of background. Using the tagging efficiencies derived from the optimization against QCD c-jets,
we present an upper bound for our sensitivity to $\mathcal{B}r(H\rightarrow c\bar{c})$. In order to evaluate the efficiency of 
the simultaneous double c-tagging identification with strategies based on 
the double application of a single c-tagger, we compare our results with those obtained applying the Atlas single charm 
tagging algorithm  JetFitterCharm. 
Sec.~\ref{sec:THDMA} is devoted to the study of the decay channel 
 $H(\rightarrow A(\rightarrow c \bar{c}) Z)+\hbox{jets}$, with $A$  the THDM  CP odd scalar. 
Finally in Sec.~\ref{sec:concl} we 
conclude. The discussion is complemented with the correlation matrices among the event shapes used
as well as with the distributions for the leading ones in each one of our studies.
A brief description of most of the observables considered is included in the appendix.

\section{\label{sec:GenandSel}Event Generation and Event selection}

The signal channels are $pp\rightarrow H(\rightarrow c\bar{c}) Z$ and 
$pp\rightarrow H(\rightarrow A(\rightarrow c \bar{c}) Z) + \hbox{jets}$. Here
$H$ is the SM Higgs boson and $A$ denotes
the CP odd THDM scalar. As background channel we include $pp\rightarrow Z + \hbox{jets}$.
In all cases we consider $Z\rightarrow l^+l^-$, for $l=e,\mu$. We take into account two possible values for the mass 
of the scalar $A$, $m_A=4~\hbox{GeV}$ and $m_A=10~\hbox{GeV}$. We generate our samples with SHERPA 2.2.1 \cite{Gleisberg:2008ta} at $\sqrt{s}=13.0~\hbox{TeV}$, and include parton shower, hadronization and underlying event contributions. For
the jet reconstruction we use the jet finding package FastJet 3.2.1 \cite{Cacciari:2011ma}. The event selection is performed with
the version 2.4.2 of the RIVET analysis framework \cite{Buckley:2010ar}.\\

Our selection strategy is based on the identification of the Higgs and a $Z$ boson in the highly boosted regime,
when both particles have a large transverse momentum and are back-to-back.  In order to reconstruct
the $Z$ boson we require two isolated leptons $l^+l^-$ (for $l=e,\mu$) with a combined mass
satisfying $80.0~\hbox{GeV}   < m_{ll} < 100.0~\hbox{GeV}$.  A lepton $l$ will be considered isolated if the following inequality 
is satisfied  $E_l/E_R<0.1$, where $E_l$ is the energy of $l$ and $E_R$ is the total energy  inside a cone of radius $R=0.3$ around $l$. The 
identification of the $Z$ boson concludes by imposing a cut $p_{T}>200.0~\hbox{GeV}$ over the combined transverse momentum of the
pair $l^+l^-$. We proceed with the next steps only if the Z boson has been successfully reconstructed as described in this paragraph.\\

A boosted Higgs decaying into a pair
of quarks $q \bar{q}$ produces a jet with a relatively large active area $R_{q \bar{q}}$, and thus is 
 commonly referred to as fat jet.
As a matter of fact, in the boosted regime, the radius
of the jet depends on the mass and the transverse momentum of the Higgs ($m_H$ and $p_{T,H}$) as well as on the momentum
fractions of the quark and the antiquark  (z  and $1-z$) according to  $R_{q \bar{q}}=m_H/(p_ {T, H} \sqrt{z(1-z)} )$.
Thus, for a Higgs boson of mass $m_H\simeq 125~\hbox{GeV}$ and a transverse momentum $p_T\simeq 200~\hbox{GeV}$
decaying symmetrically into a pair charm-anticharm, we expect an angular separation of the Higgs decay products of $R_{c \bar{c}} \simeq 1.25$. In practice we demand jets 
with radius $R=1.2$ and a transverse momentum  $p_T>200~\hbox{GeV}$
reconstructed, with the anti-$k_T$ algorithm and select the Jet with the highest $p_T$. We translate all the constituents of this
jet to the plane $\eta=0$ by taking $p_z=0$ and replacing their total energy by their corresponding transverse energy \cite{Barnard:2016qma}.\\

From NLO-QCD calculations, in the SM the decay fractions of c-quarks into leptons obey with good approximation 
\cite{Lenz:2013}
$\mathcal{B}r(c\rightarrow \bar{l} \nu_l X) =(21.74 \pm 3.90)\%$ and 
$\mathcal{B}r(c\rightarrow  X') =100\% -\mathcal{B}r(c\rightarrow \bar{l} \nu_l X)$ where 
$\bar{l}=\bar{e}, \bar{\mu}$ and $X, X'$ denote quark final states. Hence, if we consider jets originated from the
hadronization process of $c\bar{c}$ pairs, we can  expect to find  0, 1 and 2 leptons with the following probabilities
$61.24\%$, $34.03\%$ and $4.73\%$ respectively. For each one of our analyses we perform three independent studies: 
non-leptonic, single leptonic and double leptonic, 
if zero, one and two non-isolated leptons are found inside the fat jet respectively. A cut in the transverse momentum of the leptons
of  $p_{T,~l}\geq 2.0~\hbox{GeV}$ allow us to reproduce these numbers with good approximation. Nevertheless, we consider
this to be a relatively soft cut, hence in practice we impose the constraint $p_{T,~l}\geq 5.0~\hbox{GeV}$.\\
 
If an event is selected, we probe the substructure of the highest $p_T$ fat jet by applying  a collection of different event shapes on its constituents, thereby promoting the event shapes to jet shapes.
We follow this procedure separately for each one of the leptonic categories introduced in the previous paragraph. To evaluate the signal efficiency and mis-tag rate of our observables
we use a multivariate analysis implemented in the TMVA package \cite{Hocker:2007ht} and consider a  Boosted Decision Tree (BDT) as our classifier.
In addition to the event shapes and for the single-leptonic and double-leptonic categories, we also include the value of the  transverse 
momentum of the highest $p_T$ light-lepton found inside the selected fat jet.

\section{\label{sec:Atlas_analysis} Standard Model Higgs  $c\bar{c}$-tagging using Event Shapes}

\subsection{Performance}
\begin{center}
\begin{figure*}
\includegraphics[height=8.0cm]{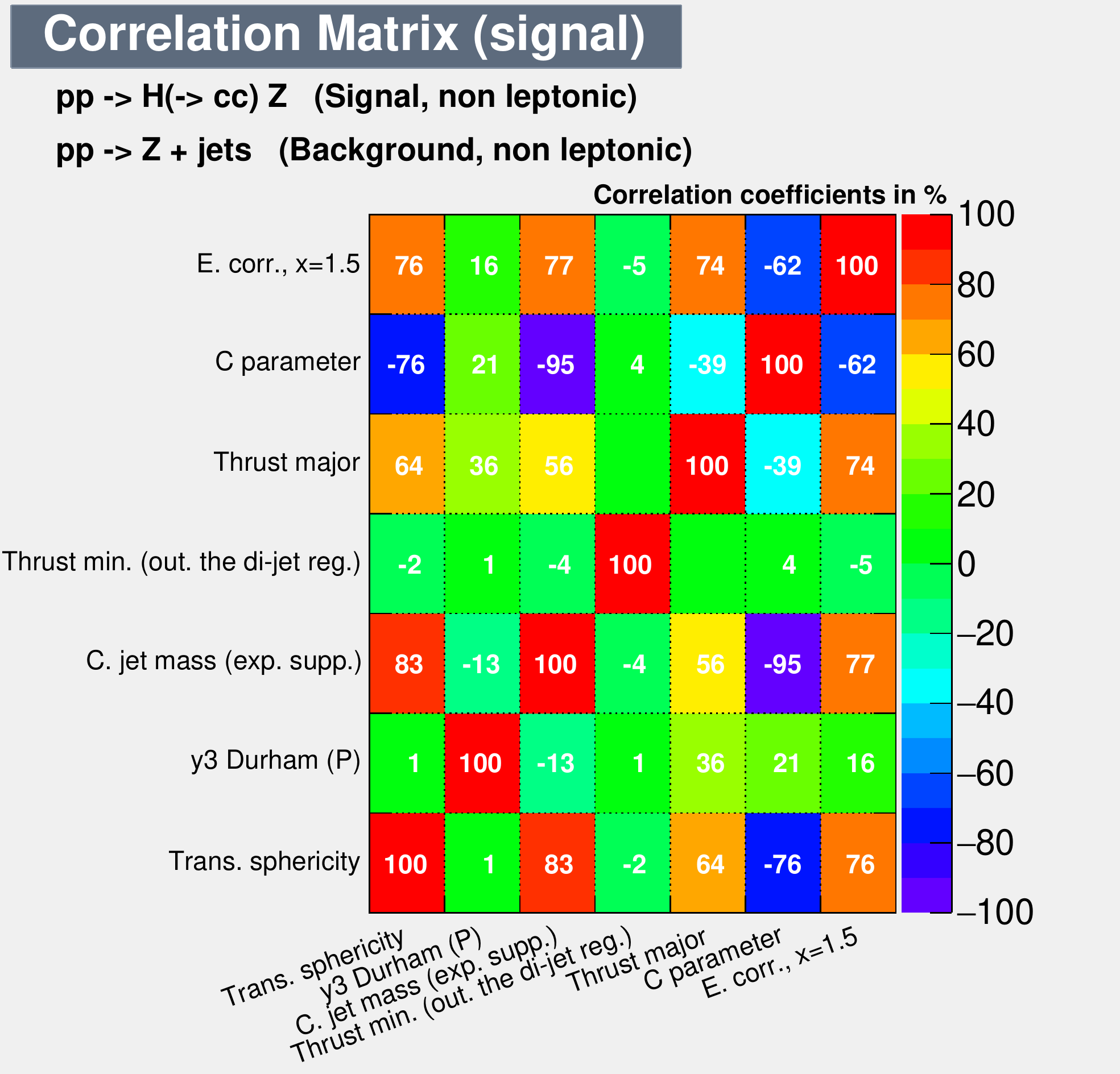}
\includegraphics[height=8.0cm]{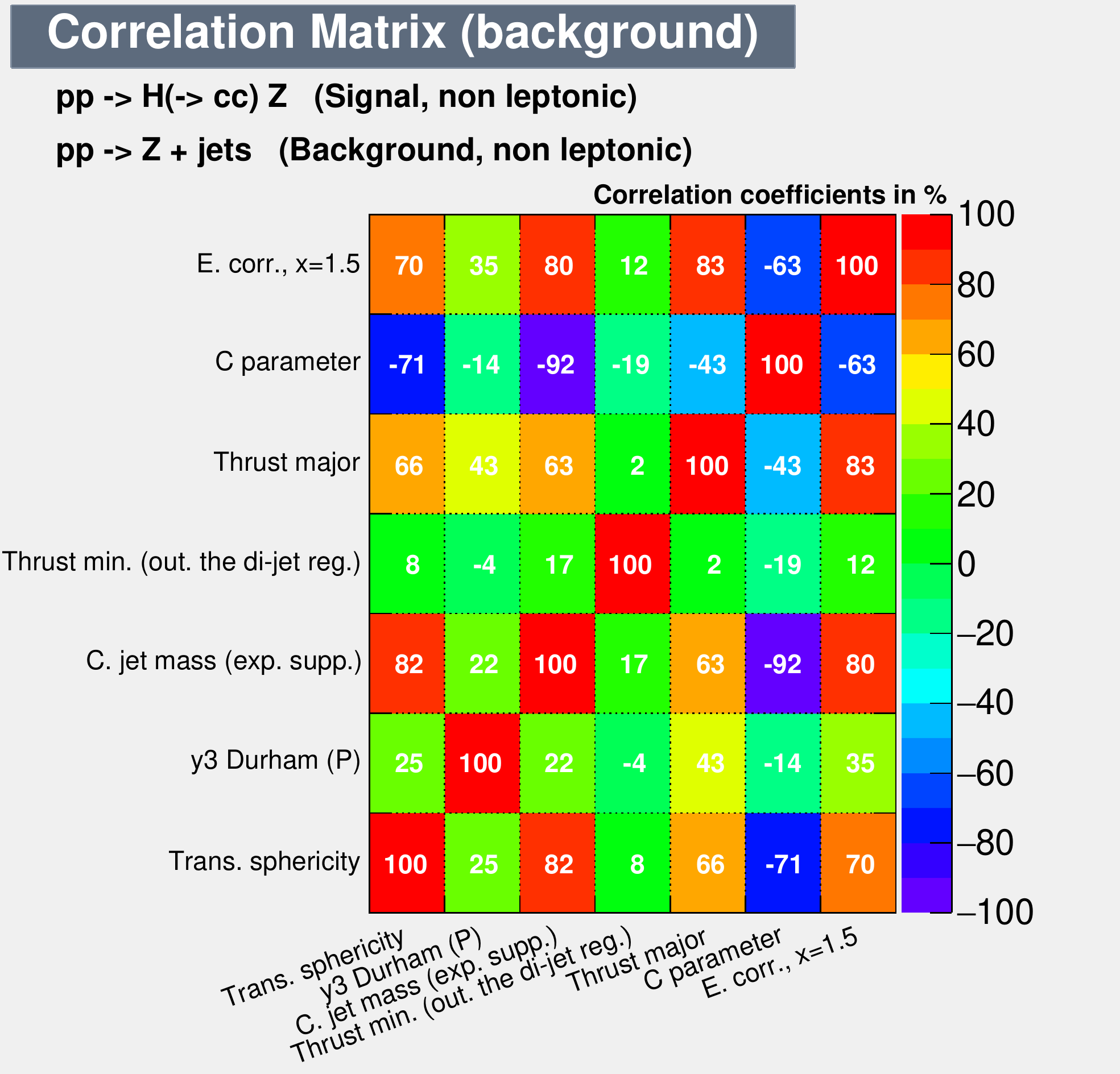}
\includegraphics[height=8.0cm]{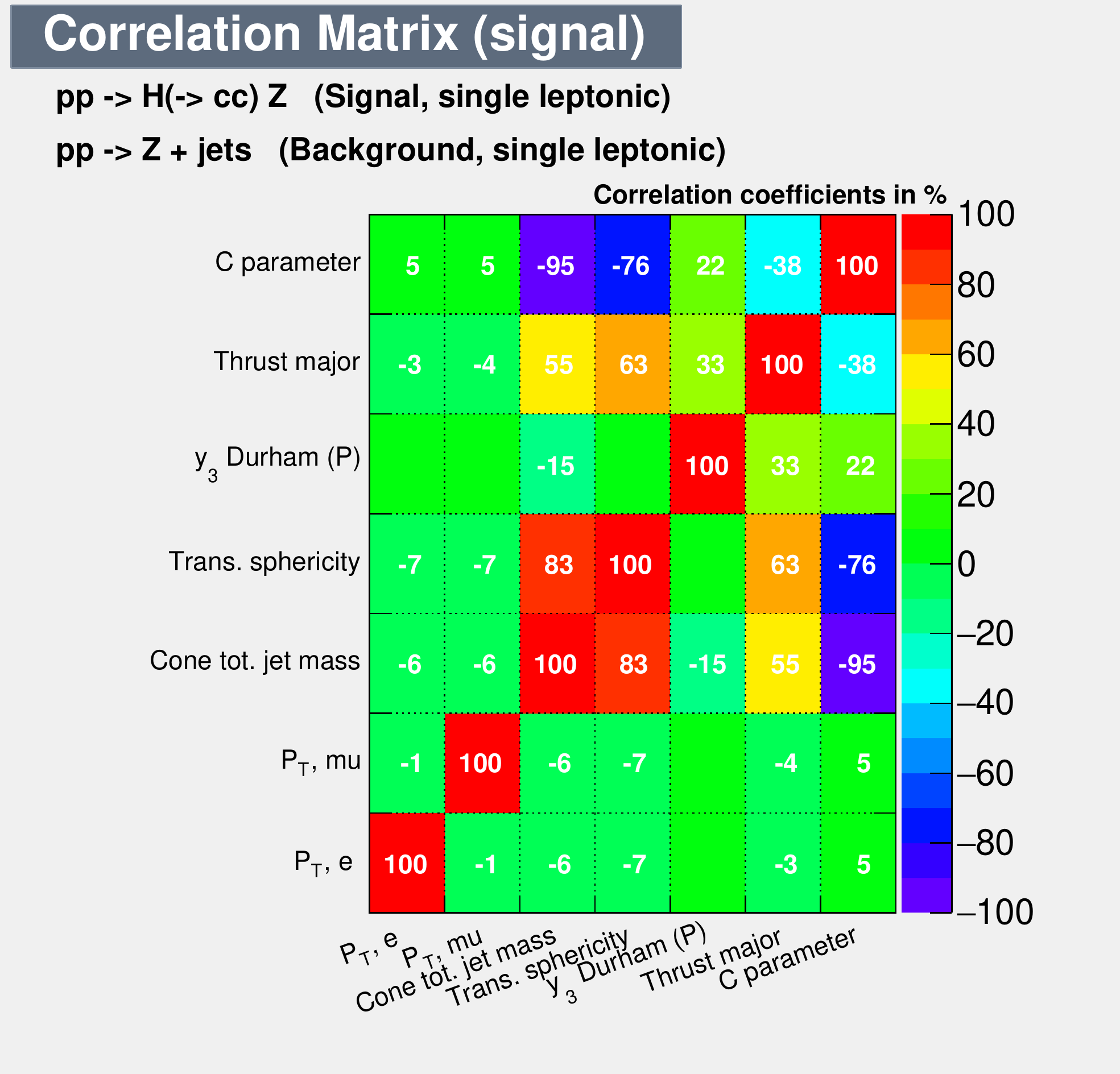}
\includegraphics[height=8.0cm]{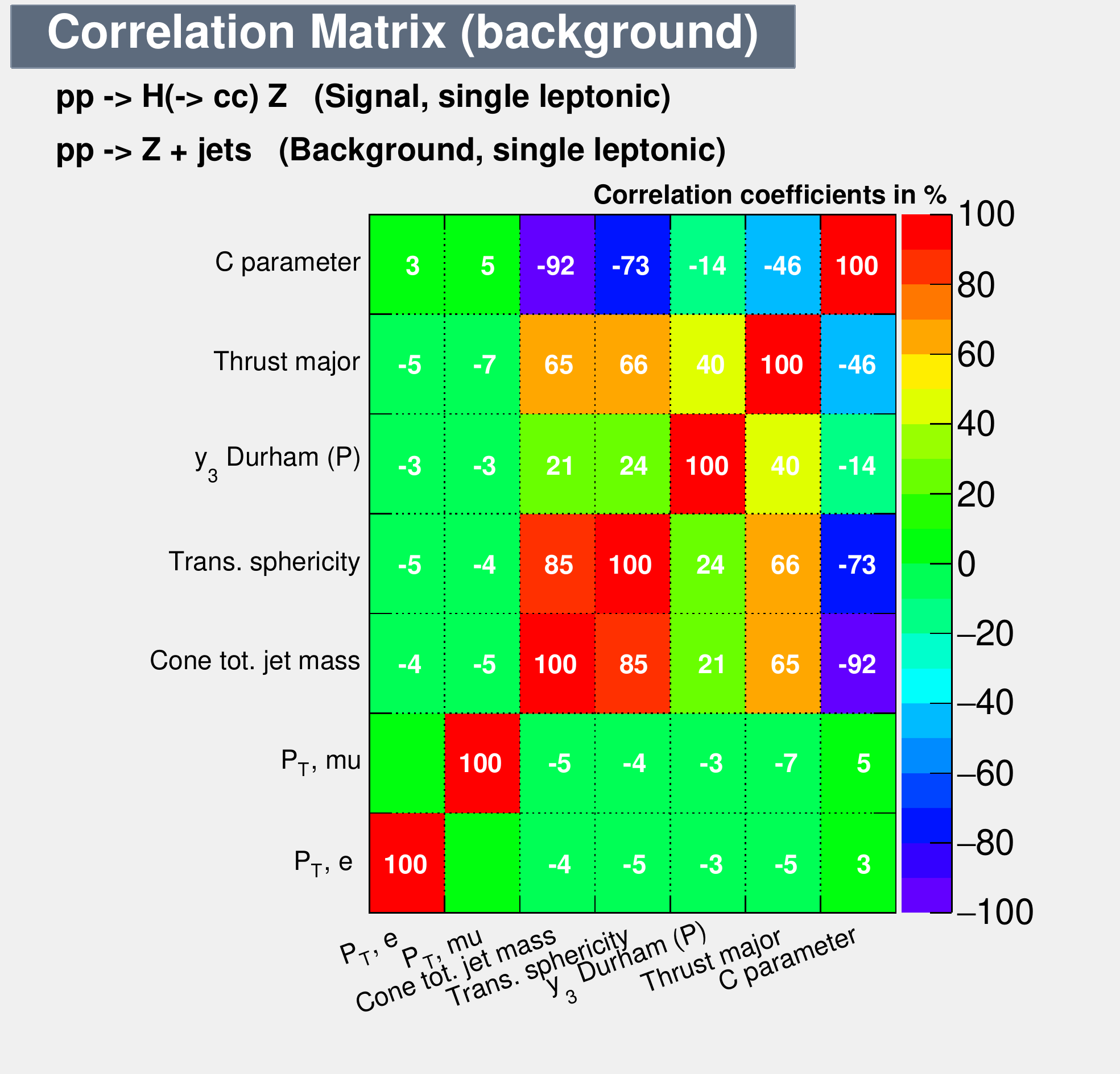}
\includegraphics[height=8.0cm]{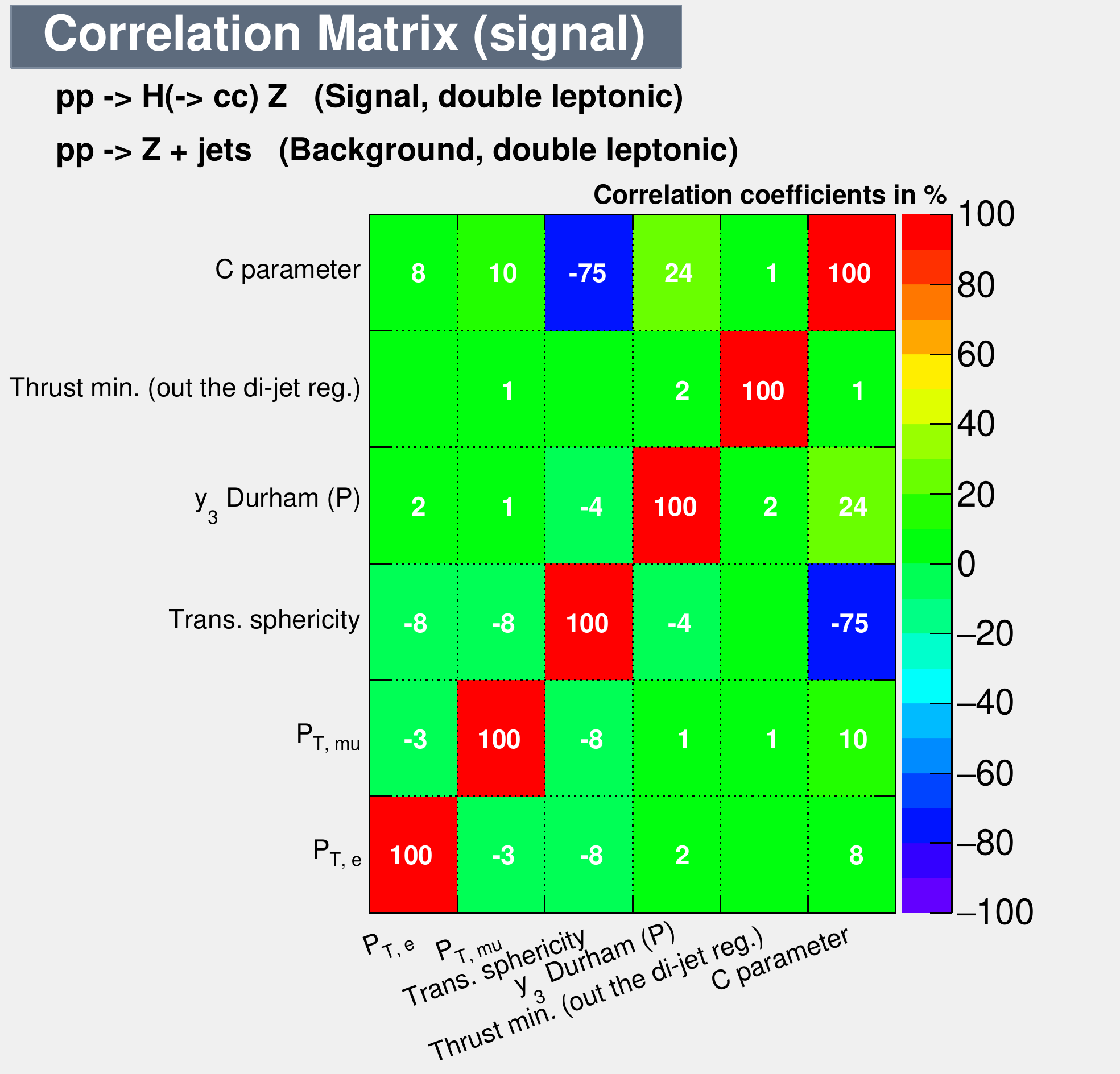}
\includegraphics[height=8.0cm]{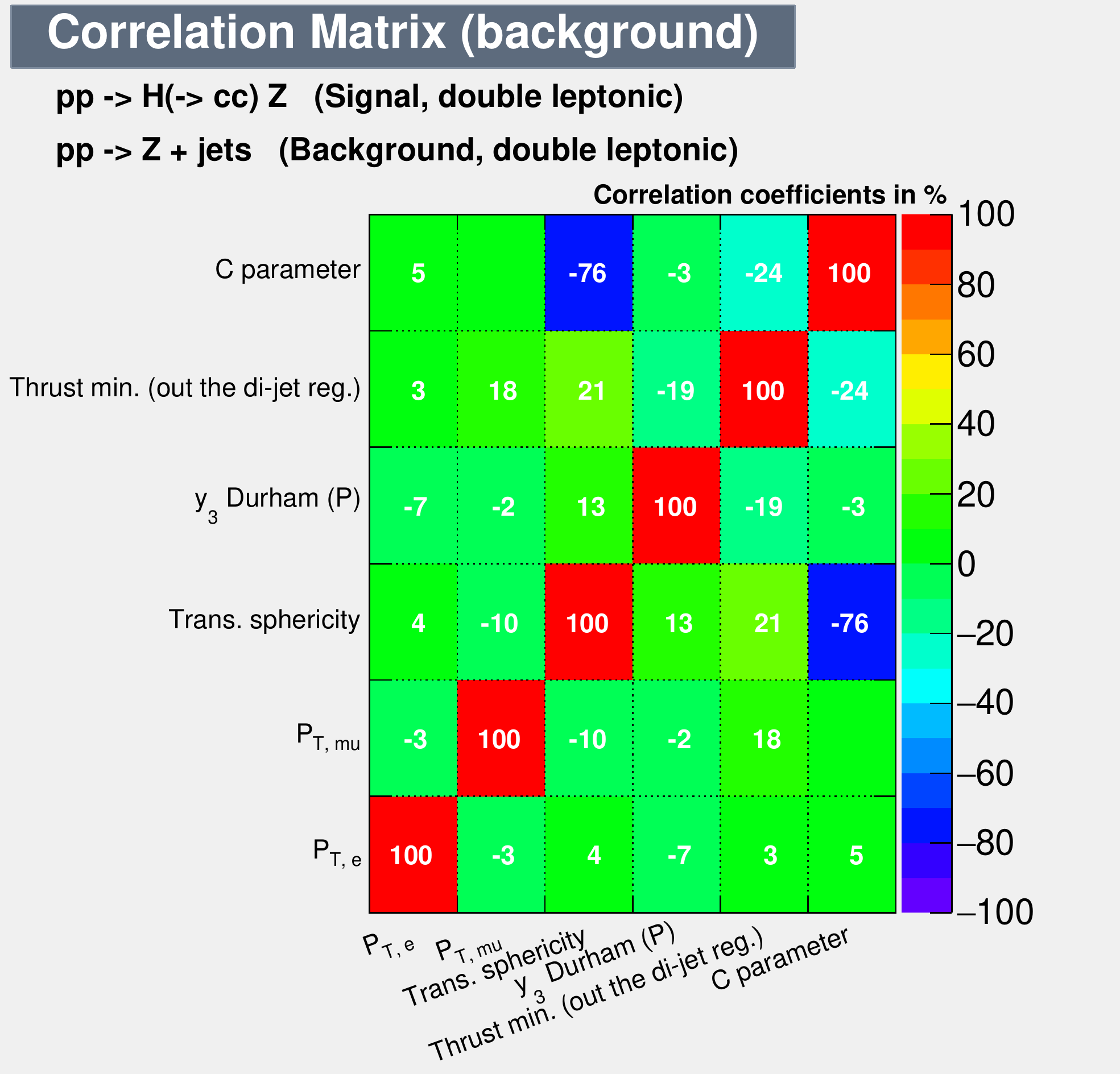}
\caption{Correlation coefficients for the observables used in the analysis $pp\rightarrow H(\rightarrow c \bar{c}) Z$ (signal) 
vs $pp\rightarrow Z + \hbox{jets}$
(background).}
\label{fig:CorrelationSMZQCDjets}
\end{figure*}
\end{center}

\begin{figure*}
\includegraphics[height=5cm]{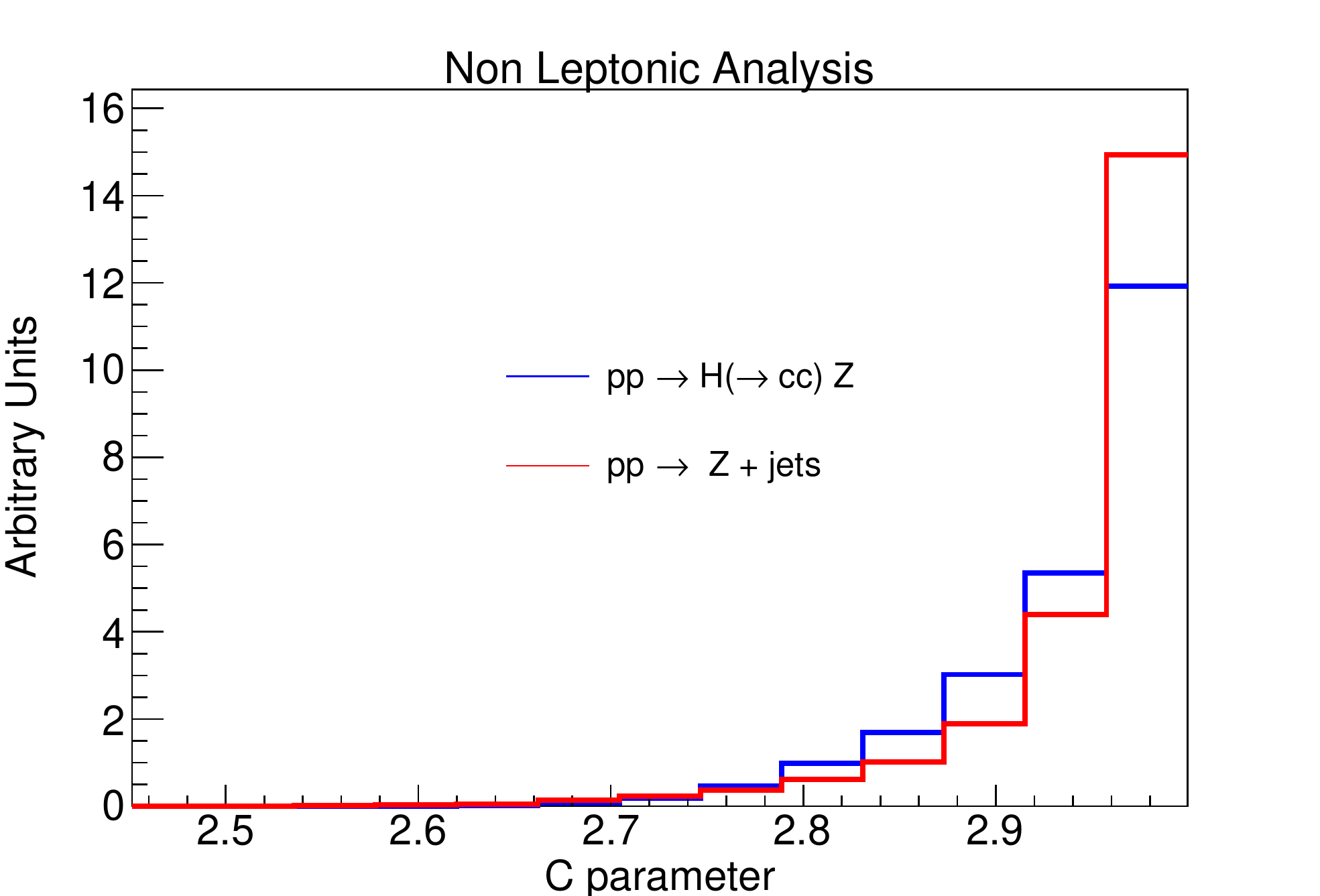}
\includegraphics[height=5cm]{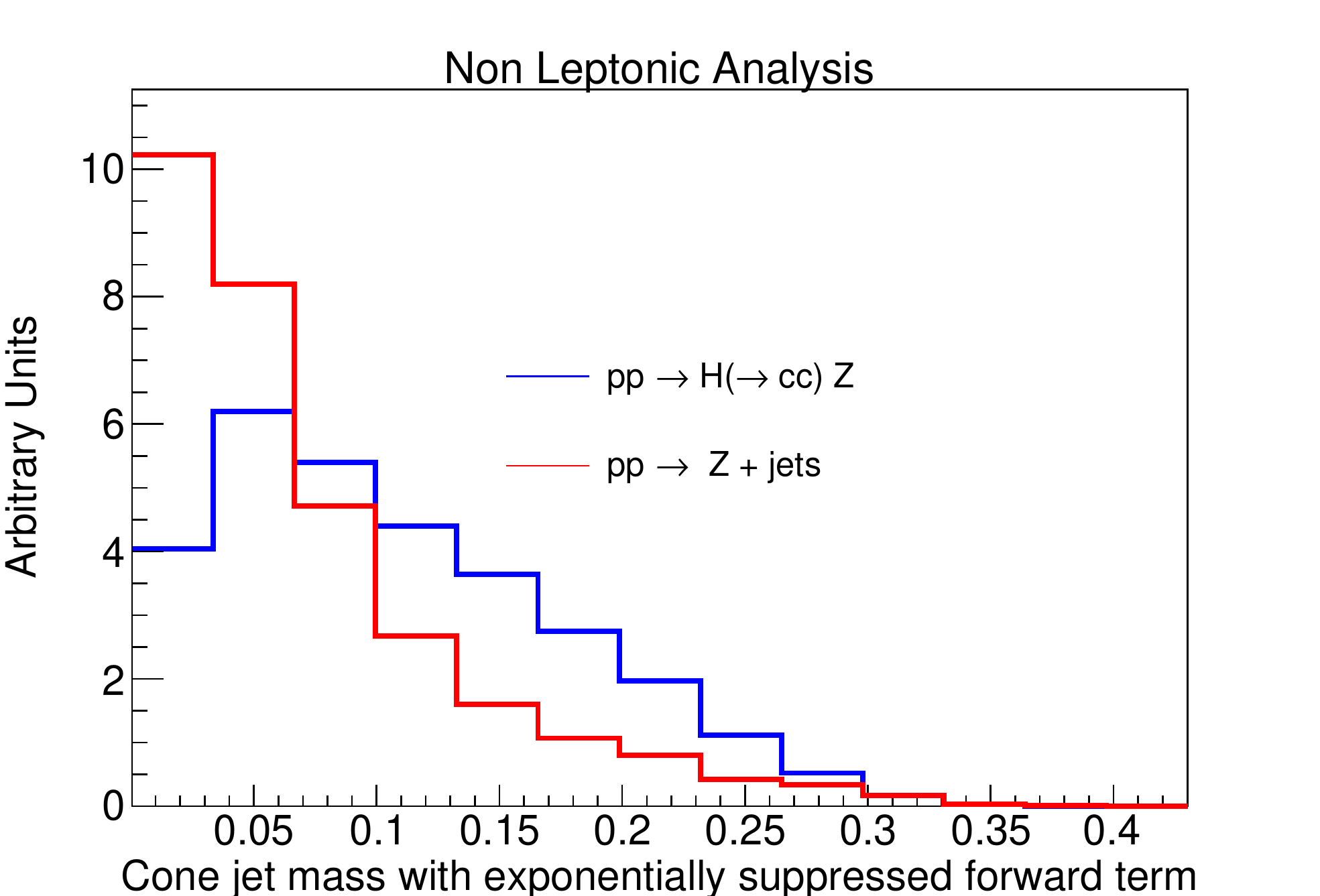}
\includegraphics[height=5cm]{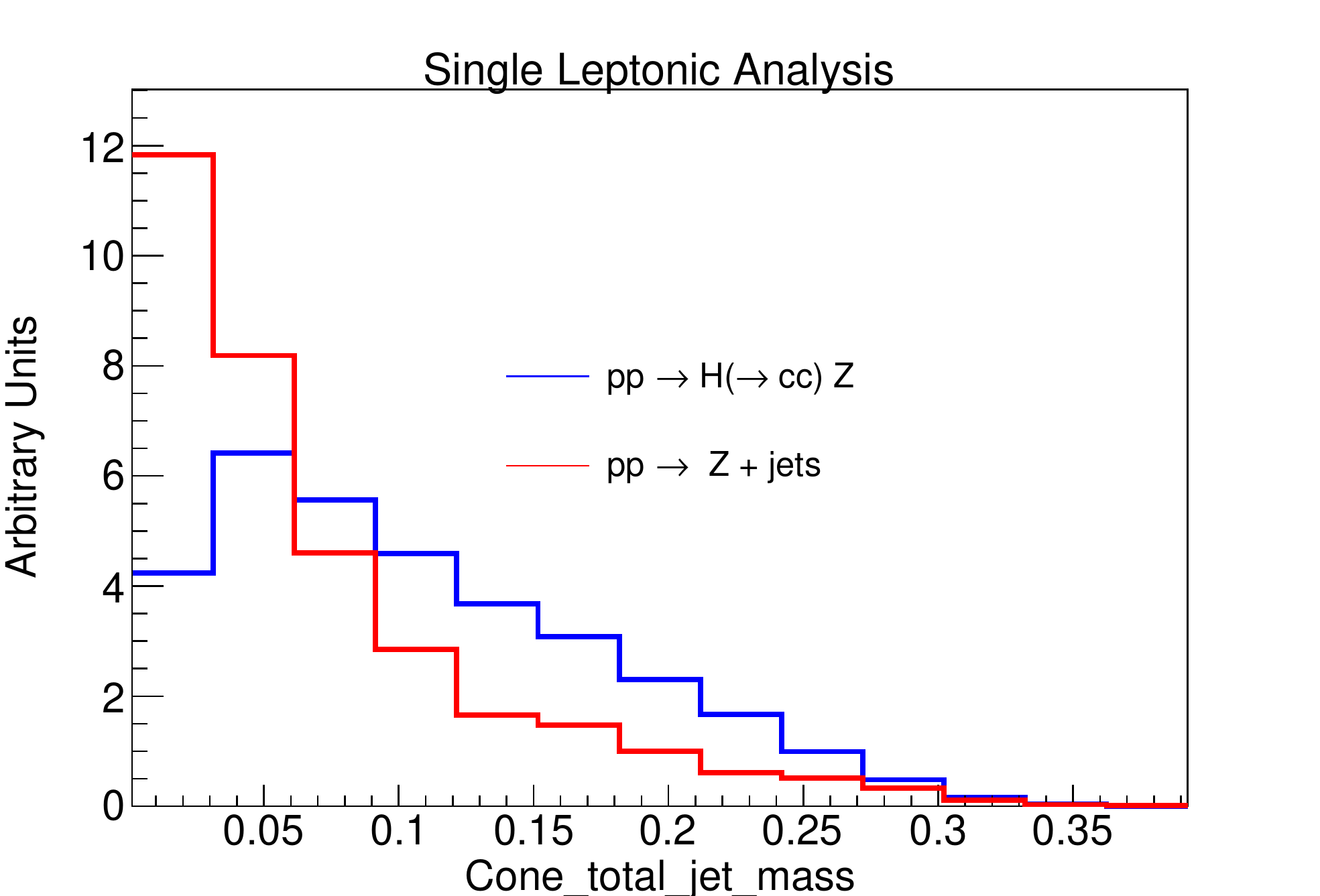}
\includegraphics[height=5cm]{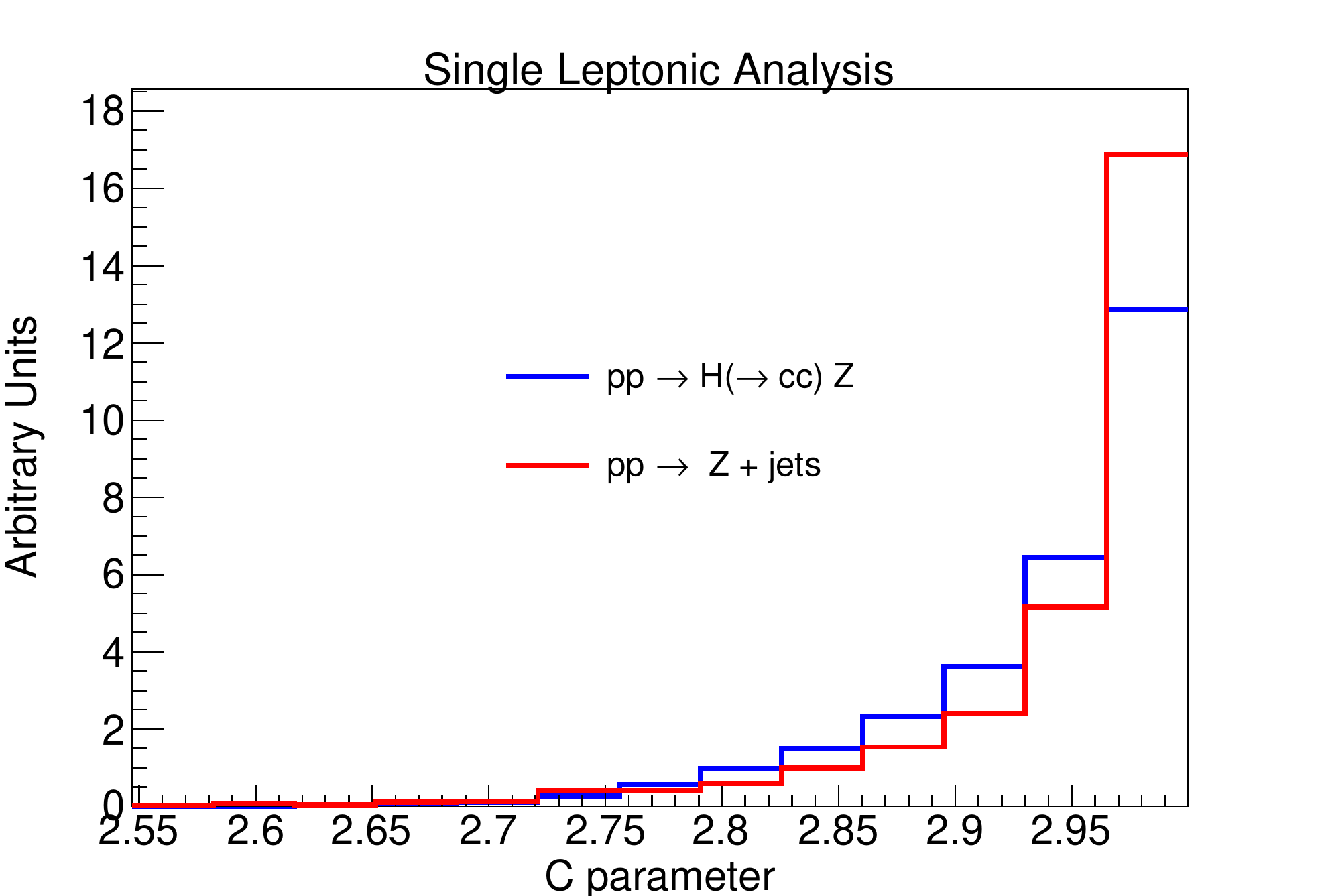}
\includegraphics[height=5cm]{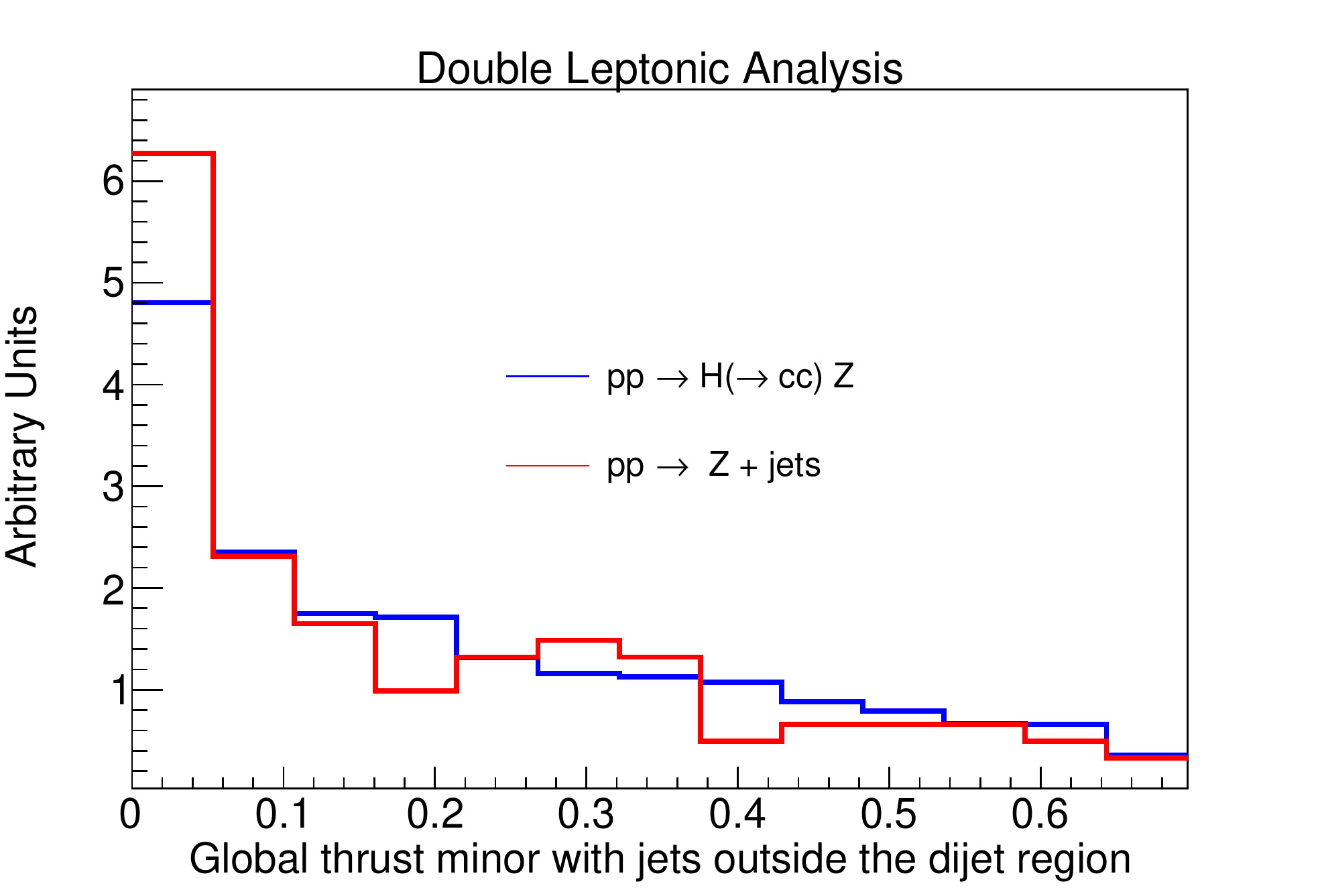}
\includegraphics[height=5cm]{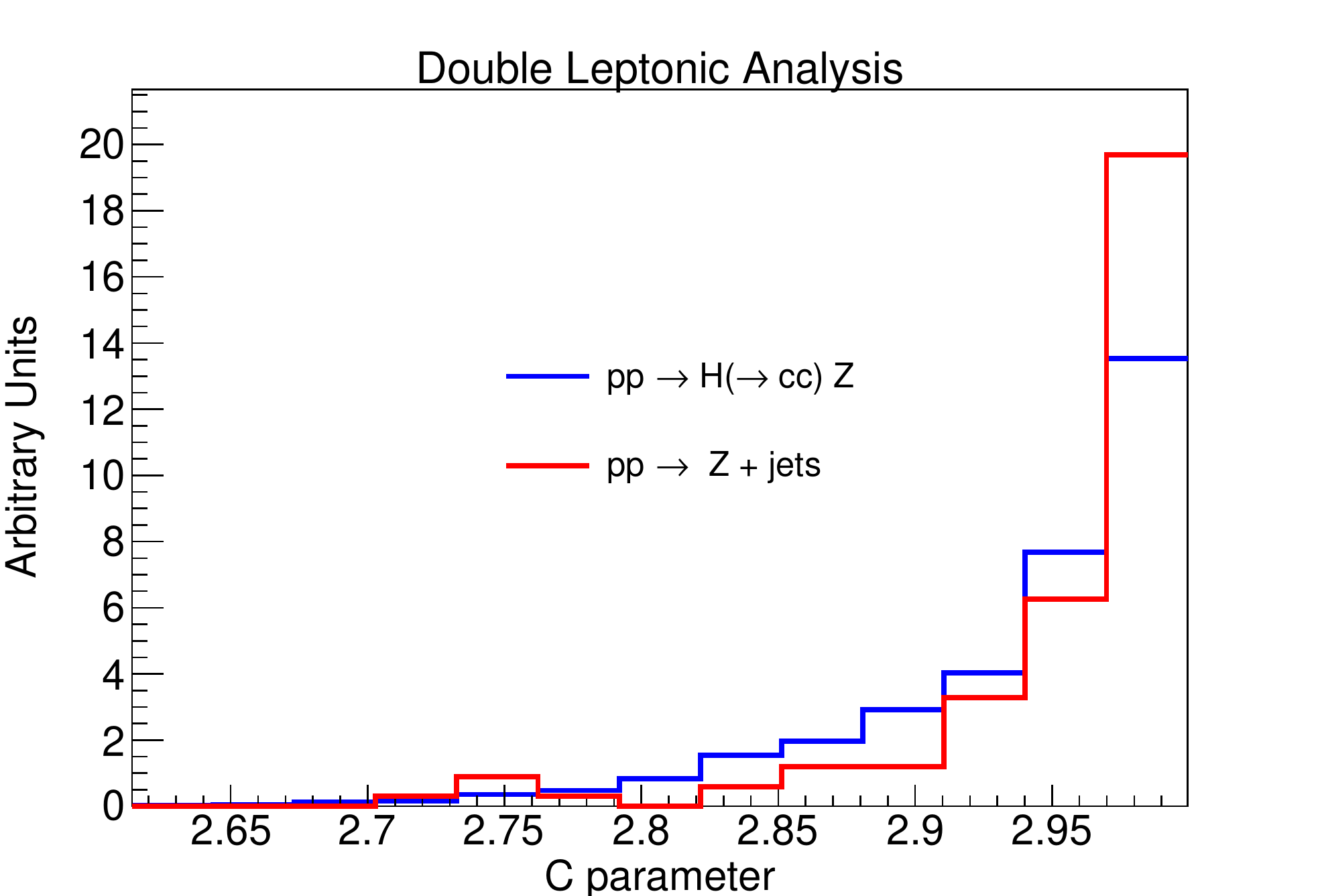}
\caption{Histograms for the main discrimination event shapes when selecting $pp\rightarrow H (\rightarrow c\bar{c})Z$ against
$pp\rightarrow  Z +  \hbox{jets}$.}
\label{fig:HistoObsZjets}
\end{figure*}

We begin by obtaining the performance of our strategy when selecting the signal channel 
$pp\rightarrow H(\rightarrow c \bar{c} ) Z$ against $pp\rightarrow Z + \hbox{jets}$. The set of observables that give us 
the best performance are presented in Table \ref{Tab:ZQCDjets} and the correlations among them are shown in Figs. 
(\ref{fig:CorrelationSMZQCDjets}), additionally we provide the distributions 
for the top two discriminating
observables in each one of the leptonic categories in Fig.~\ref{fig:HistoObsZjets}.\\

\begin{figure}
\includegraphics[height=5cm]{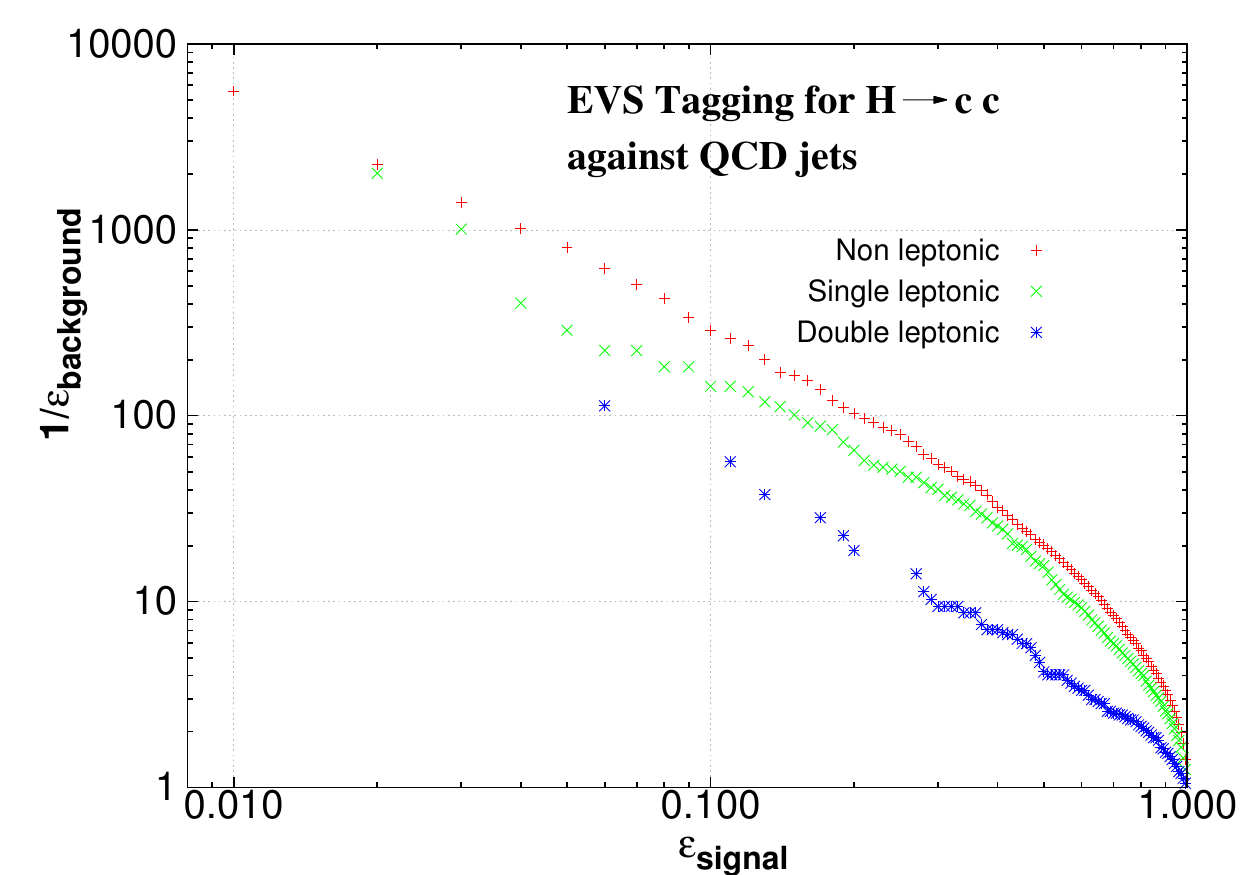}
\caption{\label{fig:SMTagger}Double $c$-jets selection efficiency (from $H\rightarrow c\bar{c}$) against the QCD-jets rejection
achieved by the Event-Shapes tagger. The curves are obtained from independent optimizations considering different subsamples
with zero (non leptonic), one (single leptonic)  and two (double leptonic) leptons inside the highest transverse momentum fat jet.}
\end{figure}

Our curves for the signal selection efficiencies as well as our
background 
fake rates in each one of the leptonic studies are shown in Fig.~\ref{fig:SMTagger}. We can combine the three leptonic studies
to obtain a single selection efficiency for signal (S) and for background (B) according to the formula
\begin{eqnarray}
\varepsilon^{Tot.}_{S./B.}&=&\varepsilon^{(0)}_{S./B.}\times f^{(0)}_{S./B.}
 + \varepsilon^{(1)}_{S./B.}\times f^{(1)}_{S./B.}\nonumber\\
&& + \varepsilon^{(2)}_{S./B.}\times f^{(2)}_{S./B.}.
\label{eq:combinationleptonic}
\end{eqnarray}

Our optimal point after the combination of the different leptonic categories corresponds to
\begin{eqnarray}
\varepsilon_{ c \bar{c}}=0.40 &&\varepsilon_{QCD, jets}=0.03
\end{eqnarray}

obtained from the following partial efficiencies
\begin{eqnarray}
\varepsilon^{(0)}_{S.}=0.37,&\varepsilon^{(1)}_{S.}=0.49,&\varepsilon^{(2)}_{S.}=0.19\nonumber\\
\varepsilon^{(0)}_{B.}=0.03,&\varepsilon^{(1)}_{B.}=0.06,&\varepsilon^{(2)}_{B.}=0.04
\end{eqnarray}

and the leptonic fractions shown in Table \ref{Tab:leptonic_fractions}.
Our branching ratio for the process $H\rightarrow c\bar{c}$ is then $Br(H\rightarrow c\bar{c})=6.1\%$ leading to the 
cross section $\sigma_{pp\rightarrow H(\rightarrow c \bar{c} ) Z}=0.08\hbox{ fb}$. In the case of background
the corresponding cross section is $\sigma_{pp\rightarrow Z + \hbox{jets}}=23.56\hbox{ fb}$. Based on these results
and considering the integrated luminosity $\int\mathcal{L}dt=3000\hbox{ fb}^{-1}$ we
can verify the 2 sigma condition for the significance $\mathcal{S}/\sqrt{\mathcal{B}}=2.0$.

\begin{table}
\begin{center}
\begin{tabular}{|c|c|c|}
\hline
\multicolumn{3}{|c|}{\bf{$Z + \hbox{jets}$} }\\
\hline
\hline
\bf{Non leptonic} & \bf{Single leptonic} & \bf{Double leptonic} \\
\hline
\hline
C parameter & Cone total & Global thrust minor \\
 & jet mass &defined with jets\\
&&outside the dijet region\\
&&dijet region\\
\hline
Cone heavy& C parameter  & C parameter\\
jet mass with& & \\
exponentially &  & \\
suppressed&  & \\
forward term&  & \\
\hline
Thrust major & Thrust major   & $P_{T, e}$\\
\hline
3-jet resolution &   3-jet resolution  & 3-jet resolution\\
$y_3$ Durham& $y_3$ Durham &$y_3$ Durham  \\
(P-scheme)&(P-scheme) & (P-scheme)\\
\hline
Fractional energy & Transverse & Transverse\\
correlation $x=1.5$ &sphericity  & sphericity\\
\hline
Global thrust minor&$P_{T, \mu}$& $P_{T, \mu}$  \\
defined with jets &&\\
outside the &&\\
 dijet region&&\\
\hline
Transverse  &$P_{T, e}$&  \\
sphericity&& \\
\hline
\end{tabular}
\caption{Top observables determined by the Multivariate Analysis to
discriminate the process $ pp \rightarrow H (\rightarrow c\bar{c}) Z$  against $pp \rightarrow Z + \hbox{jets}$.
}
\label{Tab:ZQCDjets}
\end{center}
\end{table}

\begin{table}
\begin{center}
\begin{tabular}{|c|c|c|}
\hline
\bf{Fraction of events} & $pp\rightarrow H(\rightarrow c \bar{c} ) Z$ & $pp\rightarrow Z + \hbox{jets}$ \\
\hline
$f^{(0)}$ (0 leptons) & $73.62\%$ & $84.0\%$   \\
\hline
$f^{(1)}$ (1 lepton)& $24.47\%$ & $15.15\%$ \\
\hline
$f^{(2)}$ (2 leptons)& $1.90\%$ & $0.85\%$ \\
\hline
\end{tabular}
\caption{Fraction of events in each one of the leptonic categories for the samples $pp\rightarrow H(\rightarrow c \bar{c} ) Z$ and
$pp\rightarrow Z + \hbox{jets}$ after the selection cuts.}
\label{Tab:leptonic_fractions}
\end{center}
\end{table}

\subsection{Comparison against the ATLAS JetFitterCharm algorithm}

In order to evaluate the performance of our double charm identification approach against more conventional life-time
based single charm tagging procedures, we provide a ``naive'' comparison with the ATLAS JetFitterCharm algorithm \cite{Atlas:2015}.
In the ATLAS study two main sources of backgrounds are considered, the first one are light-flavor jets, i.e. jets arising from the hadronization
of $g, u, d, s, \bar{u}, \bar{d}, \bar{s}$; the second background is heavy-flavor jets, in this context $b$-jets. \\

From \cite{Atlas:2015} we extract the JetFitterCharm single selection
efficiencies $\epsilon_c$, $\epsilon_b$ and  $\epsilon_{light}$ for the charm-jets, b-jets  and light-jets respectively.
The double tagging coefficients for each category are calculated as  $\varepsilon_{2c}=\epsilon^2_c$,  $\varepsilon_{2b}=\epsilon^2_b$ and
$\varepsilon_{2~light}=\epsilon^2_{light}$.\\

For the comparison of the different tagging strategies, we used the boosted Higgs search described in  Section \ref{sec:GenandSel}, where the dominant backgrounds are light-flavor-jets $+ Z$ 
and $b\bar{b}$ jets $+ Z$. As the JetFitterCharm efficiencies are not provided in terms of separate analyses for the different leptonic
categories introduced in Sec \ref{sec:GenandSel}, we combine the selection efficiencies achieved in our approach for the non-leptonic, single-leptonic and double-leptonic studies
for a given background according to Eq. (\ref{eq:combinationleptonic}).\\

\begin{figure}
\includegraphics[height=5cm]{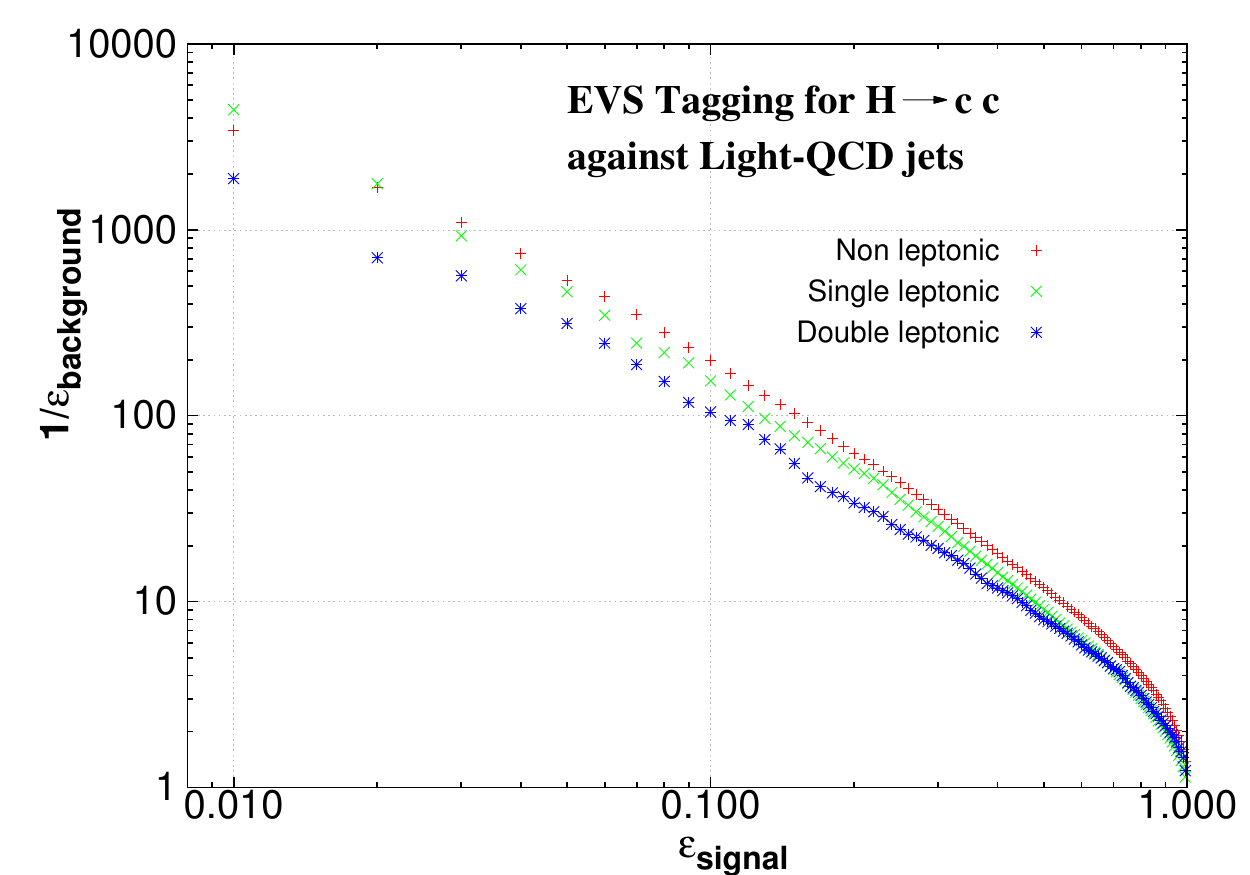}
\caption{\label{fig:LightvsAtlas}Double $c$-jets selection efficiency against the double light jets rejection
achieved by the Event-Shapes tagger per leptonic study.}
\end{figure}

\begin{figure}
\includegraphics[height=5cm]{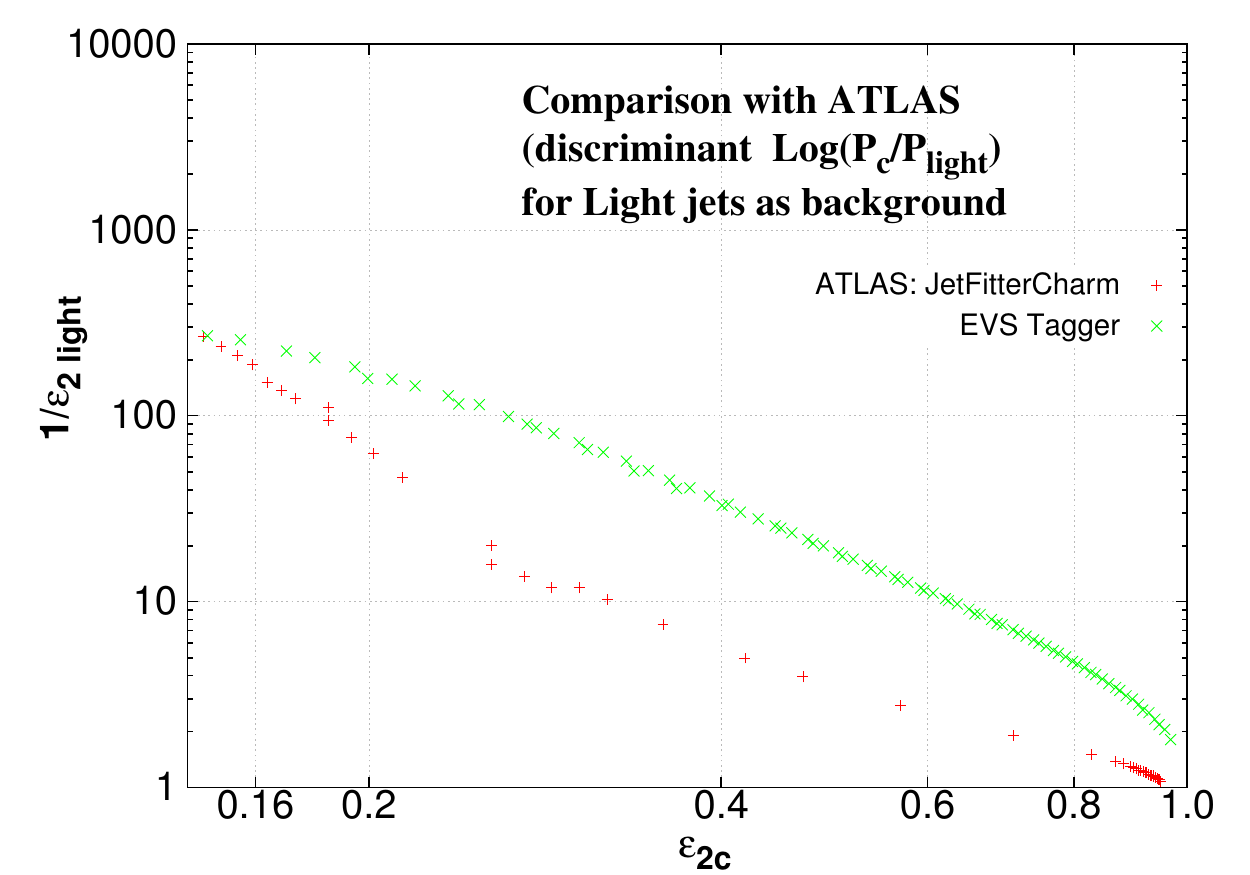}
\caption{\label{fig:LightvsAtlas2}Double $c$-jets selection efficiency against the double light jets rejection
achieved by the ATLAS JetFitterCharm tagger and the combined Event-Shapes tagger.}
\end{figure}

We find the best results in rejecting light-flavor jets, which have in this analysis a cross section that is at least an order of magnitude bigger than the $b\bar{b}$ background. In Fig.~\ref{fig:LightvsAtlas} we show the ROC curves for 
the different leptonic analyses and in Fig.~\ref{fig:LightvsAtlas2} we present the performance obtained from the combination
of the leptonic categories. Without access to the ATLAS detector simulation a direct comparison between the two approaches is not feasible. However, it can be inferred from Fig.~\ref{fig:LightvsAtlas2} that for $0.16 > \varepsilon_{2c}$ the jet-shapes strategy shows a strong performance and is likely to add to the tagging strategy employed by ATLAS. Consequently, using event shapes it is possible to outperform the double application of a charm-tagger by a single application of a double-charm tagger. This is achieved by looking at the full radiation profile inside a fat jet; without disentangling the radiation signatures of
the $c$-quark and the $\bar{c}$-quark independently.

\begin{table}
\begin{center}
\begin{tabular}{|c|c|c|}
\hline
\multicolumn{3}{|c|}{\bf{Light quark Jets} }\\
\hline
\hline
\bf{Non leptonic} & \bf{Single leptonic} & \bf{Double leptonic} \\
\hline
\hline
Fox Wolfram-like  &Fractional energy& Fractional energy   \\
$n=1/4$  &correlation $x=1.5$   &correlation $x=1.5$ \\
\hline
3-jet resolution &C parameter&$P_{T,e}$ \\
$y_3$ Durham     &           &         \\
(P-scheme)&&\\
\hline
C parameter  &3-jet resolution  &3-jet resolution  \\
             &$y_3$  Jade            &$y_3$  Jade \\
             & (E-scheme)            &  (E-scheme)  \\
\hline
Thrust of          &4-jet resolution & 3-jet resolution \\
$e^{-|\eta|}$             &$y_4$  Durham        &  $y_3$  Geneva   \\
momenta                &(P-scheme)&  (P-scheme) \\
\hline
3-jet resolution  & $P_{T,\mu}$   &C parameter \\
 $y_3$ Jade & &\\
(E-scheme)   & & \\
\hline
3-jet resolution  &$P_{T,e}$ & 3-jet resolution \\
$y_3$ Geneva              &          & $y_3$ Durham                  \\
(P-scheme)  && (P-scheme)\\
\hline
3-jet resolution   &3-jet resolution& $P_{T,\mu}$  \\
$y_3$ Jade  &$y_3$  Geneva& \\
(E0-scheme)  &(P-scheme) & \\
\hline
\end{tabular}
\caption{Top observables determined by the Multivariate Analysis to
discriminate the process $ pp \rightarrow H (\rightarrow c\bar{c}) Z$  against $pp \rightarrow Z + \hbox{light jets}$.}
\label{Tab:Z_jets}
\end{center}
\end{table}

\section{\label{sec:THDMA}CP odd THDM Scalar}

\begin{figure}
\includegraphics[height=5cm]{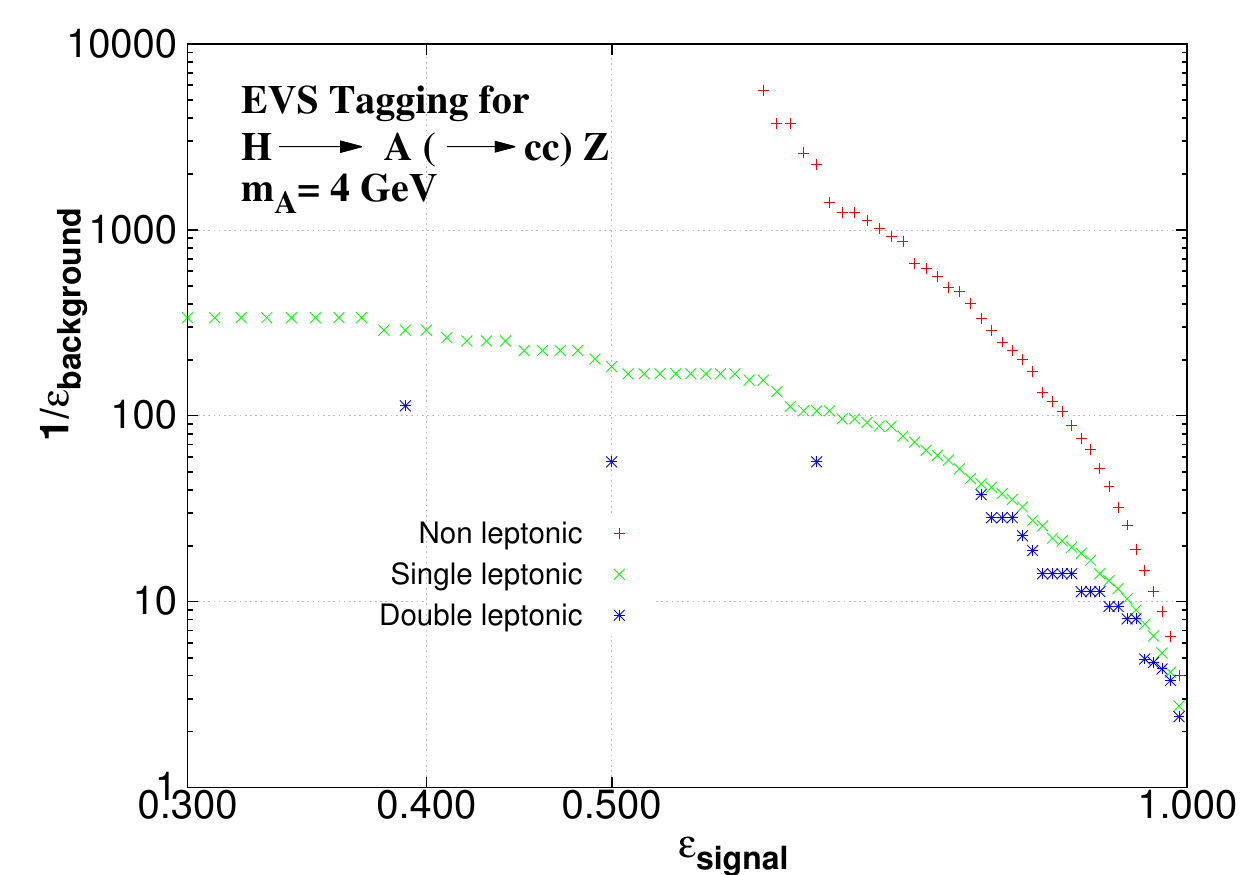}
\caption{\label{fig:A4ROC}Double $c$-jets selection efficiency (from $A\rightarrow c\bar{c}$) against the QCD-jets rejection
achieved by the Event-Shapes tagger per leptonic study. Here we are considering $m_A=4~\hbox{GeV}$.}
\end{figure}

The coupling between the CP odd THDM scalar $A$ and the pair $c\bar{c}$ is directly proportional 
to the charm quark mass  $m_c$ and inversely proportional to the THDM vacuum ratio $\tan\beta$. As shown in \cite{Dermisek:2010mg}, 
the decay channel
$A\rightarrow c \bar{c}$ is expected to be dominant for $4.0~\hbox{GeV} \lesssim  m_{A} \lesssim  10.0~\hbox{GeV}$ and low values of
$\tan\beta$.  Here we determine a  $95\%$ C.L. upper bound for the branching ratio 
$\mathcal{B}r(H\rightarrow A (\rightarrow c \bar{c}) Z)$ in this mass range. 
 Our signal is the process 
$p p \rightarrow H (\rightarrow A(\rightarrow c \bar{c}) Z) + \hbox{jets}$ and our background is given by
$p p \rightarrow Z + \hbox{jets}$. For $m_A=4.0~\hbox{GeV}$ the combination of observables that give the best performance are presented 
in Table \ref{Tab:ccA4}, from here the ROC's corresponding to the different leptonic categories are determined, 
see Fig.~\ref{fig:A4ROC}.
The optimal selection efficiency point is

\begin{eqnarray}
\varepsilon_{c\bar{c}, m_A=4~{\tiny \hbox{GeV}}}=0.81 && \varepsilon_{ {\tiny \hbox{QCD}}, {\tiny \hbox{jets}}}=0.01
\end{eqnarray}

resulting from the efficiencies

\begin{eqnarray}
\varepsilon^{(0)}_{S.}=0.83,&\varepsilon^{(1)}_{S.}=0.69,&\varepsilon^{(2)}_{S.}=0.39\nonumber\\
\varepsilon^{(0)}_{B.}=0.01,&\varepsilon^{(1)}_{B.}=0.01,&\varepsilon^{(2)}_{B.}=0.01
\end{eqnarray}

combined with the leptonic fractions presented in Table \ref{Tab:leptonic_fractions_mA_4} as given in Eq. (\ref{eq:combinationleptonic}). 
Thus, we get the following $95\%$ C.L. upper
 limit for the branching ratio
$\mathcal{B}r(H\rightarrow A (\rightarrow c\bar{c}) Z)<0.01\%$, leading to the cross section for the signal process
$\sigma_{p p \rightarrow H (\rightarrow A (\rightarrow c \bar{c}) Z) + \hbox{jets} }=0.02~\hbox{fb}$. For comparison,
using track-based substructure observables and considering  $m_{A}=4.0~\hbox{GeV}$, the $95\%$ C.L. bound
$\mathcal{B}r(H\rightarrow A (\rightarrow c \bar{c}) Z)\leq 2.1\%$ has been previously  determined in \cite{Chisholm:2016fzg}.

\begin{table}
\begin{center}
\begin{tabular}{|c|c|c|}
\hline
\multicolumn{3}{|c|}{\bf{CP odd THDM scalar A (4 GeV)} }\\
\hline
\hline
\bf{Non leptonic} & \bf{Single leptonic} & \bf{Double leptonic} \\
\hline
\hline
Transverse &Transverse &Transverse \\
spherocity &spherocity &spherocity  \\
\hline
Fractional energy&Fox Wolfram-like & Fractional energy \\
correlation $x=1.5$&$n=1/4$ & correlation $x=1.5$ \\
\hline
Thrust major&Thrust major &4-jet resolution\\
&&$y_4$  Durham \\
&& (P-scheme)\\
\hline
C parameter&3-jet resolution  &3-jet resolution\\
 &$y_3$  Jade &$y_3$ Jade \\
& (E-scheme)& (E-scheme)\\
\hline
Cone $y_3$                      &3-jet resolution  &$P_{T, \mu}$\\
($k_t$, $\Delta R$, $E$-scheme) & $y_3$ Durham &\\
& (P-scheme)&\\
\hline
3-jet resolution & $P_{T, e}$  & $P_{T, e}$ \\
$y_3$ Jade &  &  \\
(P-scheme) &  &  \\
\hline
3-jet resolution &$P_{T, \mu}$ & 3-jet resolution\\
 $y_3$ Jade &&  $y_3$ Durham\\
(E-scheme)&&(E0-scheme)\\
\hline
\end{tabular}
\caption{Top observables determined by the Multivariate Analysis to
discriminate the process $ pp \rightarrow H (\rightarrow A (\rightarrow c\bar{c} ) Z ) + \hbox{jets}$  against $pp \rightarrow Z + \hbox{jets}$
for $m_A=4~\hbox{GeV}$.
}
\label{Tab:ccA4}
\end{center}
\end{table}

\begin{figure}
\includegraphics[height=5cm]{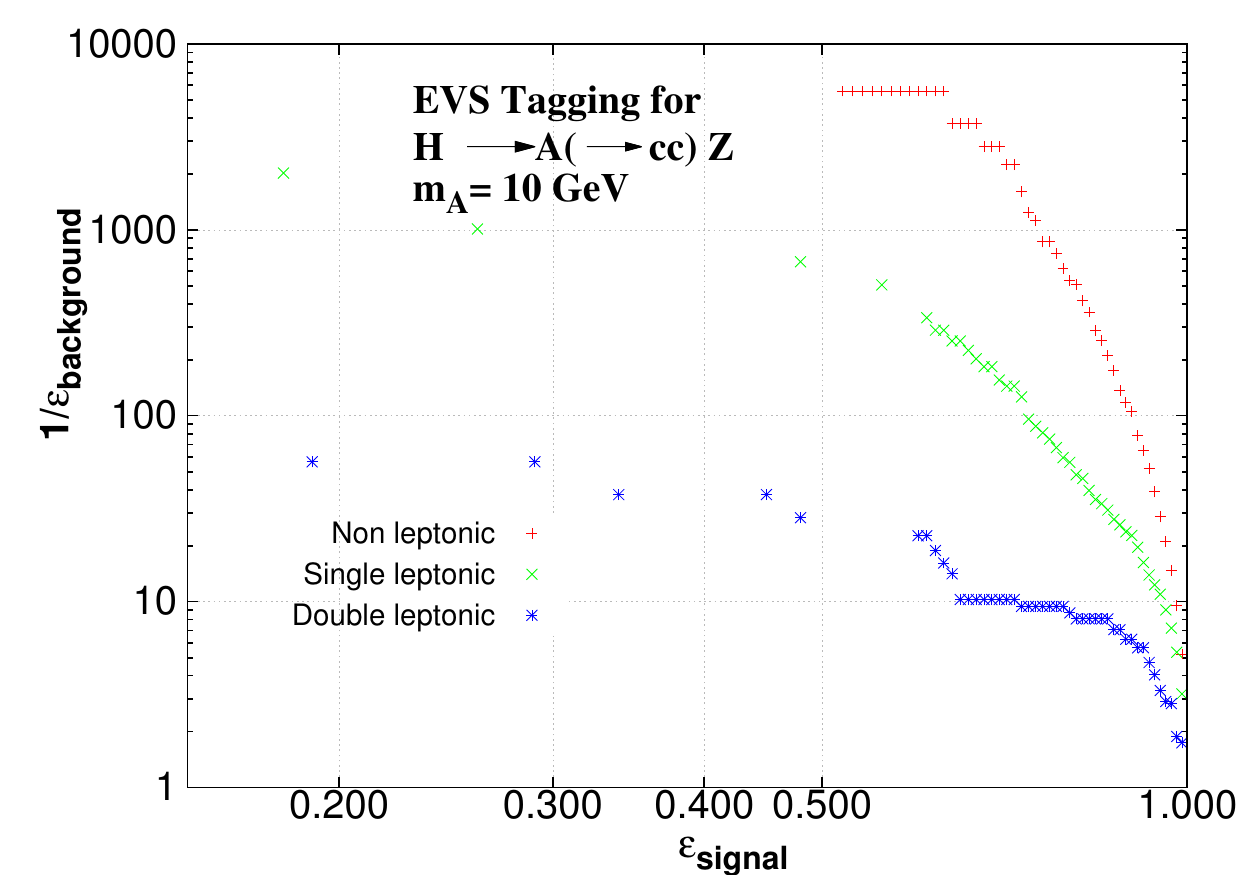}
\caption{\label{fig:A10ROC}Double $c$-jets selection efficiency (from $A\rightarrow c\bar{c}$) against the QCD-jets rejection
achieved by the Event-Shapes tagger per leptonic study. Here we are considering $m_A=10~\hbox{GeV}$.}
\end{figure}

\begin{center}
\begin{figure*}
\includegraphics[height=8.0cm]{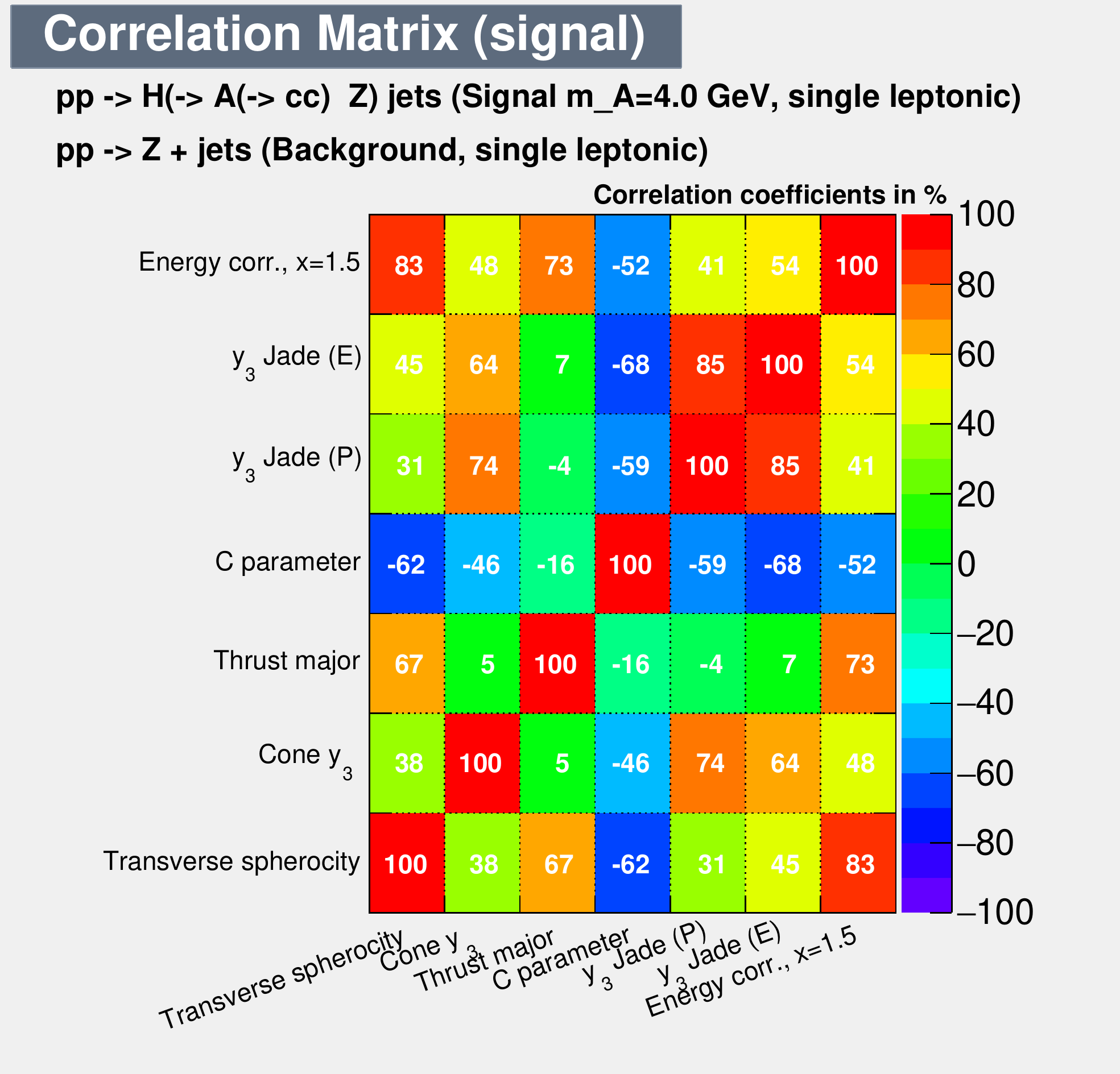}
\includegraphics[height=8.0cm]{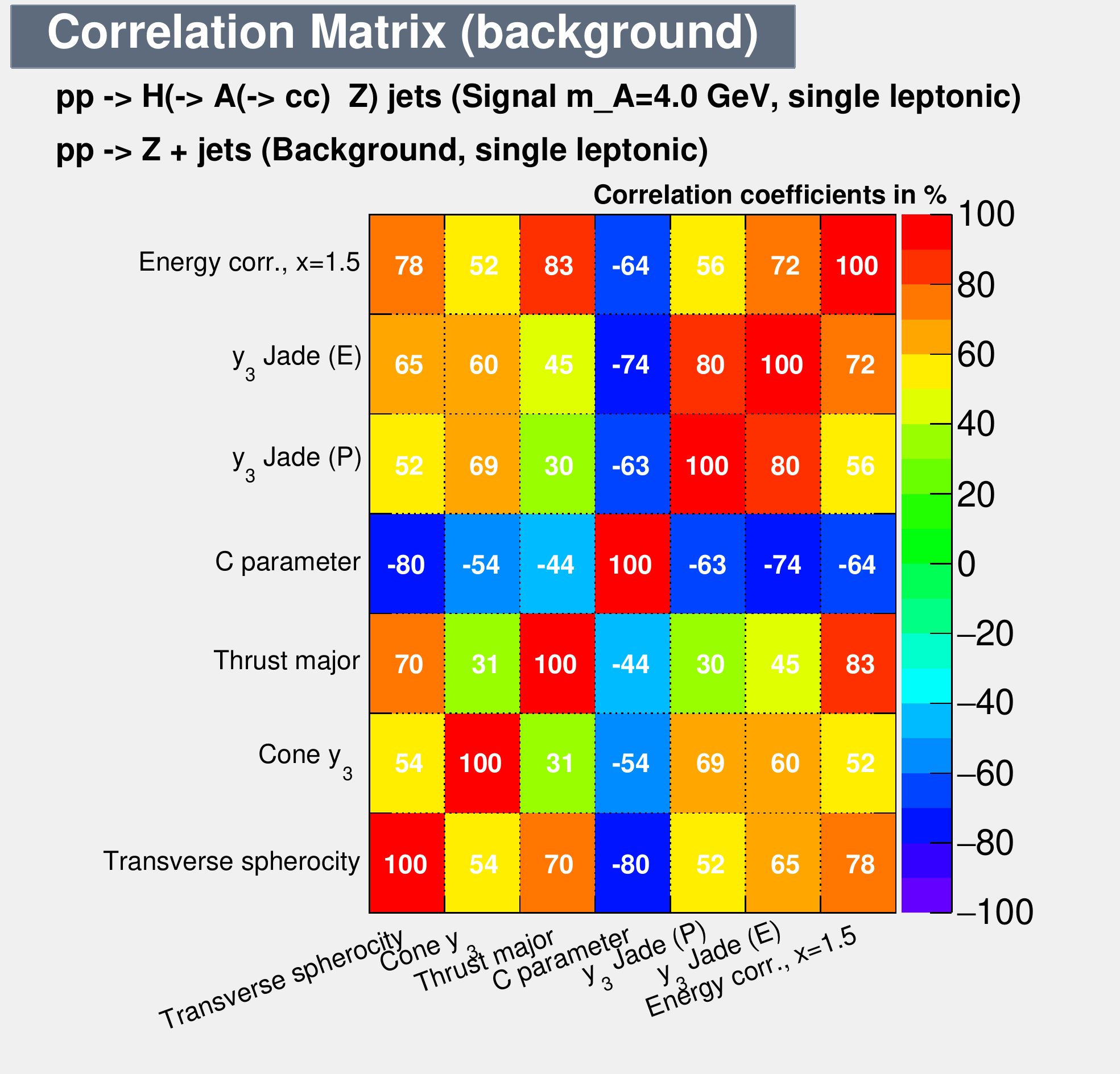}
\includegraphics[height=8.0cm]{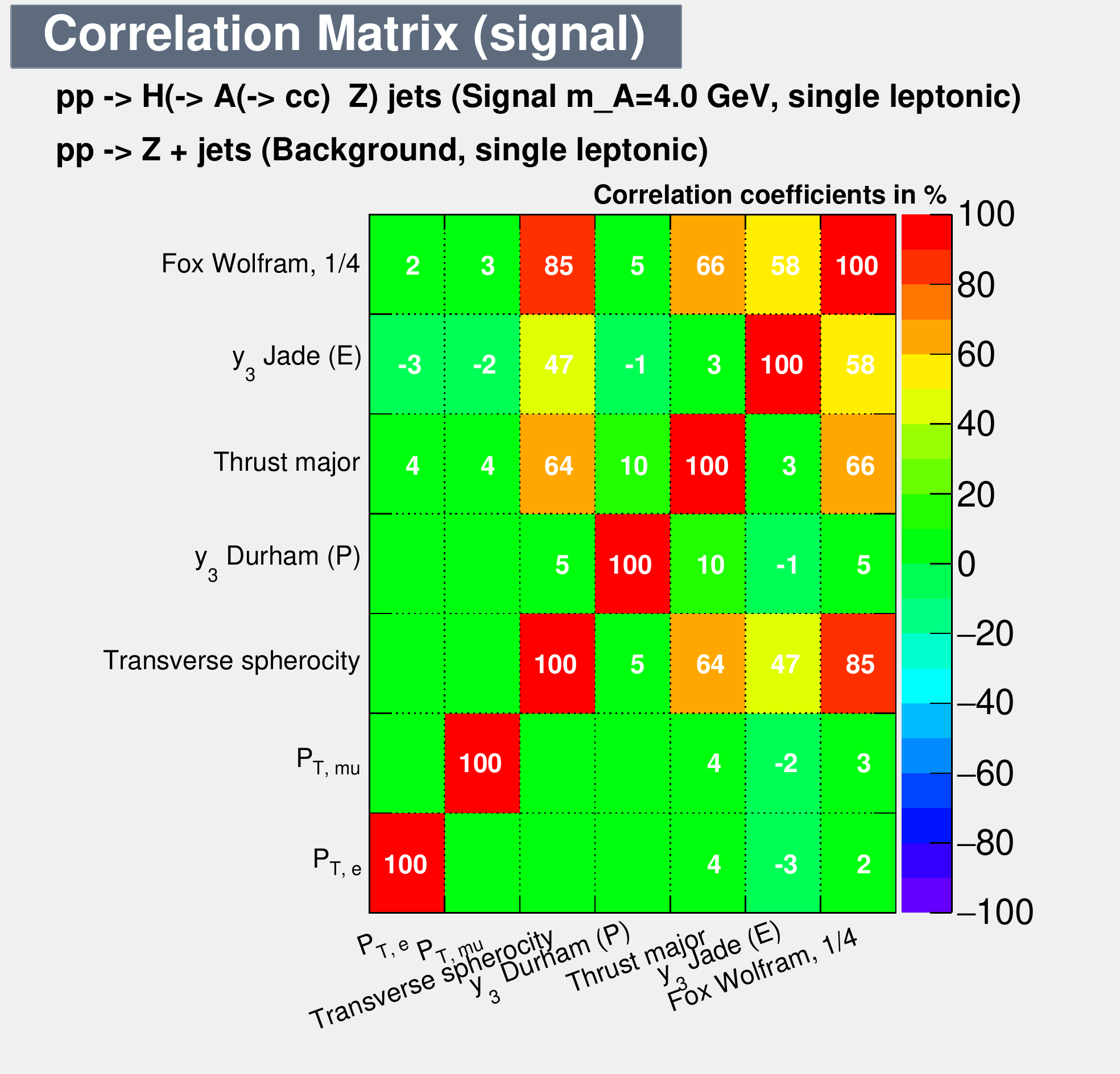}
\includegraphics[height=8.0cm]{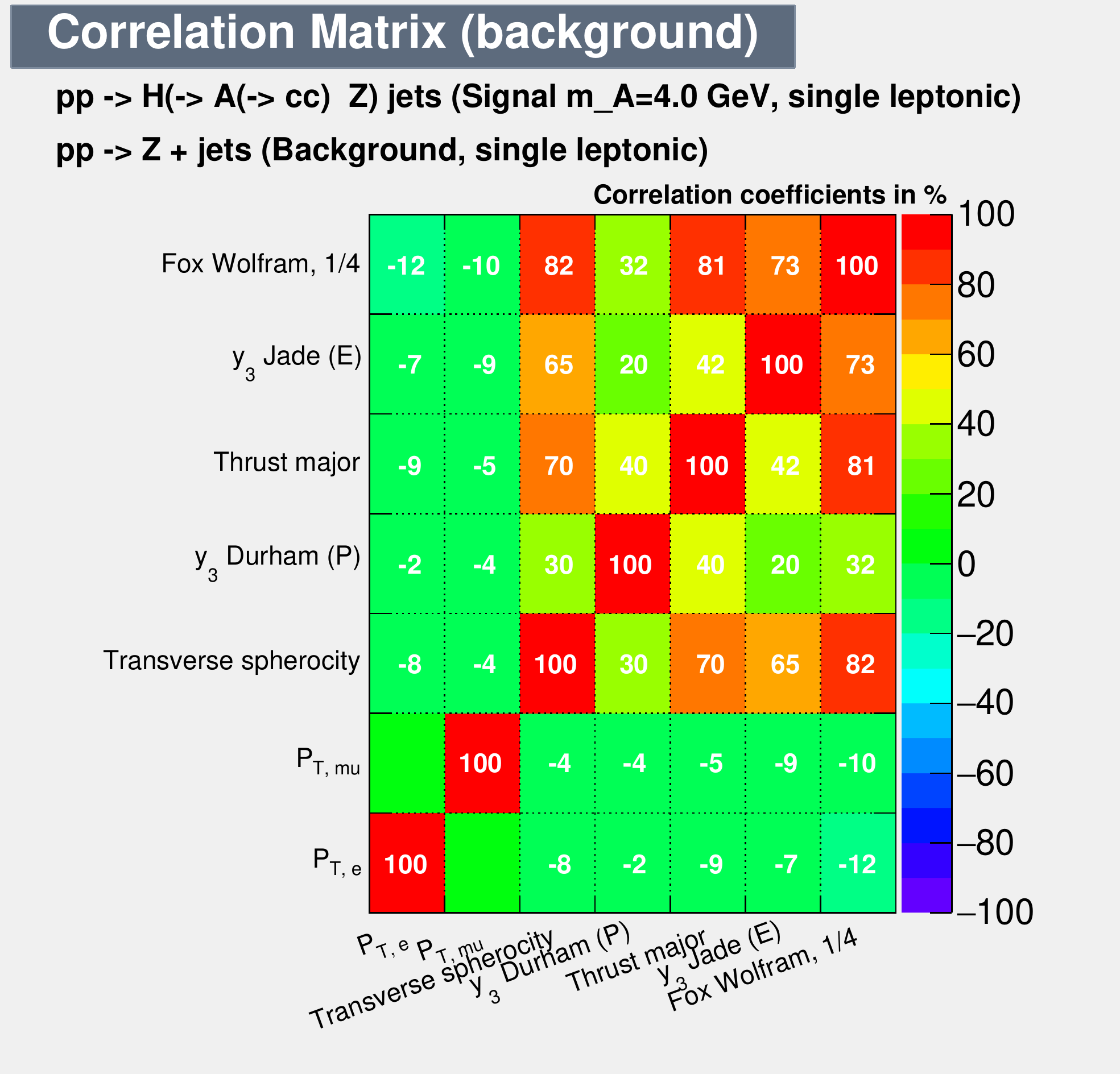}
\includegraphics[height=8.0cm]{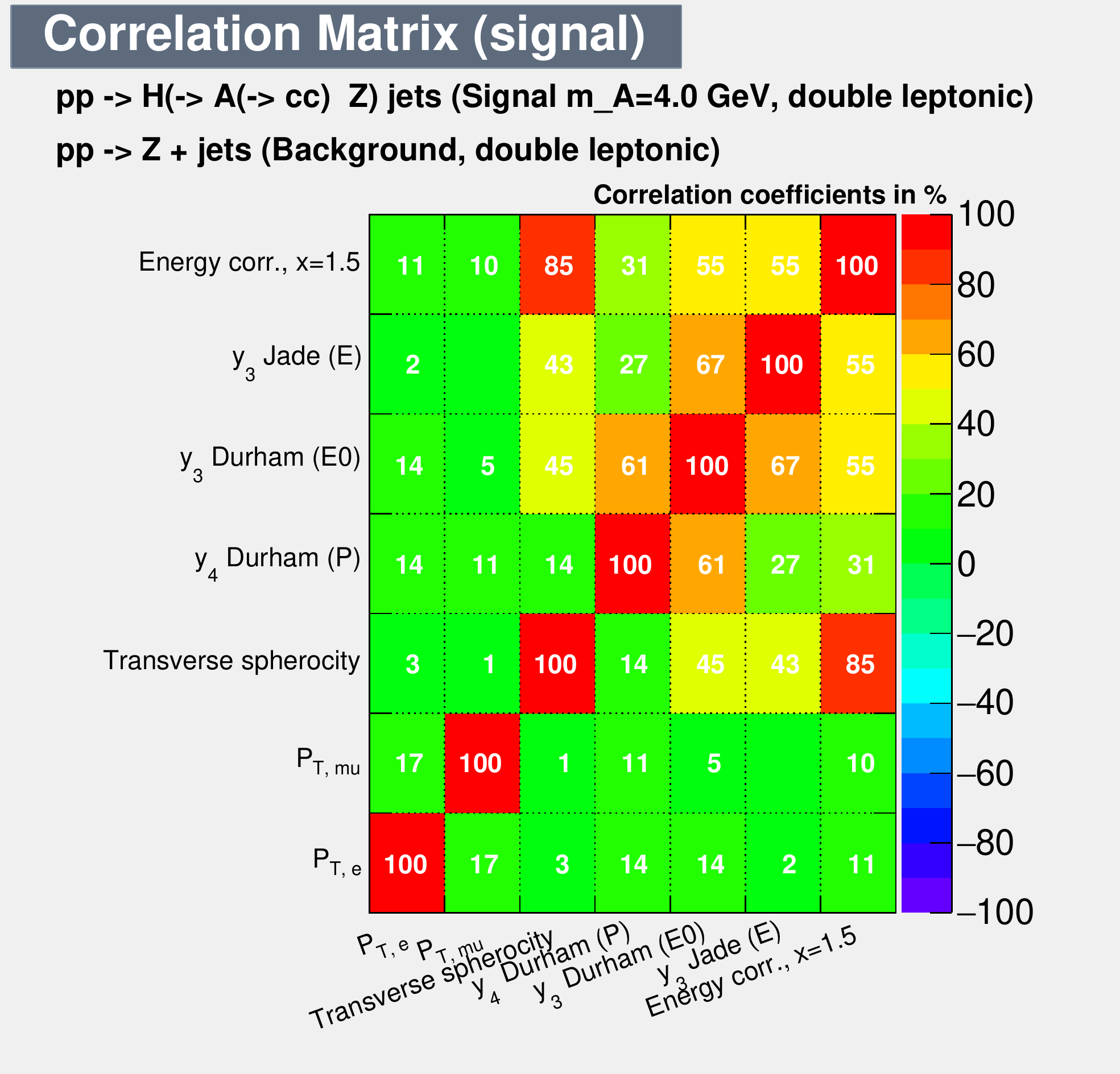}
\includegraphics[height=8.0cm]{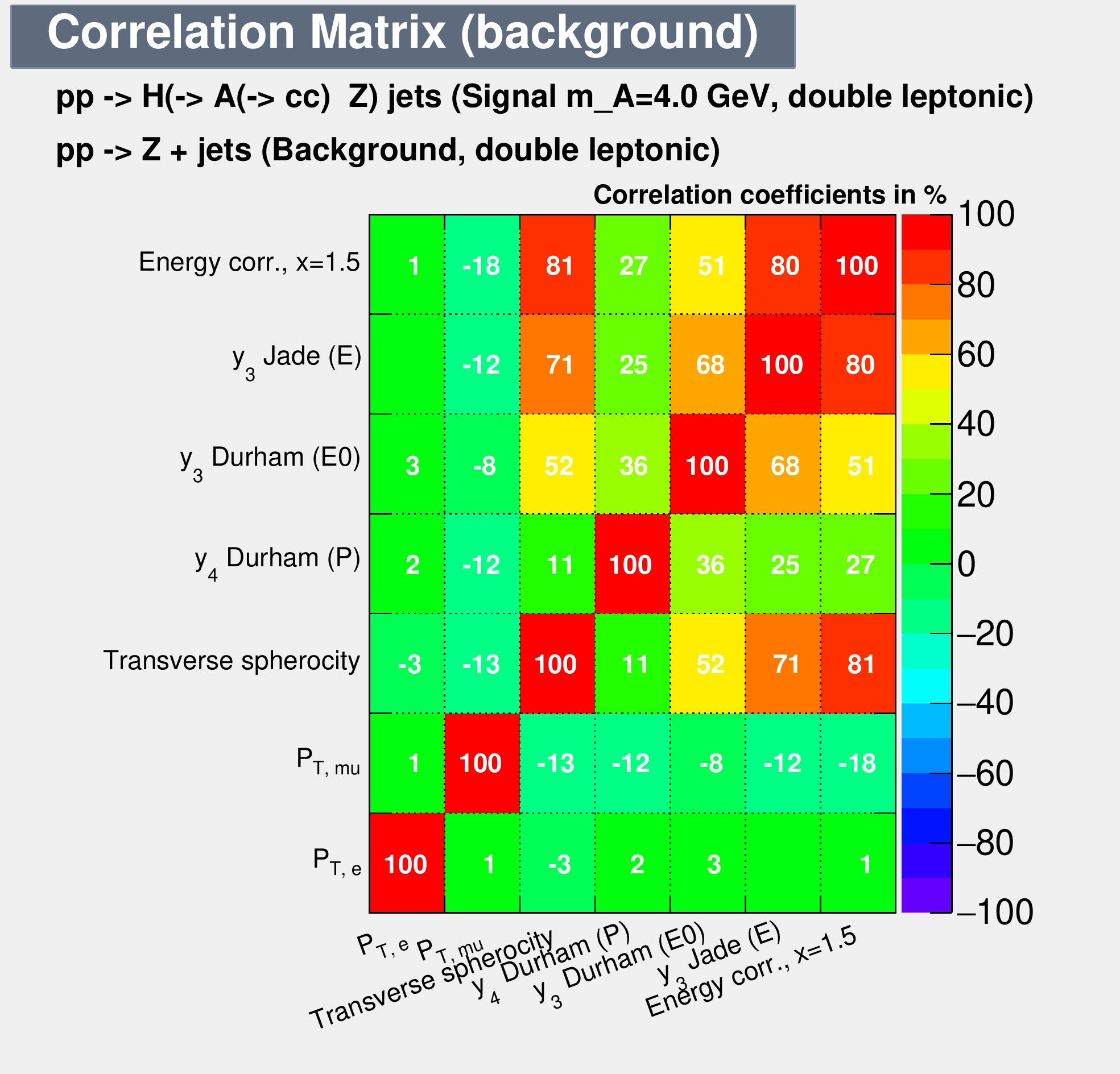}
\caption{Correlation coefficients for the observables used in the analysis of the CP odd THDM scalar $A$, for $m_A=4$ GeV.}
\label{fig:CorrelationA4}
\end{figure*}
\end{center}

\begin{center}
\begin{figure*}
\includegraphics[height=8.0cm]{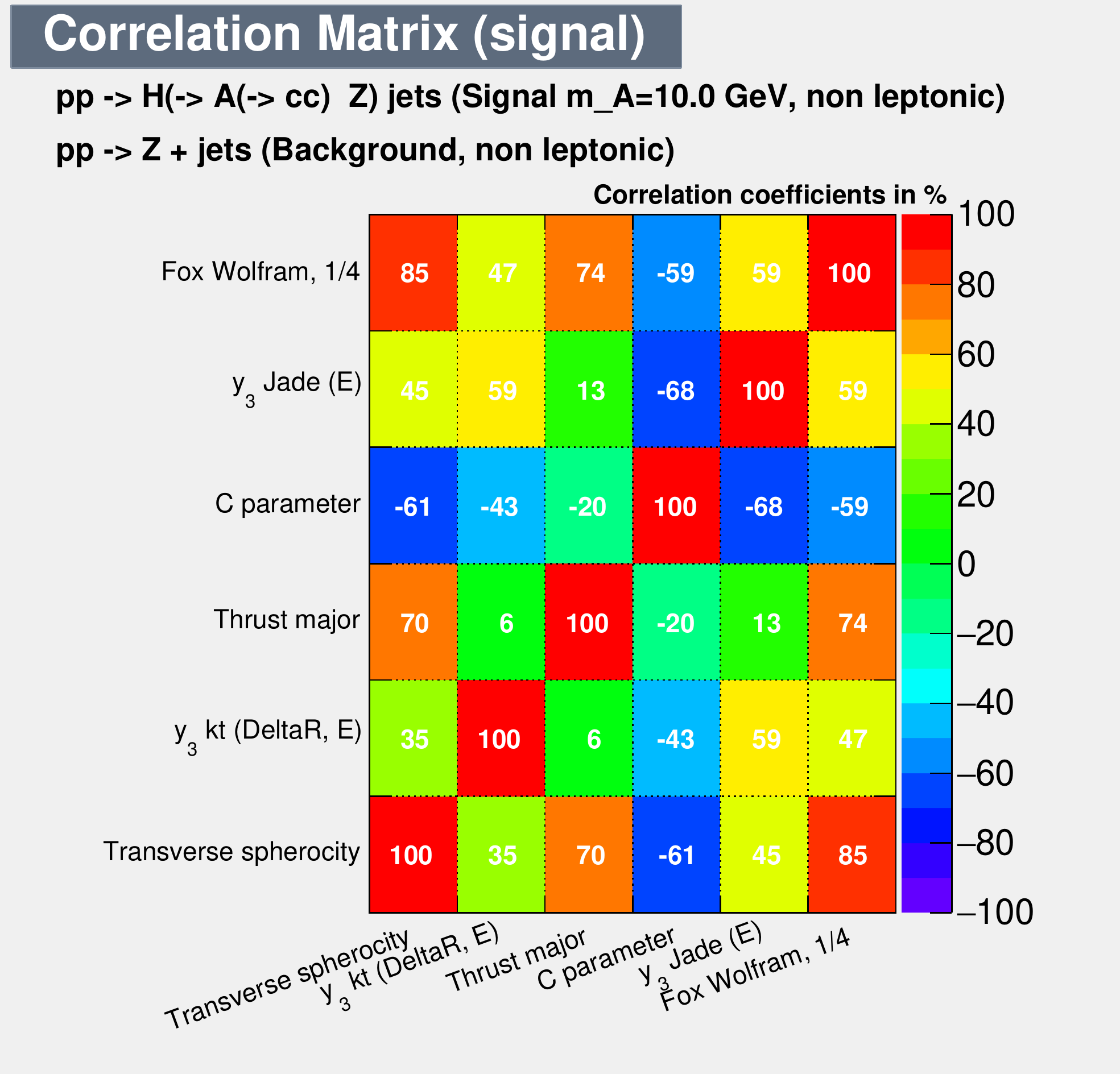}
\includegraphics[height=8.0cm]{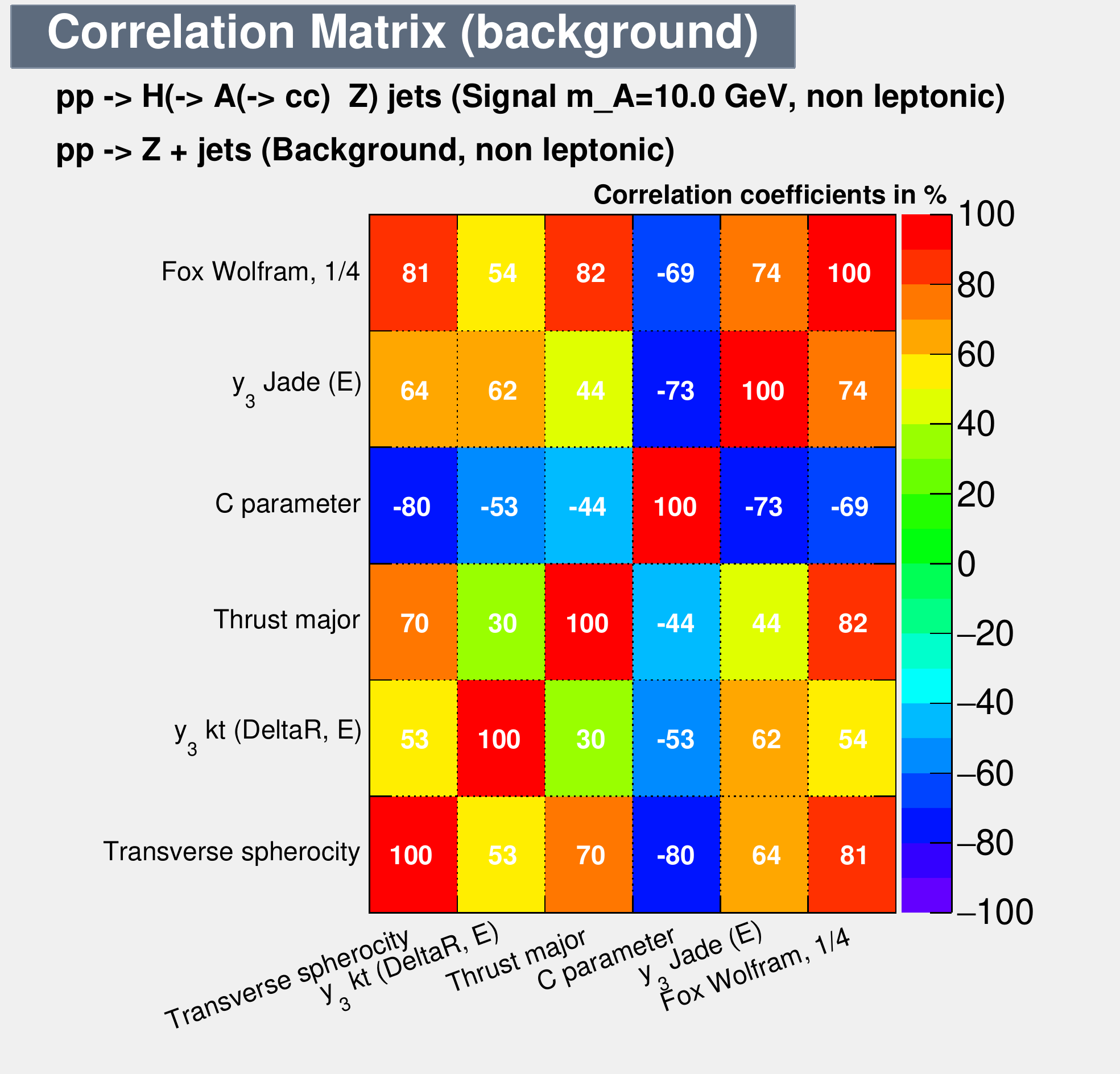}
\includegraphics[height=8.0cm]{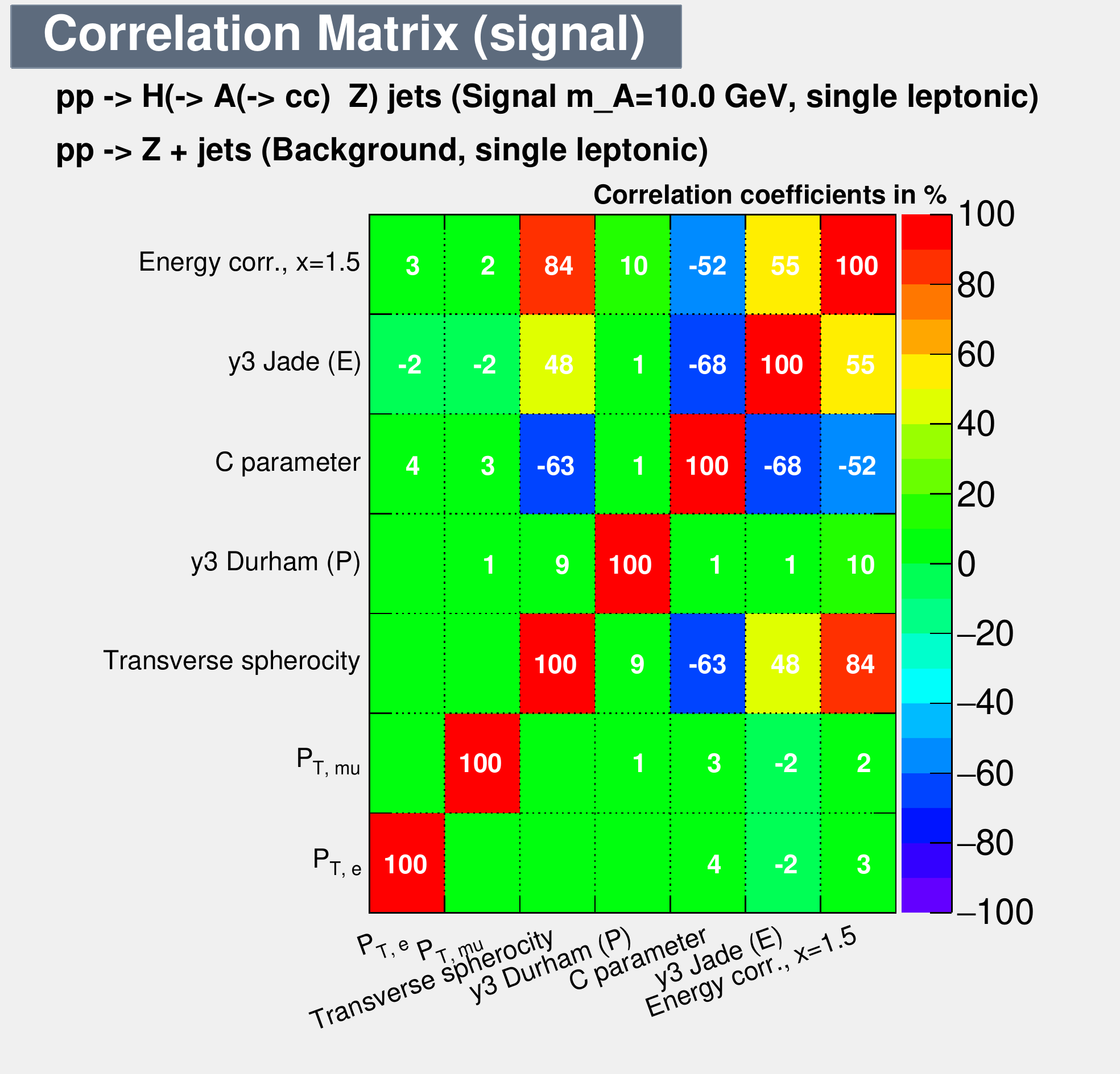}
\includegraphics[height=8.0cm]{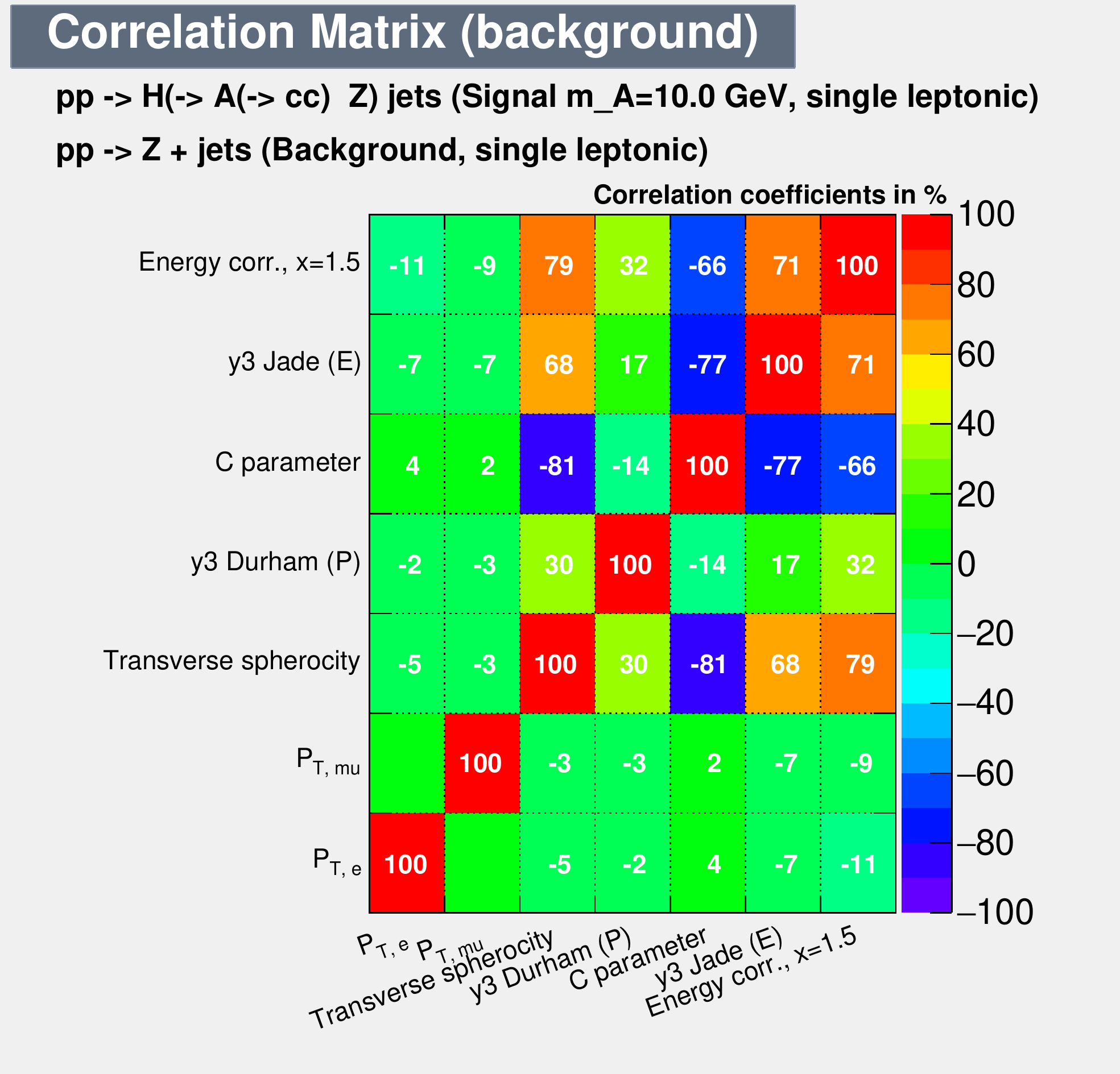}
\includegraphics[height=8.0cm]{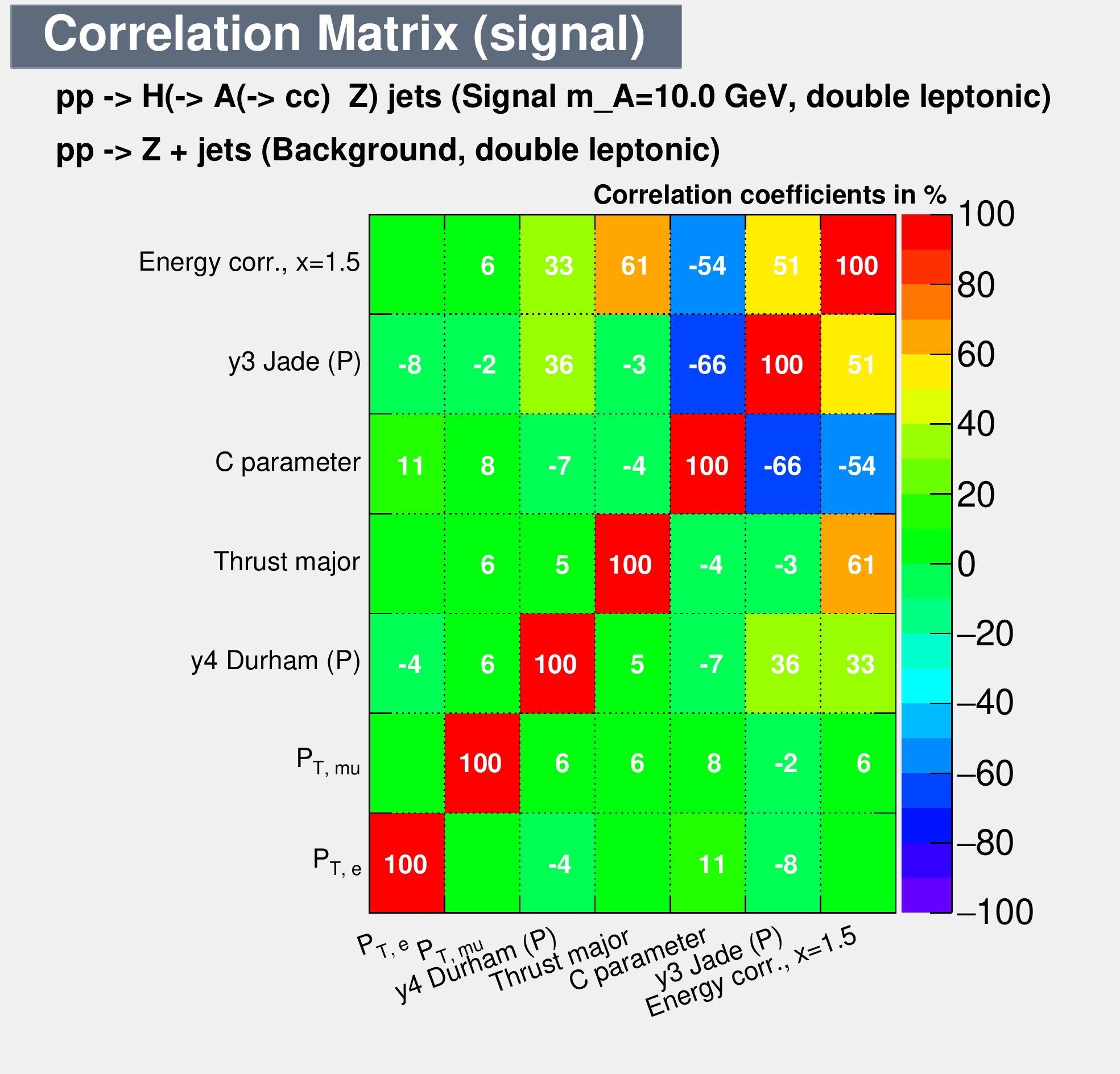}
\includegraphics[height=8.0cm]{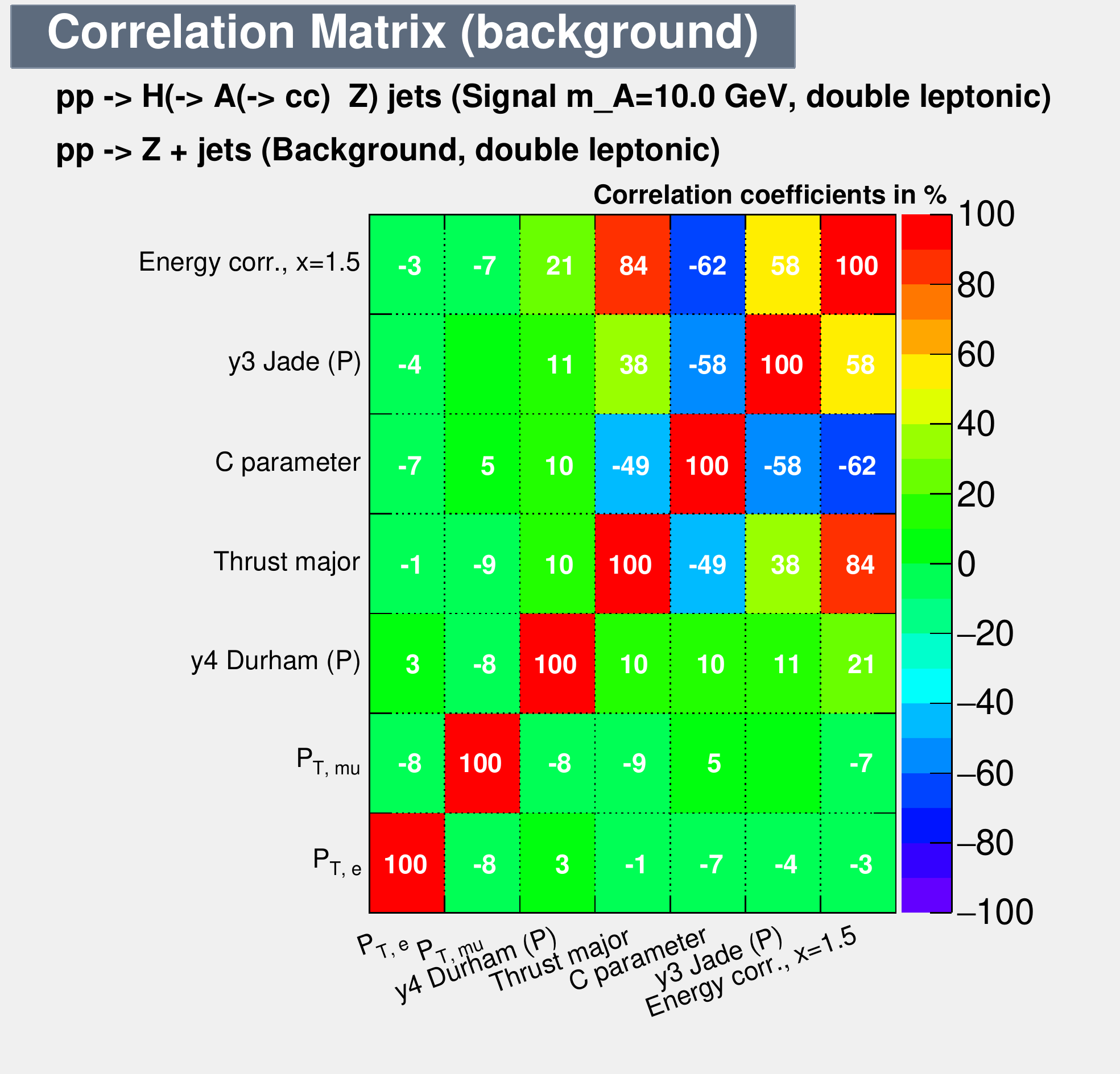}
\caption{Correlation coefficients for the observables used in the analysis of the CP odd THDM scalar $A$, for $m_A=10$ GeV}
\label{fig:CorrelationA10}
\end{figure*}
\end{center}

\begin{figure*}
\includegraphics[height=5cm]{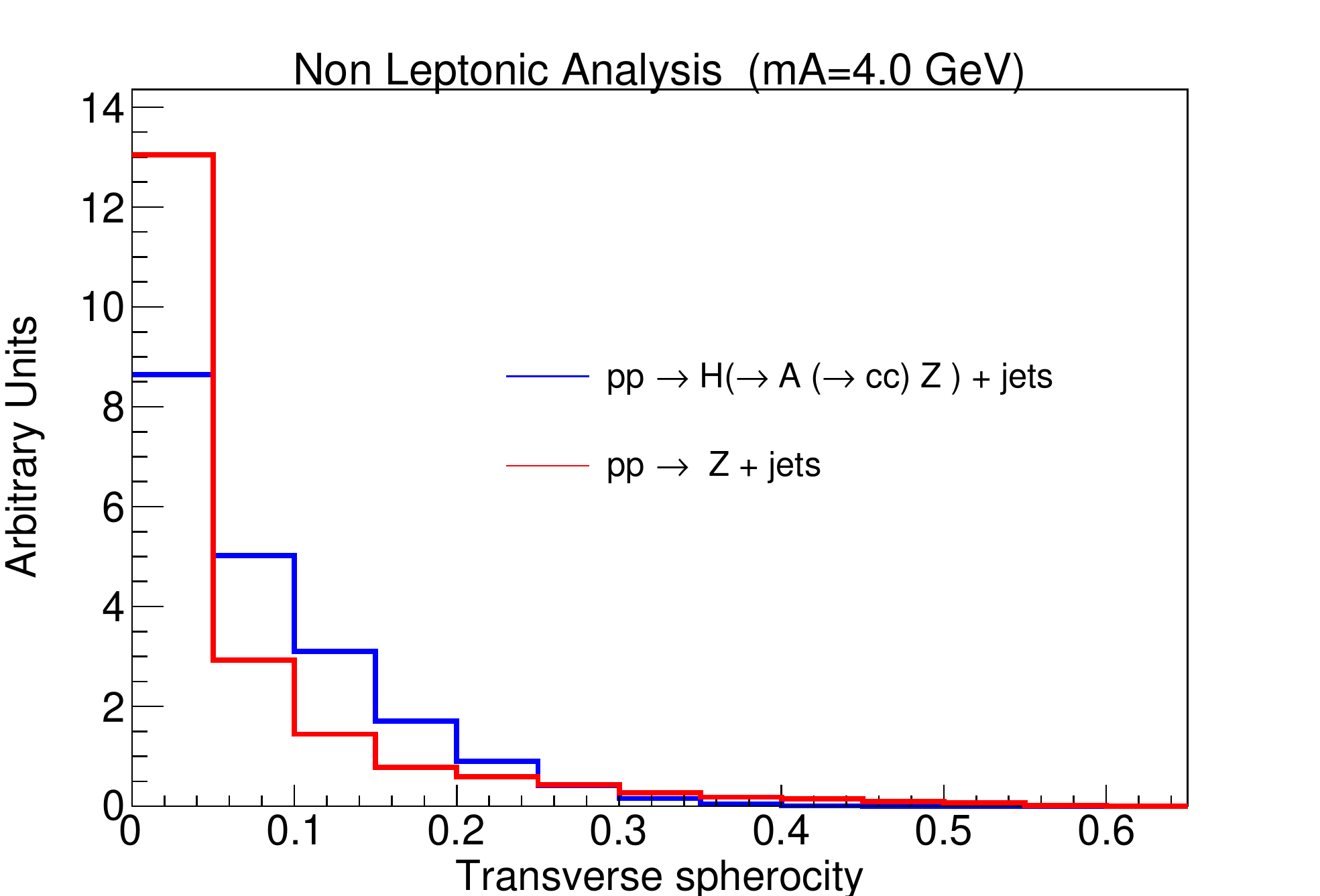}
\includegraphics[height=5cm]{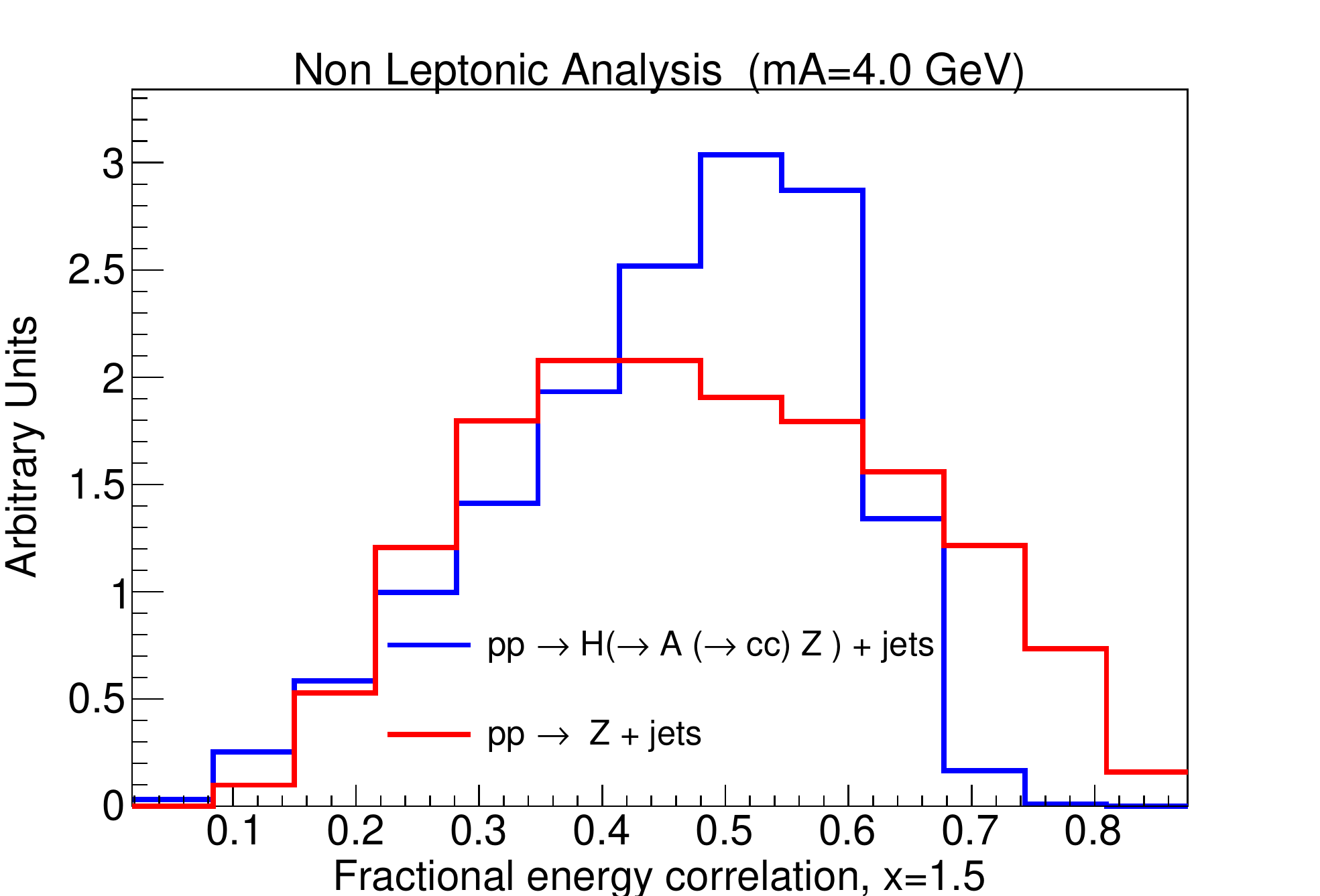}
\includegraphics[height=5cm]{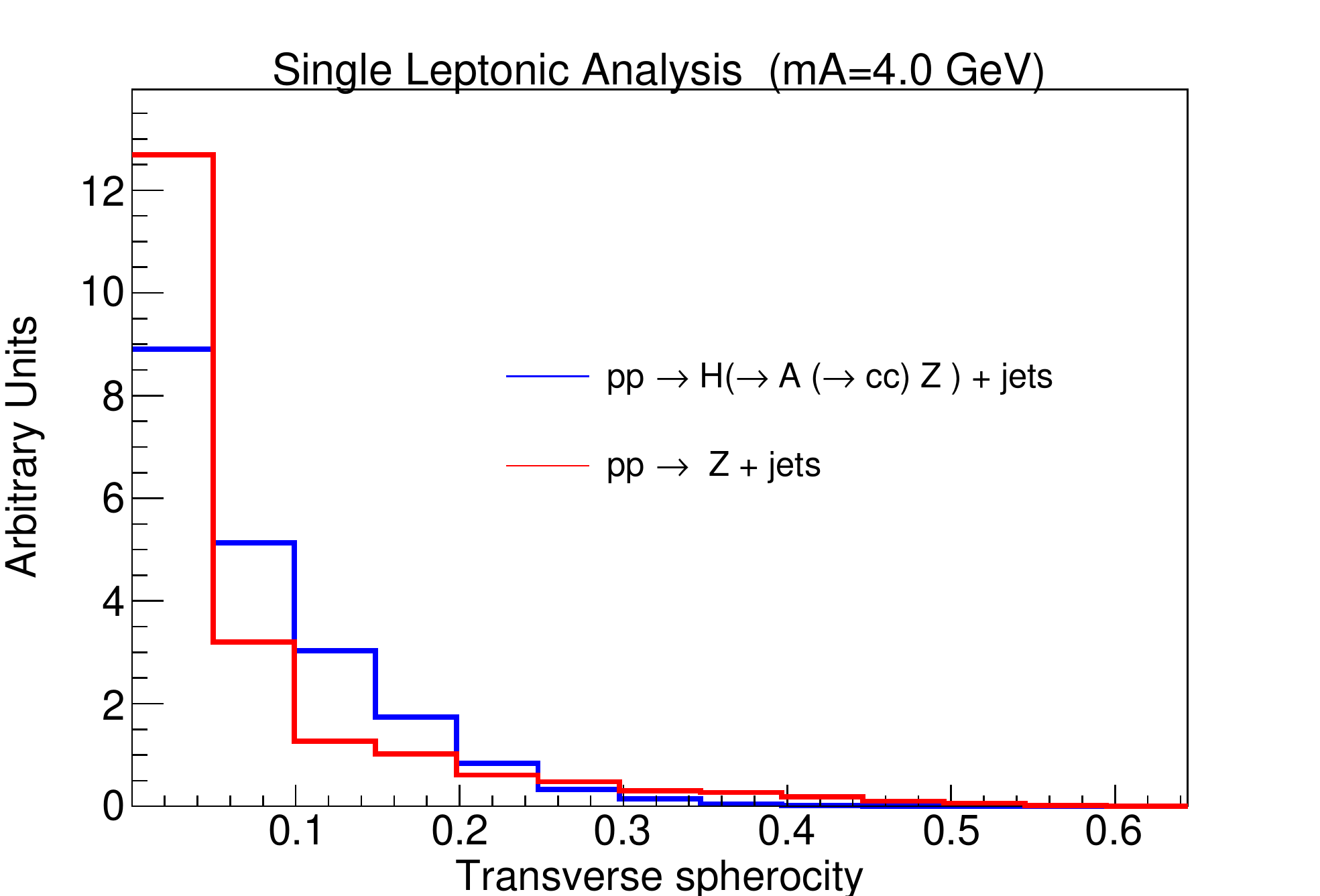}
\includegraphics[height=5cm]{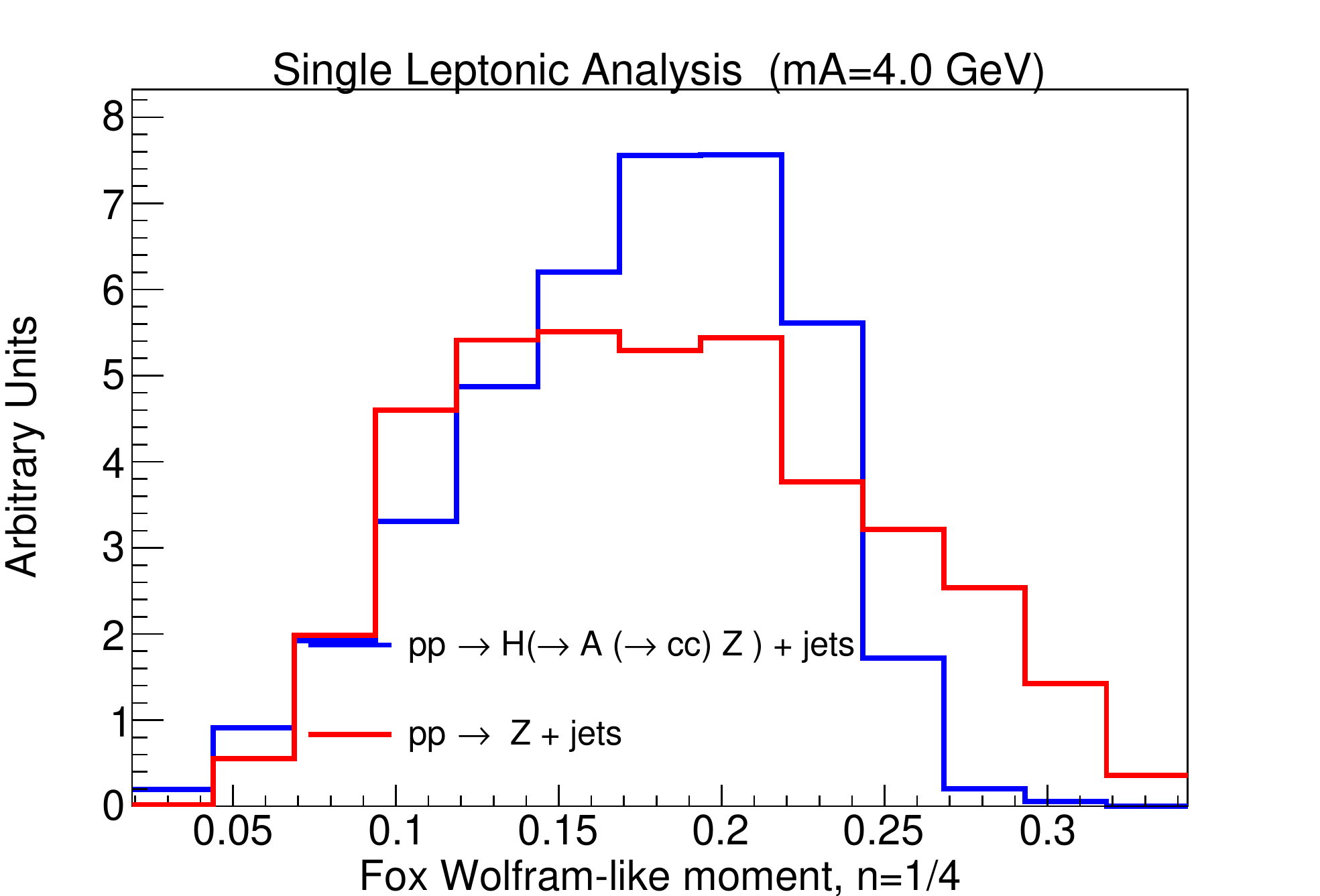}
\includegraphics[height=5cm]{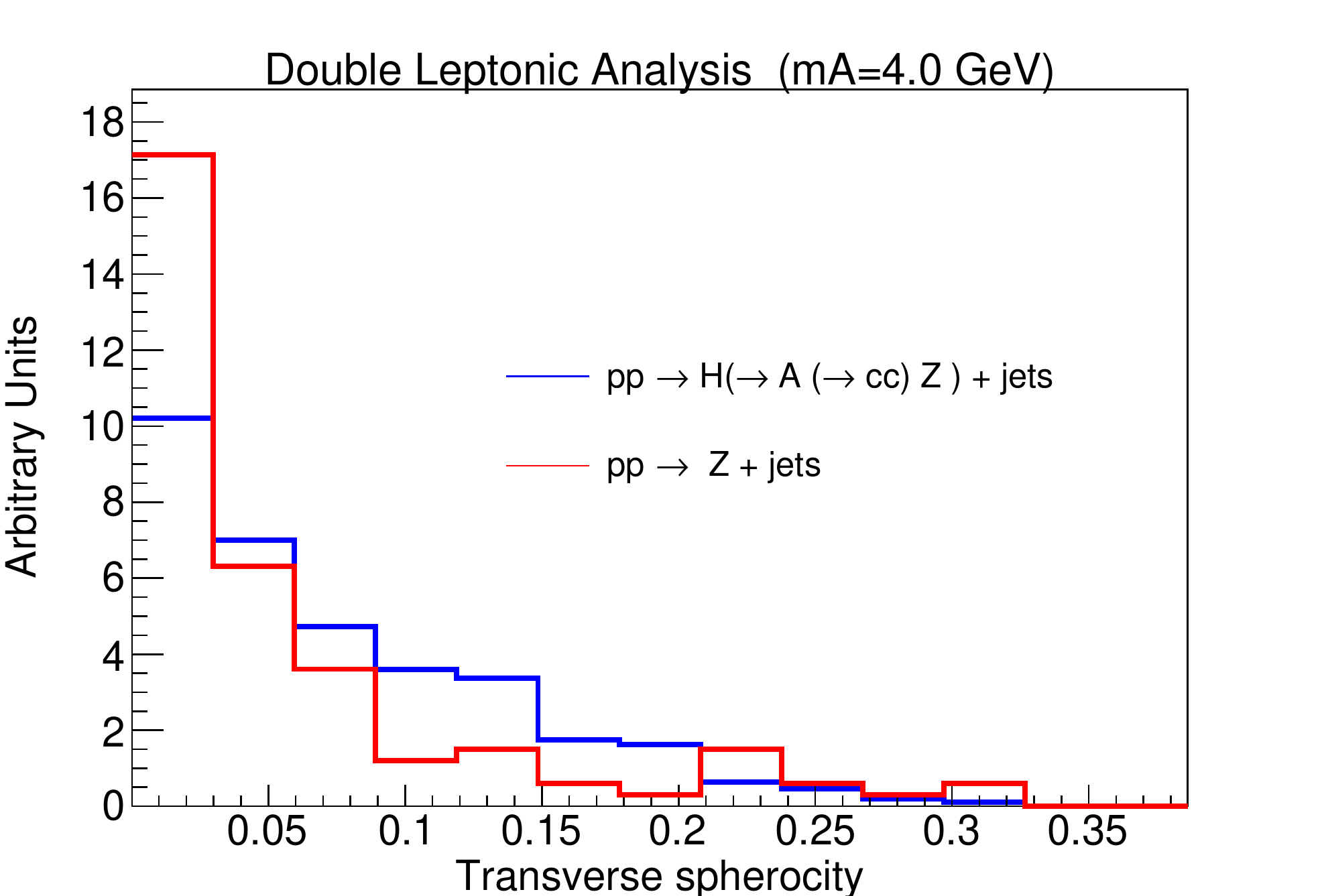}
\includegraphics[height=5cm]{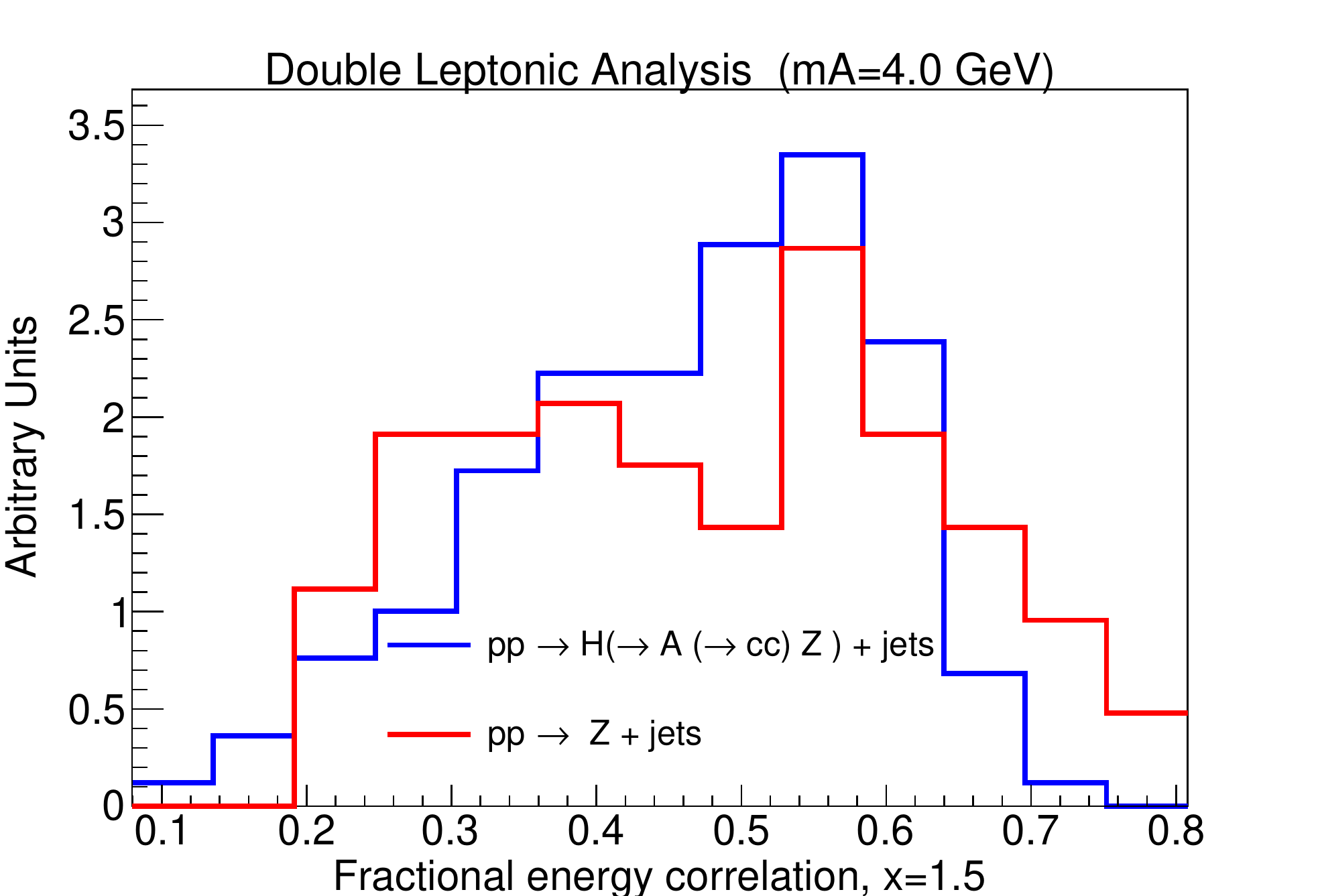}
\caption{Histograms for the main discrimination event shapes when selecting 
$pp\rightarrow H (\rightarrow A( \rightarrow  c\bar{c} )  Z ) + \hbox{jets}$ against
$pp\rightarrow  Z + \hbox{jets}$ for $m_A=4~\hbox{GeV}$.}
\label{fig:HistoObsA4}
\end{figure*}

\begin{figure*}
\includegraphics[height=5cm]{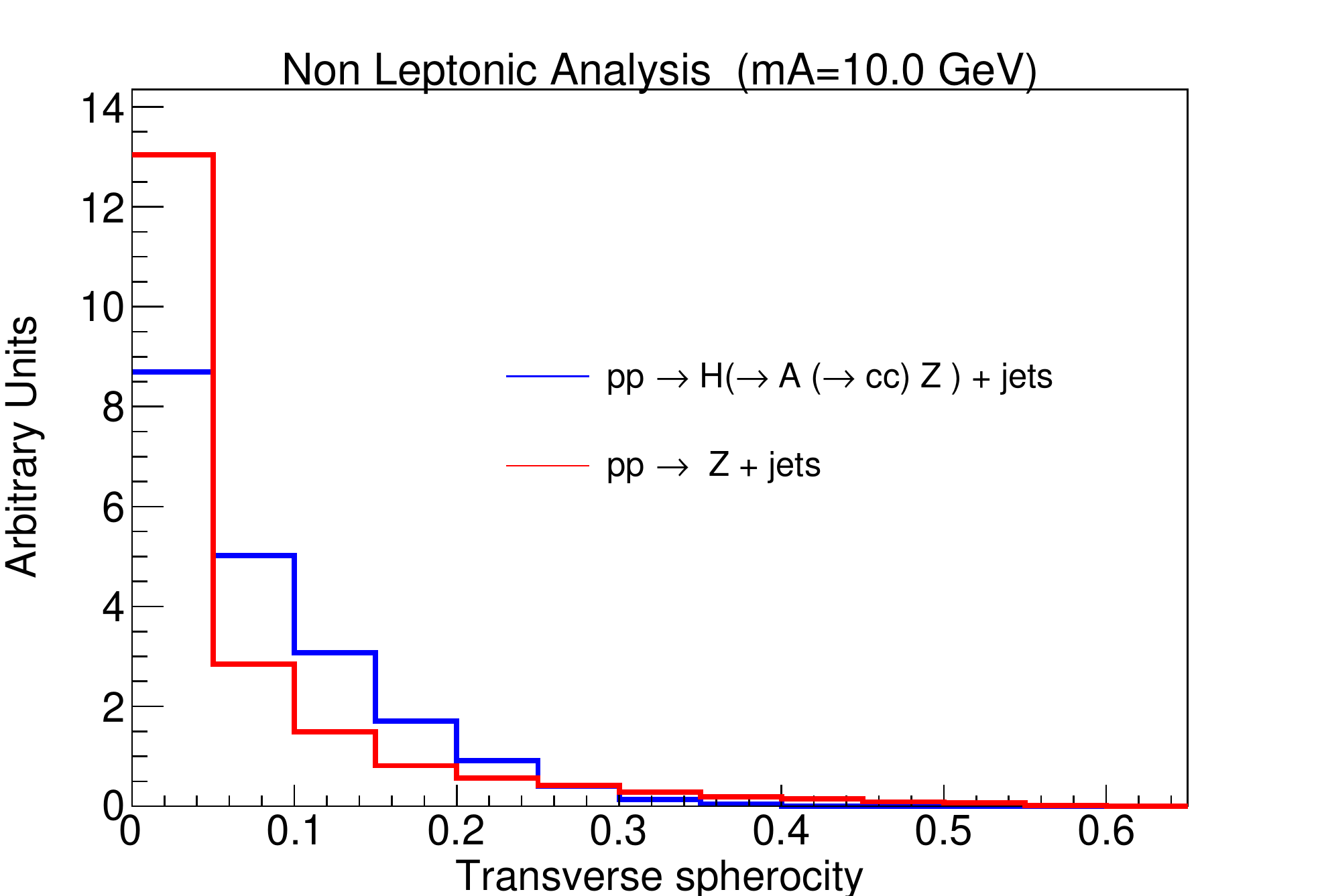}
\includegraphics[height=5cm]{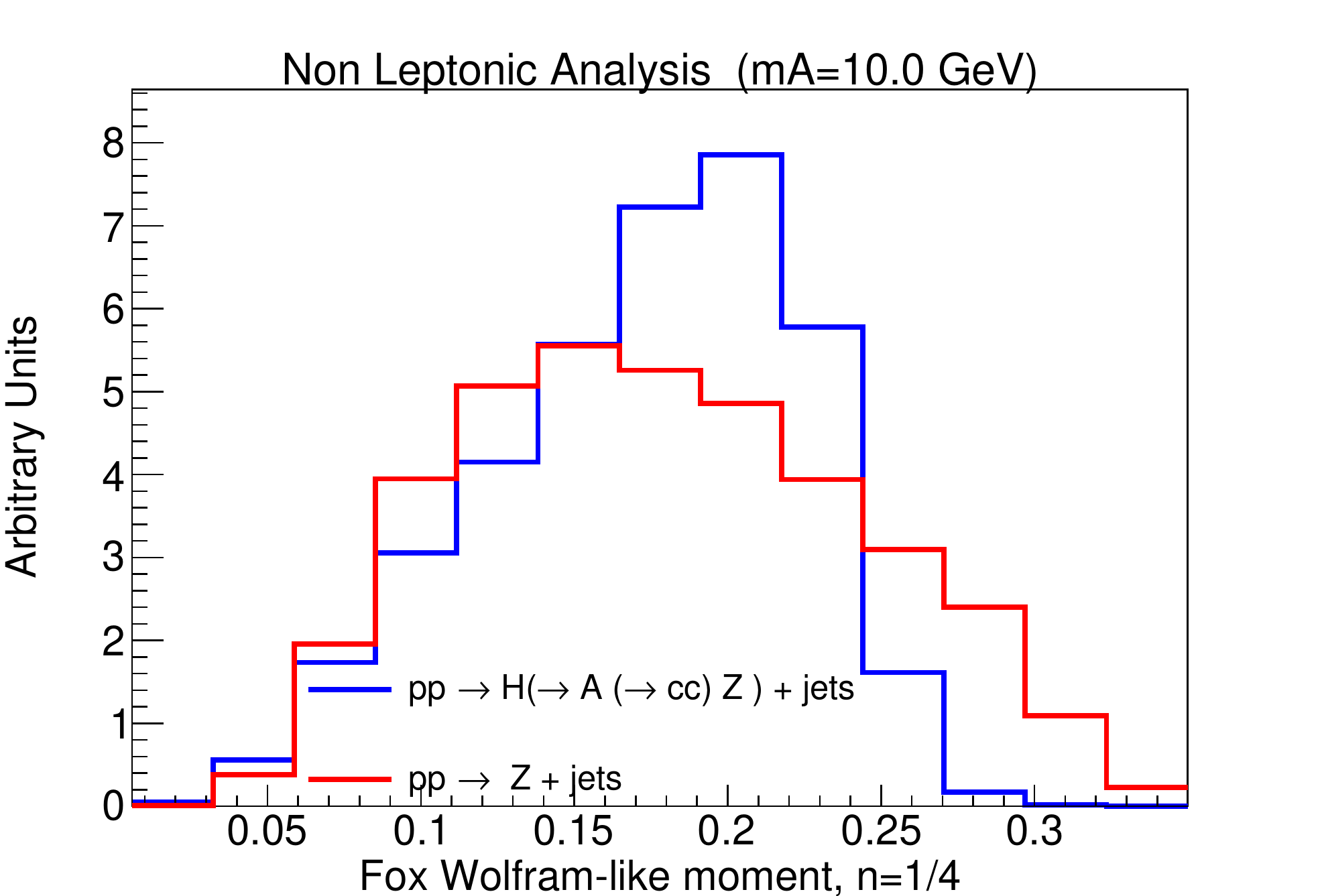}
\includegraphics[height=5cm]{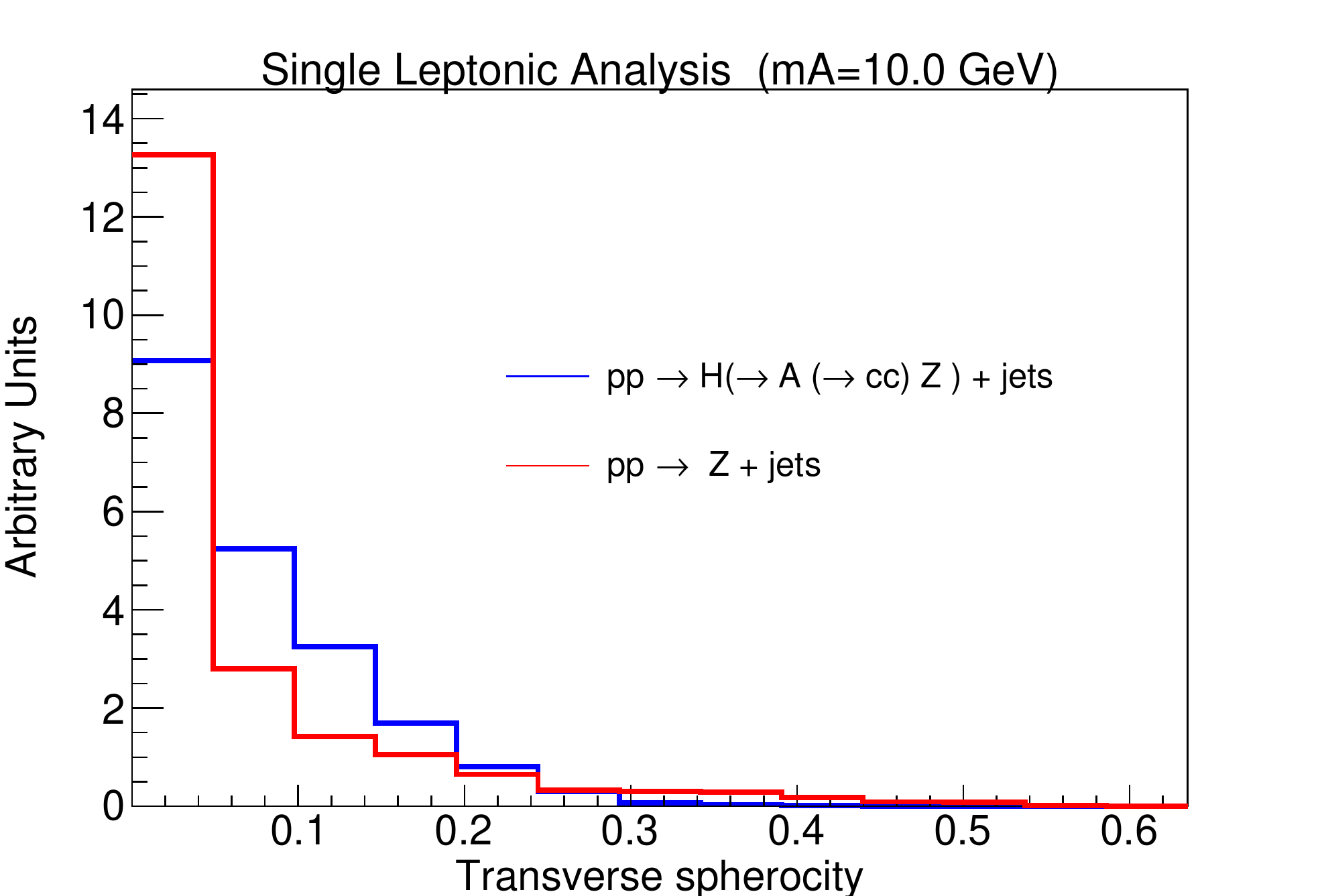}
\includegraphics[height=5cm]{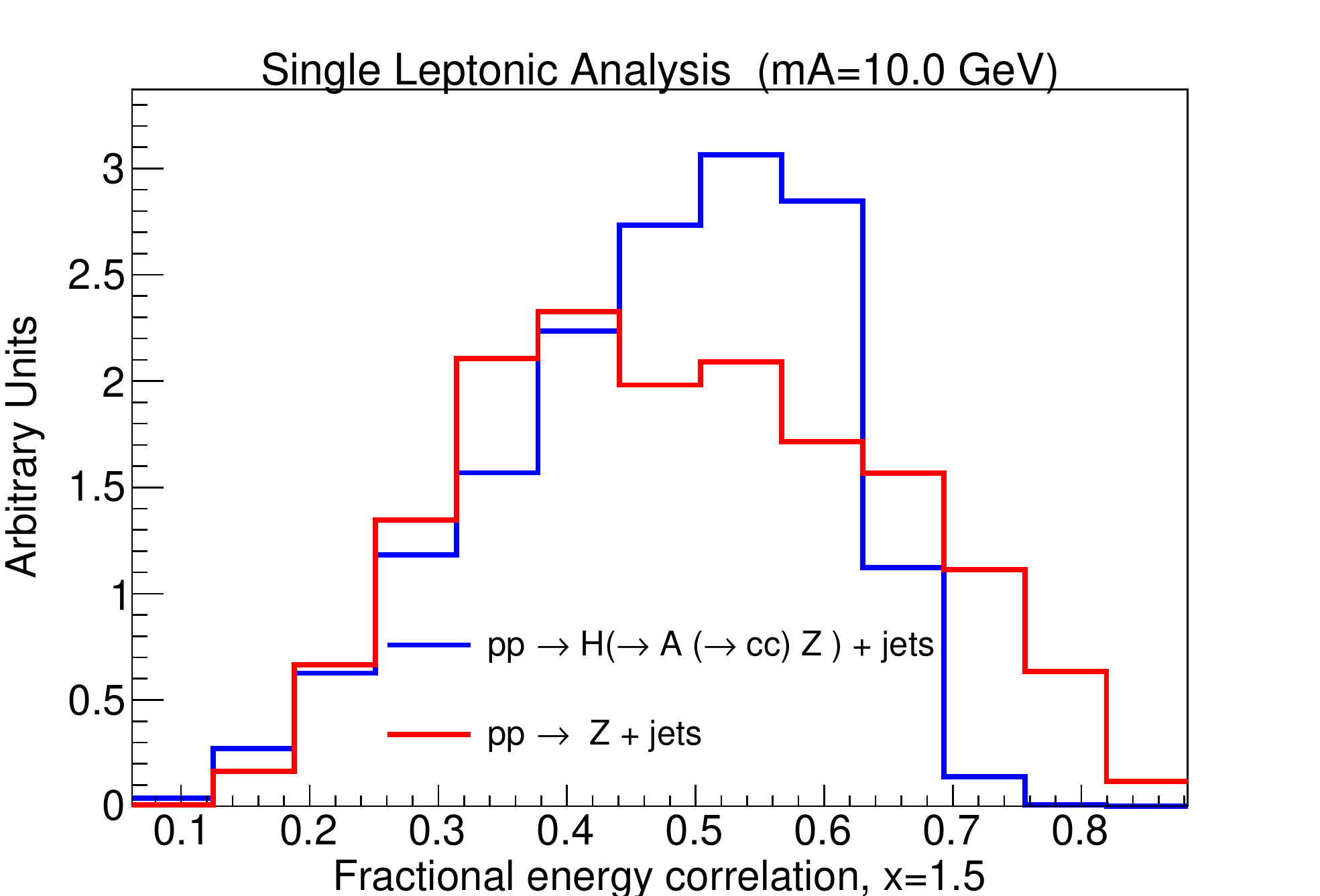}
\includegraphics[height=5cm]{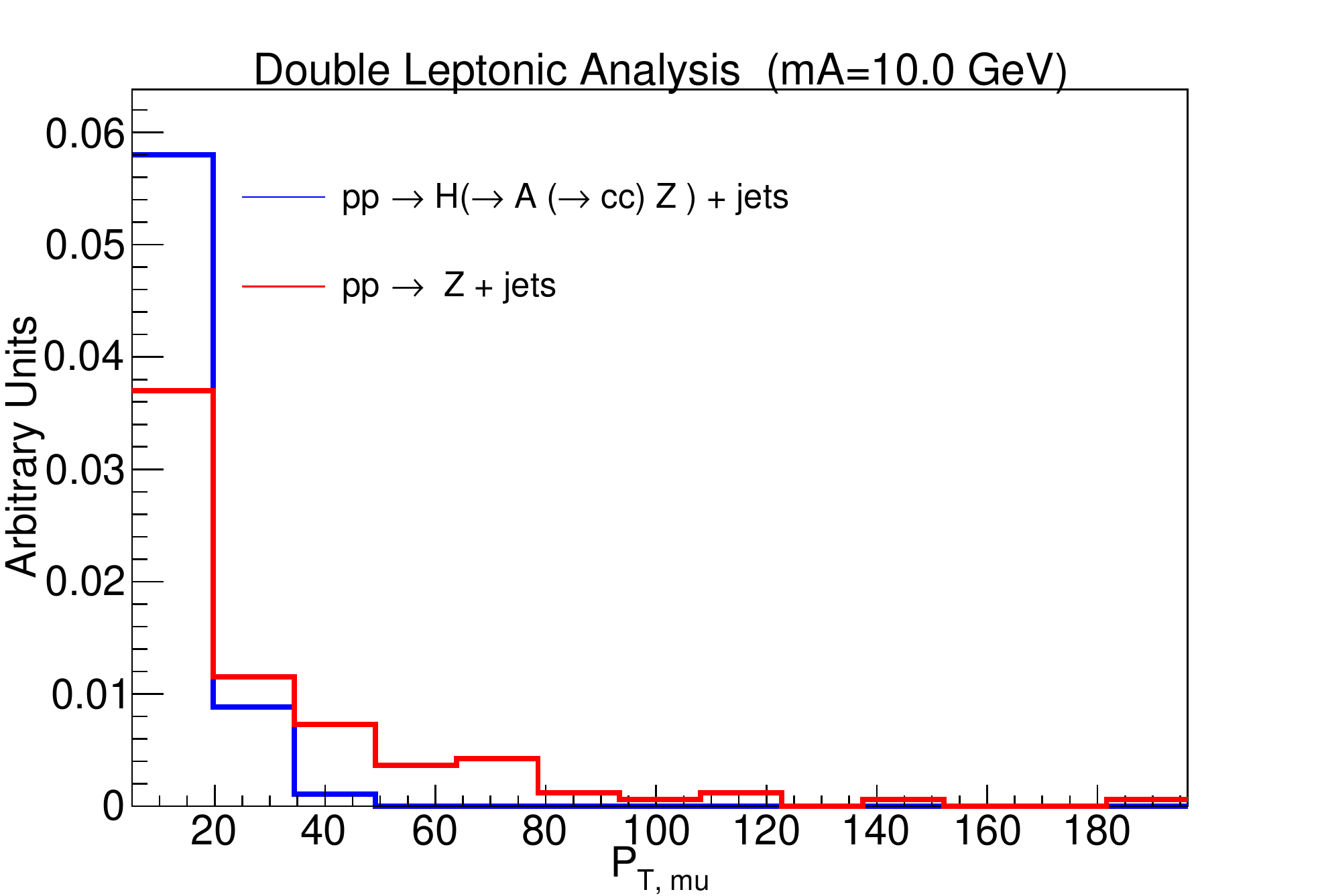}
\includegraphics[height=5cm]{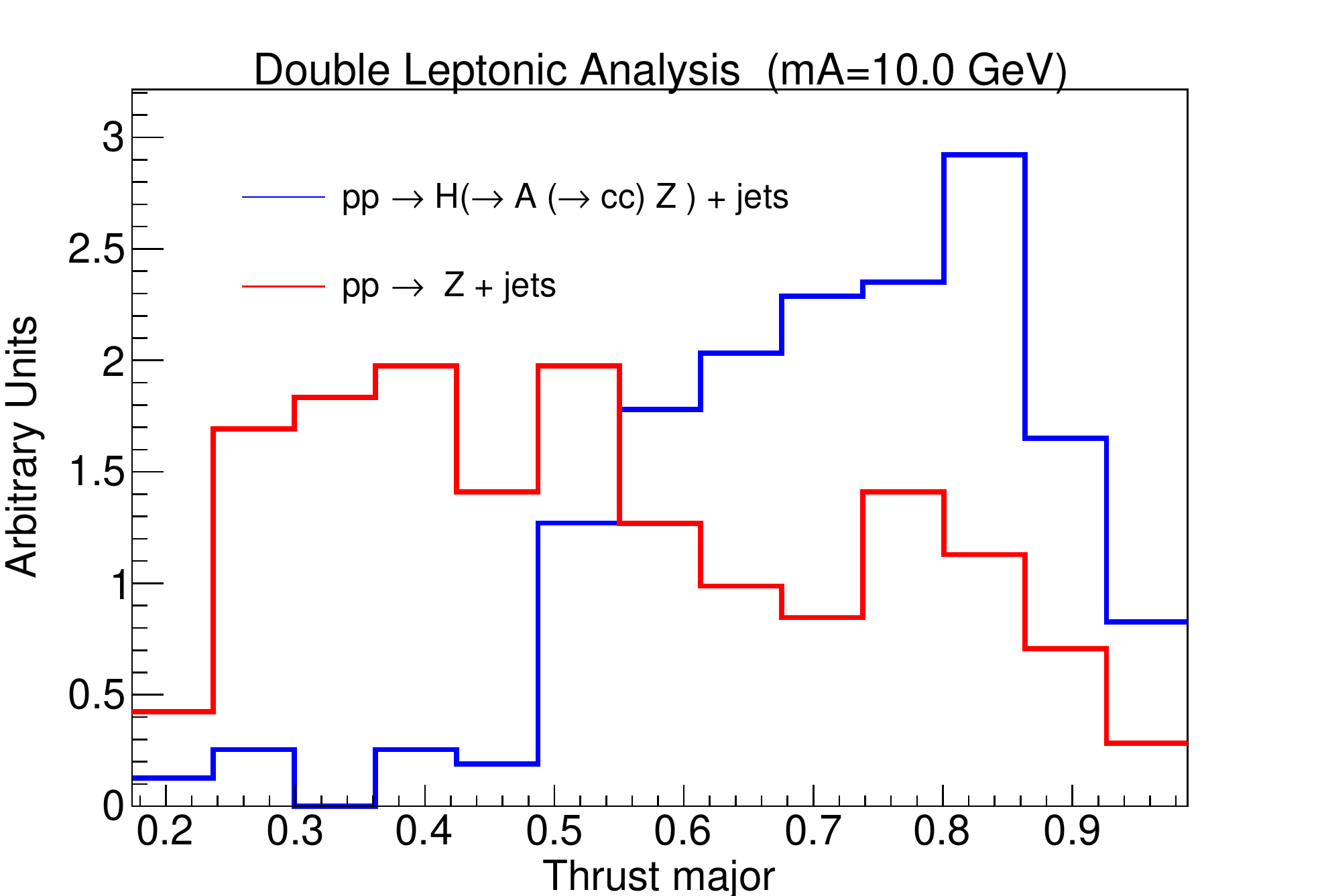}
\caption{Histograms for the main discrimination event shapes when selecting 
$pp\rightarrow H (\rightarrow A( \rightarrow  c\bar{c} )  Z ) + \hbox{jets}$ against
$pp\rightarrow  Z + \hbox{jets}$ for $m_A=10~\hbox{GeV}$.}
\label{fig:HistoObsA10}
\end{figure*}

For $m_A=10~\hbox{GeV}$ the observables 
per leptonic category that yield the best selection efficiency curves, shown in Fig.~\ref{fig:A10ROC}, are presented in 
Table (\ref{Tab:ccA10}). Our optimal result corresponds to

\begin{eqnarray}
\varepsilon_{c\bar{c}, m_A=10~{\tiny \hbox{GeV}}}=0.38 && \varepsilon_{ {\tiny \hbox{QCD}}, {\tiny \hbox{jets}}}=0.0004
\end{eqnarray}

calculated from the individual efficiencies per leptonic category 

\begin{eqnarray}
\varepsilon^{(0)}_{S.}=0.39,&\varepsilon^{(1)}_{S.}=0.29,&\varepsilon^{(2)}_{S.}=0.19\nonumber\\
\varepsilon^{(0)}_{B.}=0.9\times 10^{-4},&\varepsilon^{(1)}_{B.}=14.9\times 10^{-4},&\varepsilon^{(2)}_{B.}=177.0\times 10^{-4}\nonumber\\
\end{eqnarray}

and the partial fractions of leptonic events presented in Table (\ref{Tab:leptonic_fractions_mA_10}).
The  $95\%$ C.L.  limit on the branching ratio is $\mathcal{B}r(H\rightarrow A (\rightarrow c \bar{c}) Z)\leq 0.003\%$, 
leading to the signal cross section
$\sigma_{p p \rightarrow H (\rightarrow A (\rightarrow c \bar{c}) Z) + \hbox{jets} }=0.01~\hbox{fb}$. 
The correlation matrices for the analyses of this section and the histograms for the main discriminating observables 
are shown in Figs.~\ref{fig:CorrelationA4}-\ref{fig:CorrelationA10} and Figs.~\ref{fig:HistoObsA4}-\ref{fig:HistoObsA10} respectively.

\begin{table}
\begin{center}
\begin{tabular}{|c|c|c|}
\hline
\bf{Fraction of events} & $pp\rightarrow H + \hbox{jets}$ & $pp\rightarrow Z + \hbox{jets}$ \\
& with  $H\rightarrow A (\rightarrow c \bar{c}) Z$&\\
&$m_A=4~\hbox{GeV}$ &\\
\hline
\hline
$f^{(0)}$ (0 leptons) & $85.5\%$ & $84.0\%$   \\
\hline
$f^{(1)}$ (1 lepton)& $13.5\%$ & $15.15\%$ \\
\hline
$f^{(2)}$ (2 leptons)& $1.0\%$ & $0.85\%$ \\
\hline
\end{tabular}
\caption{Fraction of events in each one of the leptonic categories for the samples $ pp \rightarrow H (\rightarrow A (\rightarrow c\bar{c} ) Z ) + \hbox{jets}$ and
$pp\rightarrow Z + \hbox{jets}$ after the selection cuts for $m_A=10~\hbox{GeV}$.}
\label{Tab:leptonic_fractions_mA_4}
\end{center}
\end{table}

\begin{table}
\begin{center}
\begin{tabular}{|c|c|c|}
\hline
\multicolumn{3}{|c|}{\bf{CP odd THDM scalar A (10 GeV)} }\\
\hline
\hline
\bf{Non leptonic} & \bf{Single leptonic} & \bf{Double leptonic} \\
\hline
\hline
Transverse  &Transverse & $P_{T, \mu}$   \\
spherocity &spherocity &  \\
\hline
Fox Wolfram-like&Fractional energy & Thrust major \\
$n=1/4$&correlation $x=1.5$  &  \\
\hline
Directly global&C parameter  & $P_{T, e}$ \\
$y_3$  & & \\
\hline
Thrust major&$P_{T, \mu}$ &Fractional energy   \\
&& correlation $x=1.5$   \\
\hline
C parameter&3-jet resolution  &3-jet resolution\\
& $y_3$  Jade  &  $y_3$ Jade  \\
&  (E-scheme)  &(P-scheme)\\
\hline
3-jet resolution  & $P_{T, e}$ & C parameter \\
$y_3$ Jade & & \\
 (E-scheme)   &&\\
\hline
&3-jet resolution &4-jet resolution   \\
&$y_3$ Durham          &  $y_4$ Durham \\
& (P-scheme)&  (P-scheme)\\
\hline
\end{tabular}
\caption{Top observables determined by the Multivariate Analysis to
discriminate the process $ pp \rightarrow H (\rightarrow A (\rightarrow c\bar{c} ) Z ) + \hbox{jets}$  against $pp \rightarrow Z + \hbox{jets}$
for $m_A=10~\hbox{GeV}$.
}
\label{Tab:ccA10}
\end{center}
\end{table}

\begin{table}
\begin{center}
\begin{tabular}{|c|c|c|}
\hline
\bf{Fraction of events} & $pp\rightarrow H + \hbox{jets}$ & $pp\rightarrow Z + \hbox{jets}$ \\
& with  $H\rightarrow A (\rightarrow c \bar{c}) Z$&\\
&$m_A=10~\hbox{GeV}$ &\\
\hline
\hline
$f^{(0)}$ (0 leptons) & $92.1\%$ & $84.0\%$   \\
\hline
$f^{(1)}$ (1 lepton)& $7.6\%$ & $15.15\%$ \\
\hline
$f^{(2)}$ (2 leptons)& $0.3\%$ & $0.85\%$ \\
\hline
\end{tabular}
\caption{Fraction of events in each one of the leptonic categories for the samples $ pp \rightarrow H (\rightarrow A (\rightarrow c\bar{c} ) Z ) + \hbox{jets}$ and
$pp\rightarrow Z + \hbox{jets}$ after the selection cuts for $m_A=10~\hbox{GeV}$.}
\label{Tab:leptonic_fractions_mA_10}
\end{center}
\end{table}

\section{\label{sec:concl} Conclusions}

We have studied the efficiency of event shapes for tagging jets resulting from $c\bar{c}$ pairs originated
in the decay $H\rightarrow c\bar{c}$. The results obtained can be considered as an optimal limit for the performance of 
our selection strategy as we have not included detector effects. We have optimized our analysis depending
on the main backgrounds in the selected processes and have taken into account the following possibilities $pp\rightarrow q \bar{q} Z$ 
for $q=\{u, d, s, c\}$ and $pp\rightarrow g g Z$. Our signal channel is $pp\rightarrow H (\rightarrow c\bar{c}) Z$
 and we select highly boosted Higgs bosons.
Using jet shape observables as input to a BDT, we find a good performance to separate the $c\bar{c}$ signal from $b\bar{b}$ and light-flavor fat jets.\\

Thus, with this approach we can project an upper limit on 
 $\mathcal{B}r(H\rightarrow c\bar{c})\leq 6.1\%$ with SM production rates for $\int\mathcal{L}dt=3000.0~\hbox{fb}^{-1}$ and  $\sqrt{s}=13.0~\hbox{TeV}$.\\

Following an analogous strategy we have studied the CP-odd THDM scalar $A$ decaying into pairs $c\bar{c}$. In particular
we have determined $\mathcal{B}r(H\rightarrow A(\rightarrow c\bar{c}) Z)\leq 0.01\%$ by considering 
masses  for $A$ inside the range  $4.0~\hbox{GeV} \lesssim  m_{A} \lesssim  10.0~\hbox{GeV}$ where the channel $A\rightarrow c\bar{c}$ is particularly dominant for low values of $\tan\beta$.

\begin{acknowledgments}
MS is supported in part by the European Commission through the 'HiggsTools' Initial Training Network PITN-GA-2012-316704. 
We thank Silvan Kuttimalai and Nathan Hartland for useful discussions on SHERPA and multivariate analyses respectively.
\end{acknowledgments}

\begin{center}
\begin{figure*}
\includegraphics[height=8.0cm]{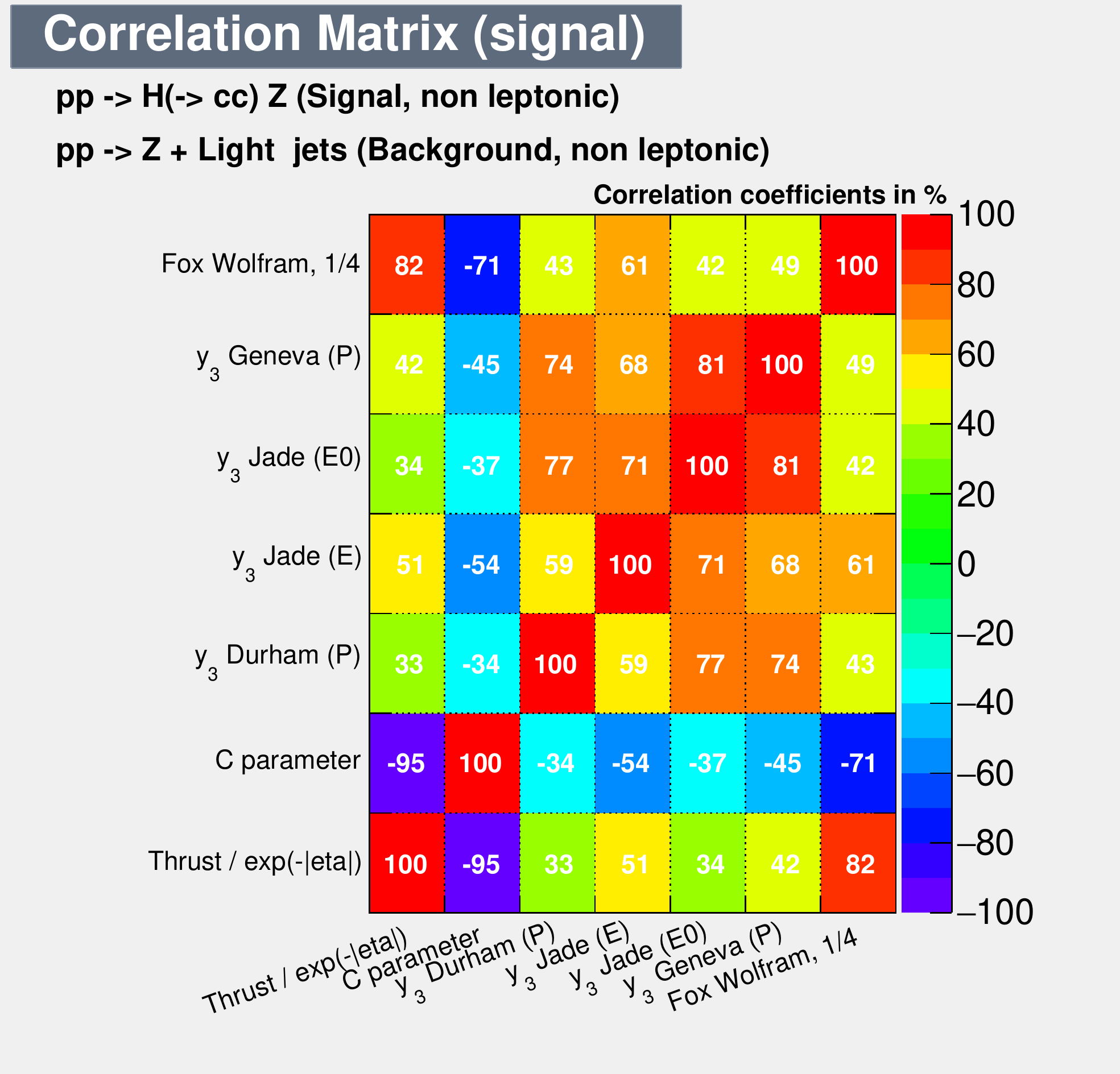}
\includegraphics[height=8.0cm]{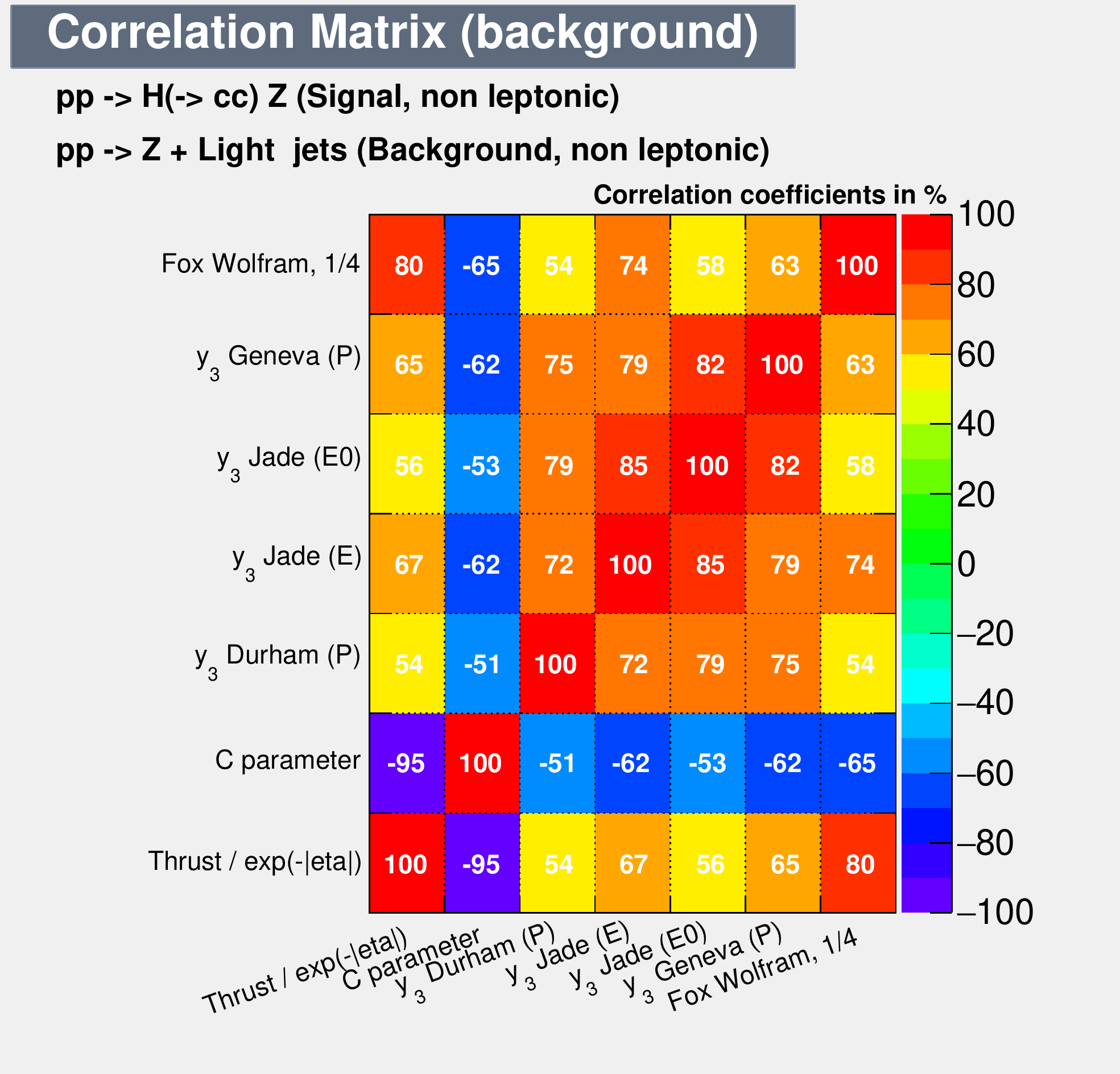}
\includegraphics[height=8.0cm]{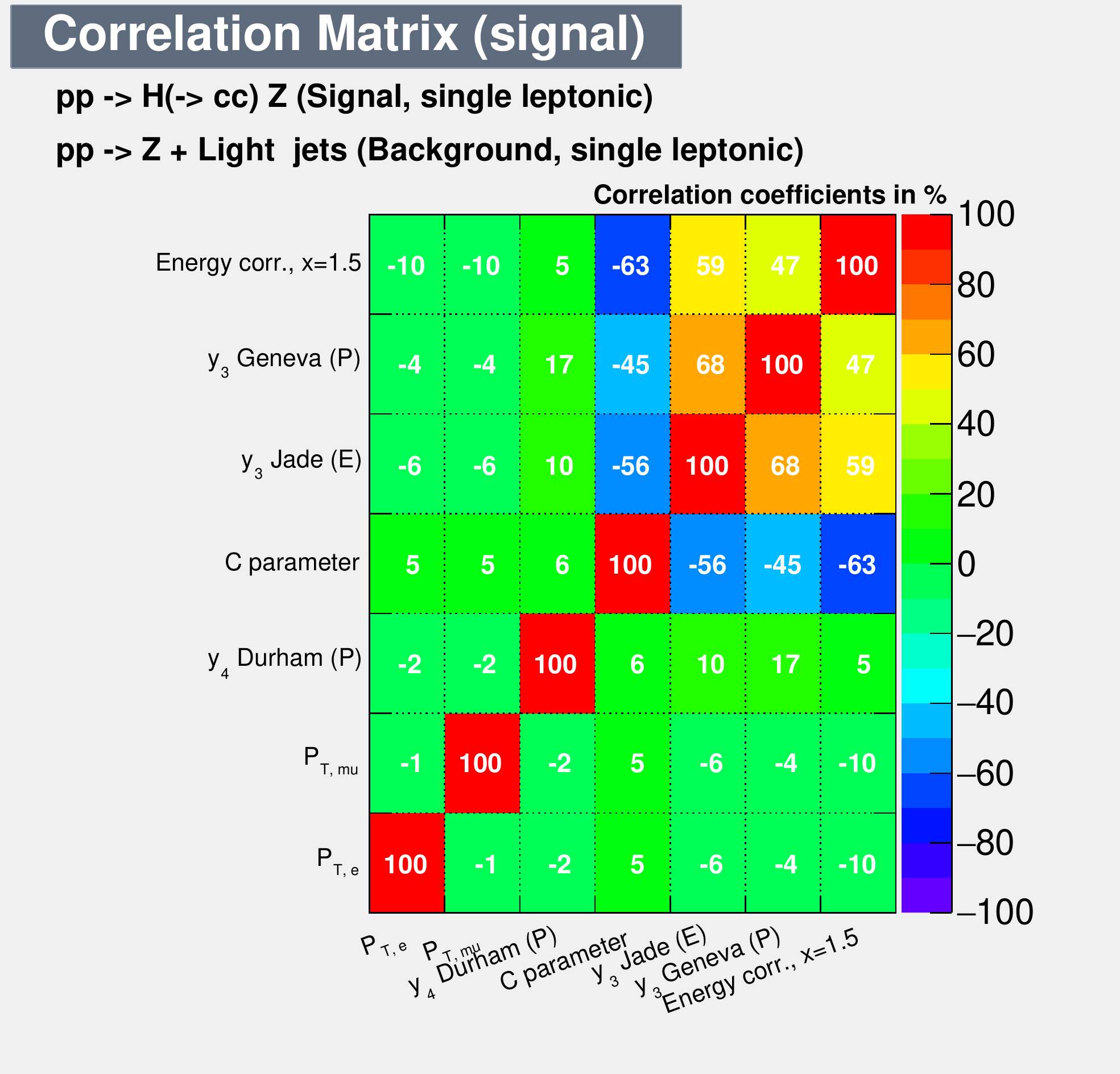}
\includegraphics[height=8.0cm]{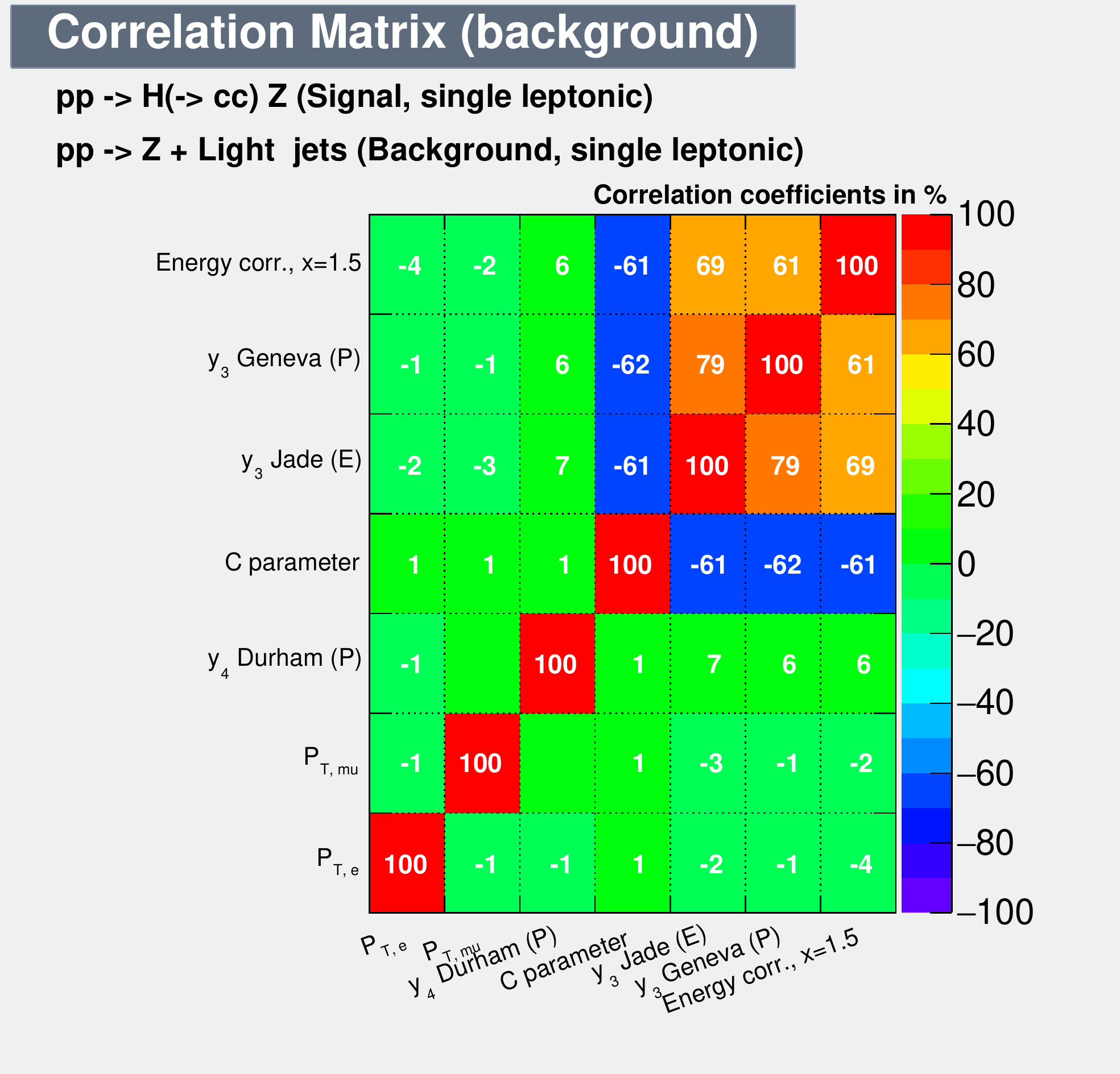}
\includegraphics[height=8.0cm]{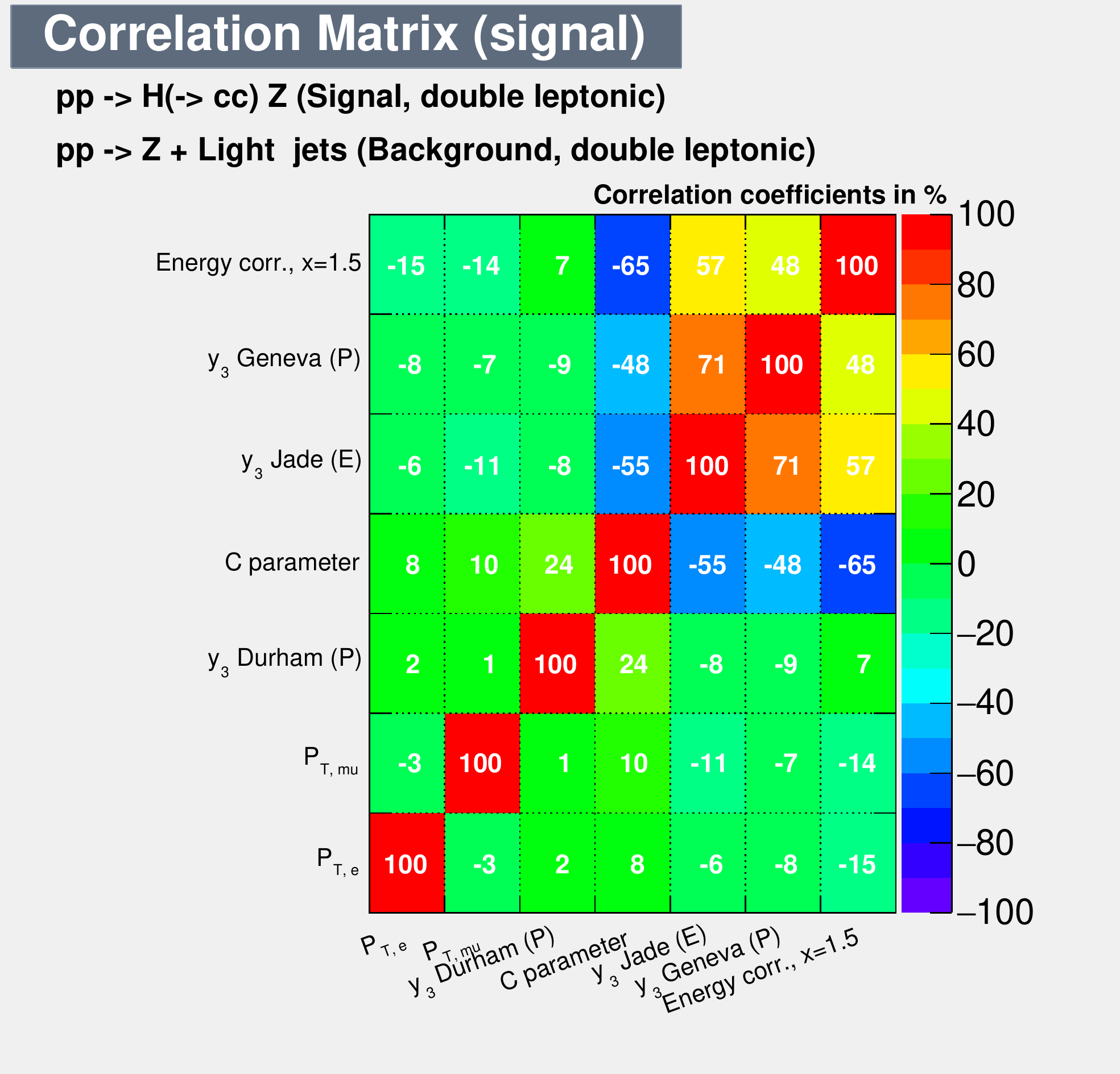}
\includegraphics[height=8.0cm]{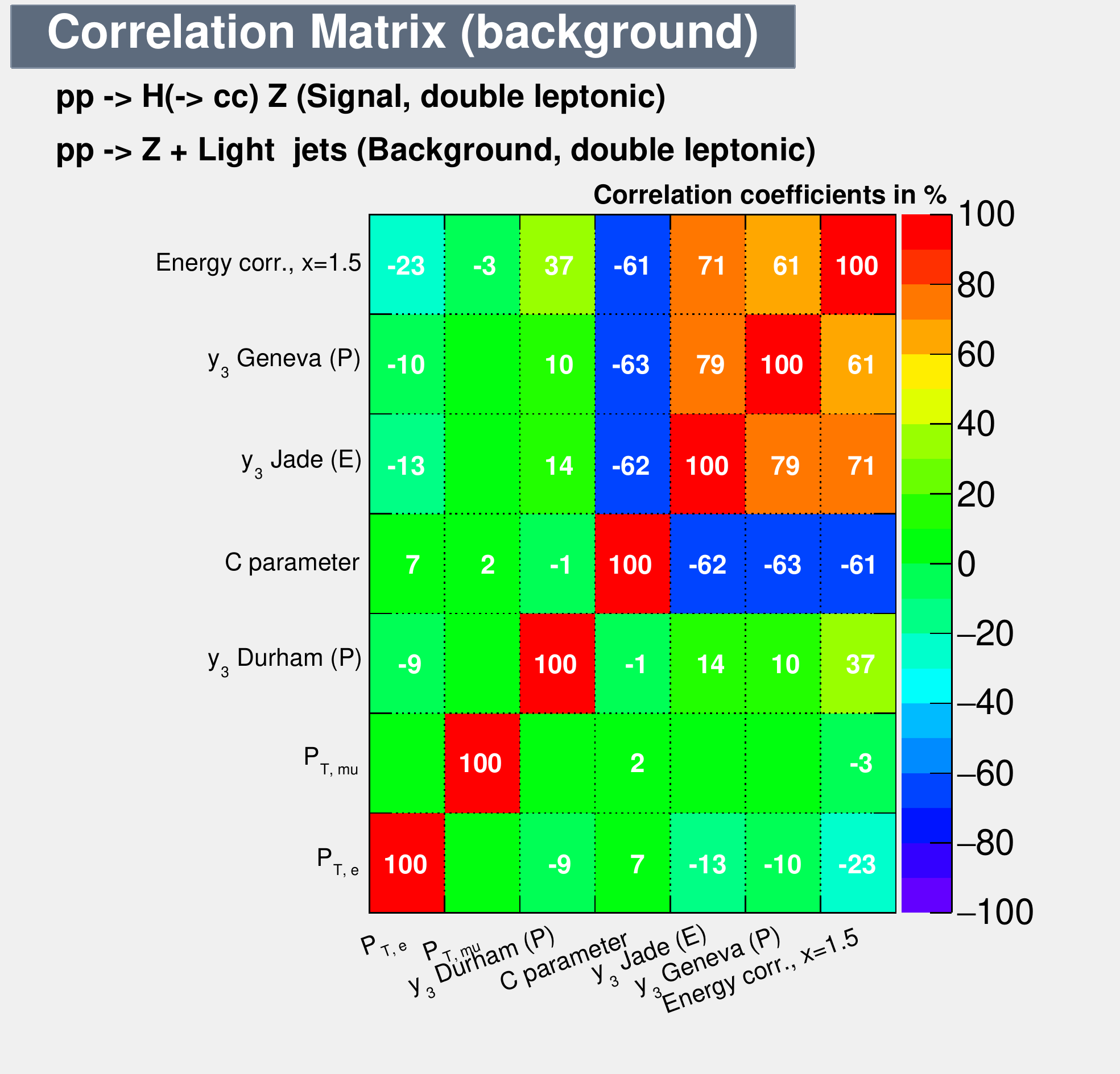}
\caption{Correlation coefficients for the observables used in the analysis $pp\rightarrow H(\rightarrow c \bar{c}) Z$ (signal)
vs $pp\rightarrow Z + \mathrm{light~jets}$
(background).}
\label{fig:CorrelationSMLightJets}
\end{figure*}
\end{center}

\begin{figure*}
\includegraphics[height=5cm]{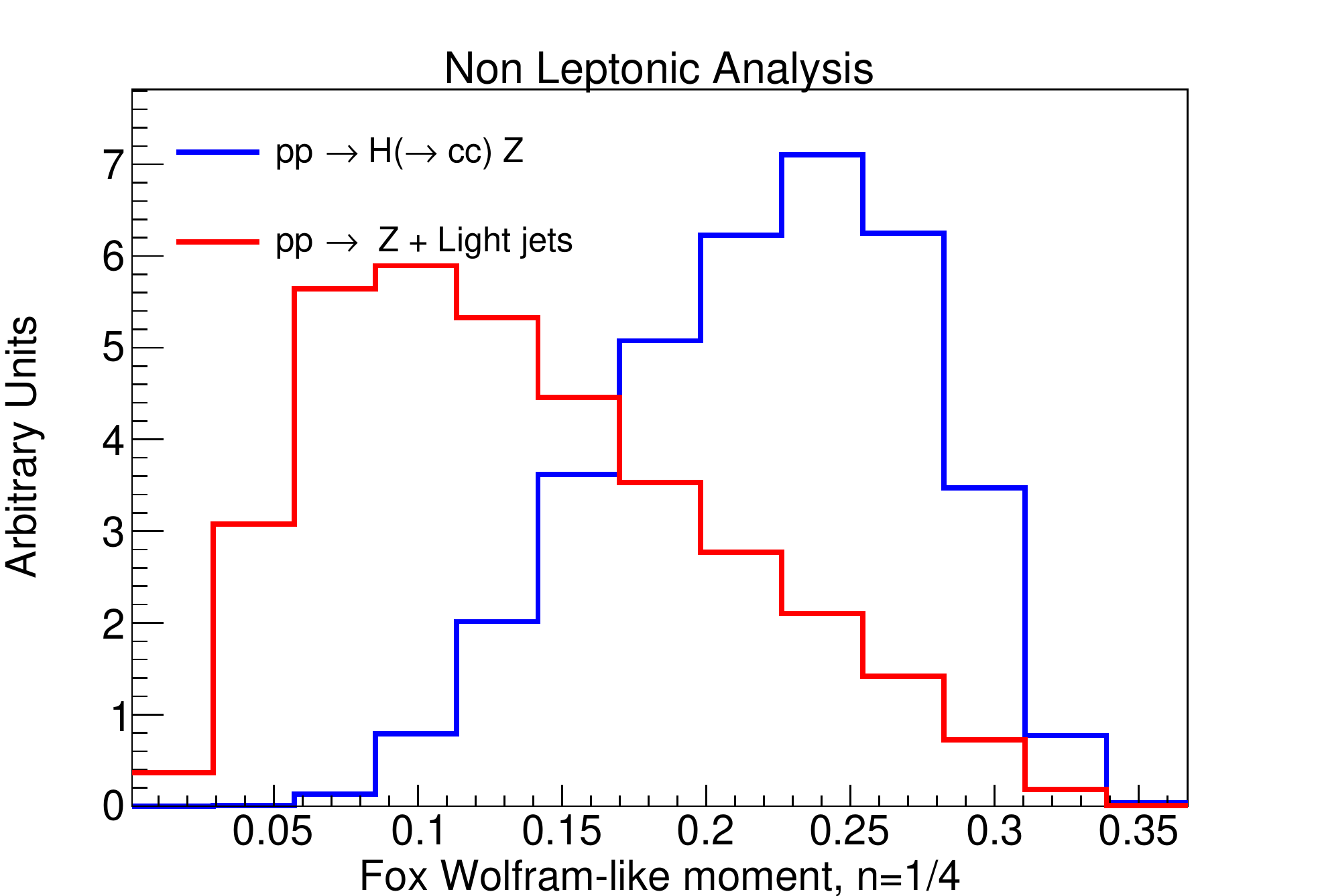}
\includegraphics[height=5cm]{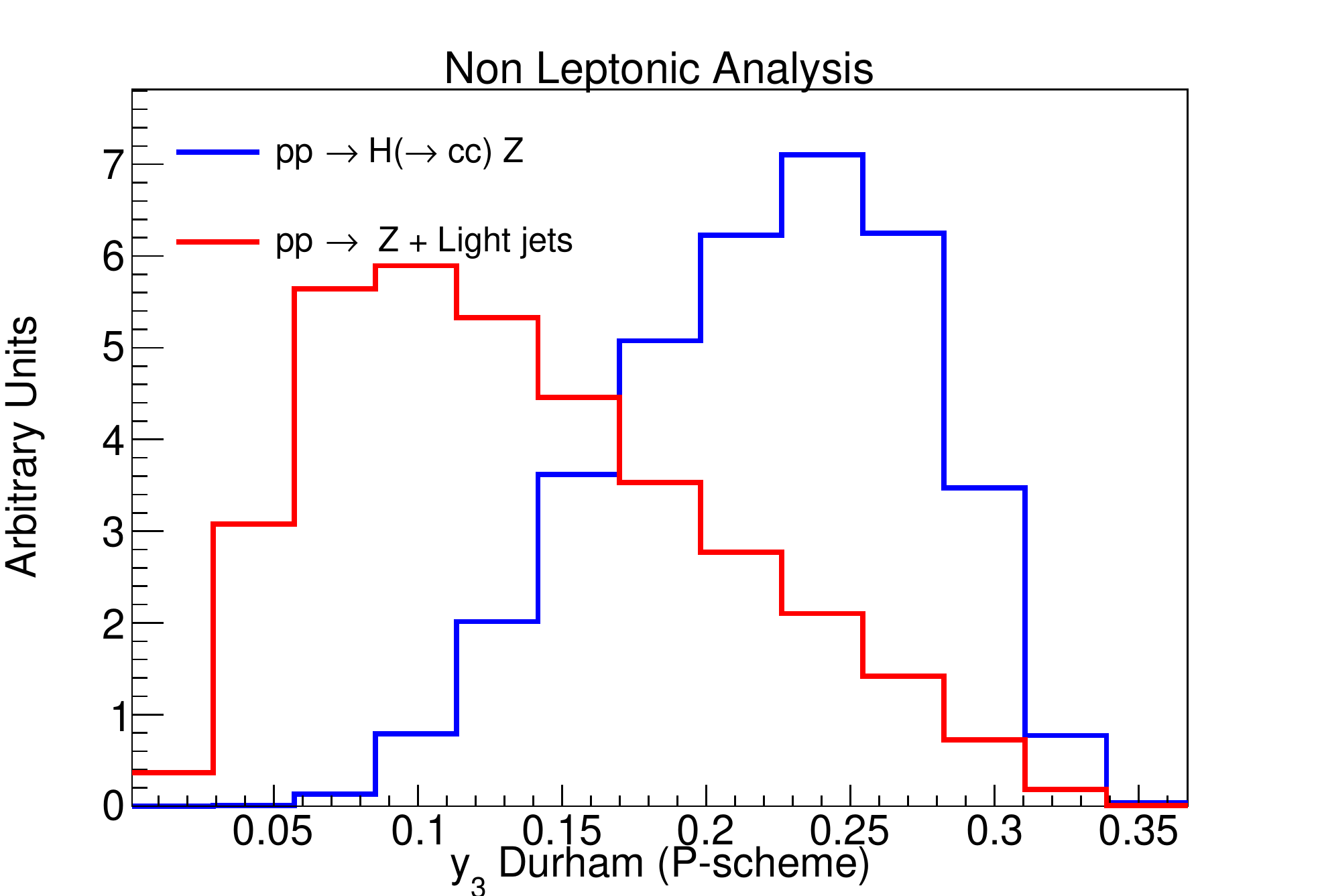}
\includegraphics[height=5cm]{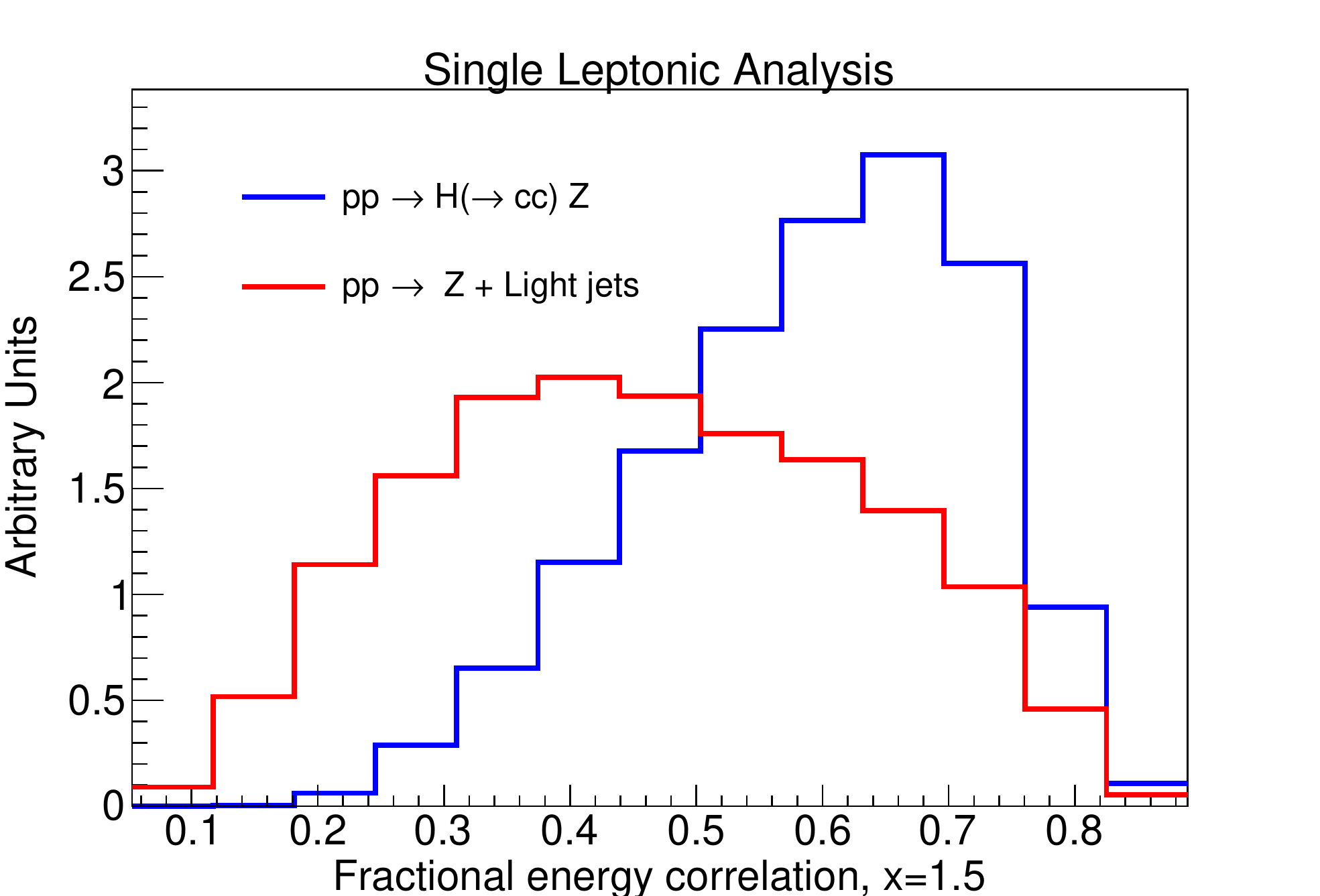}
\includegraphics[height=5cm]{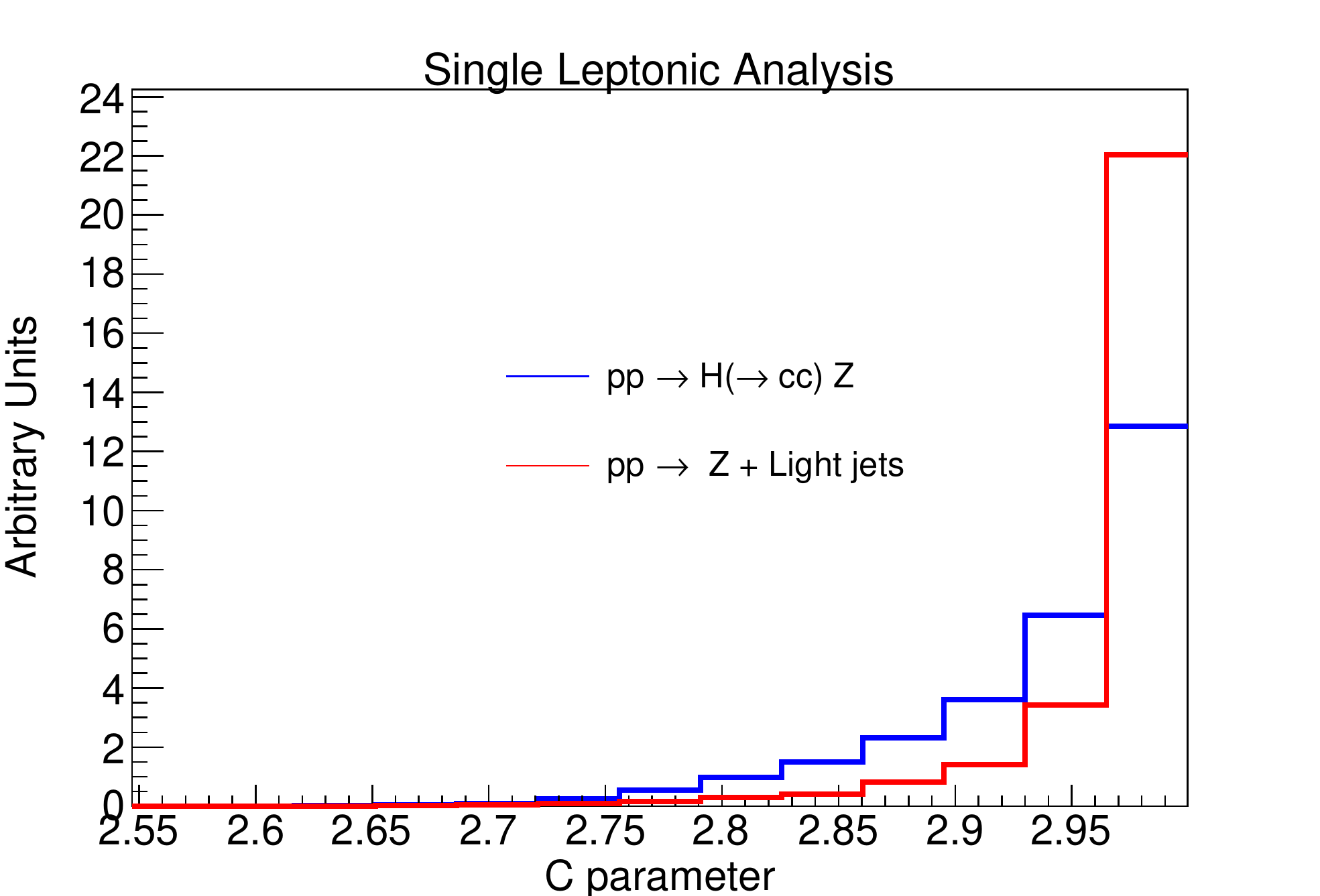}
\includegraphics[height=5cm]{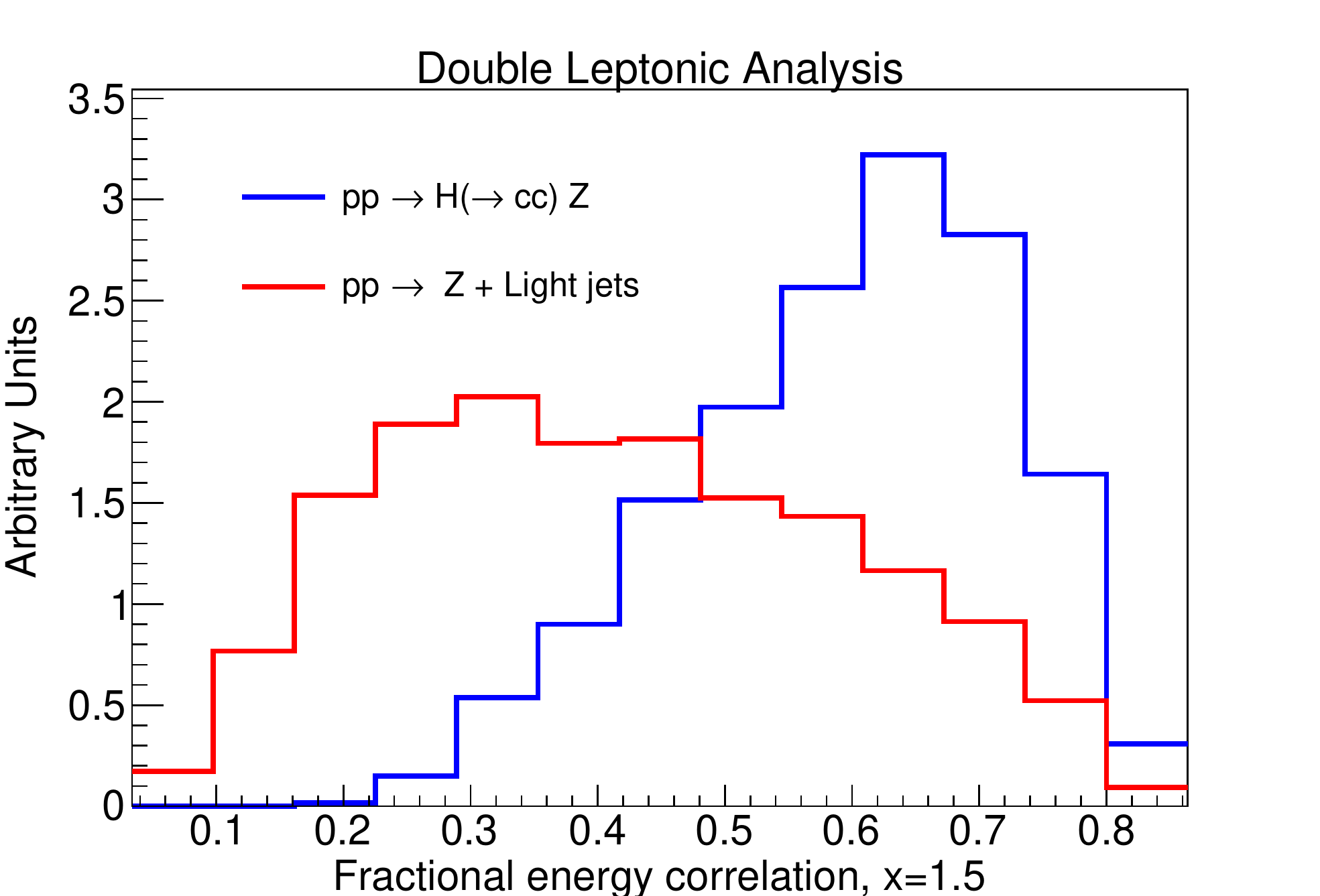}
\includegraphics[height=5cm]{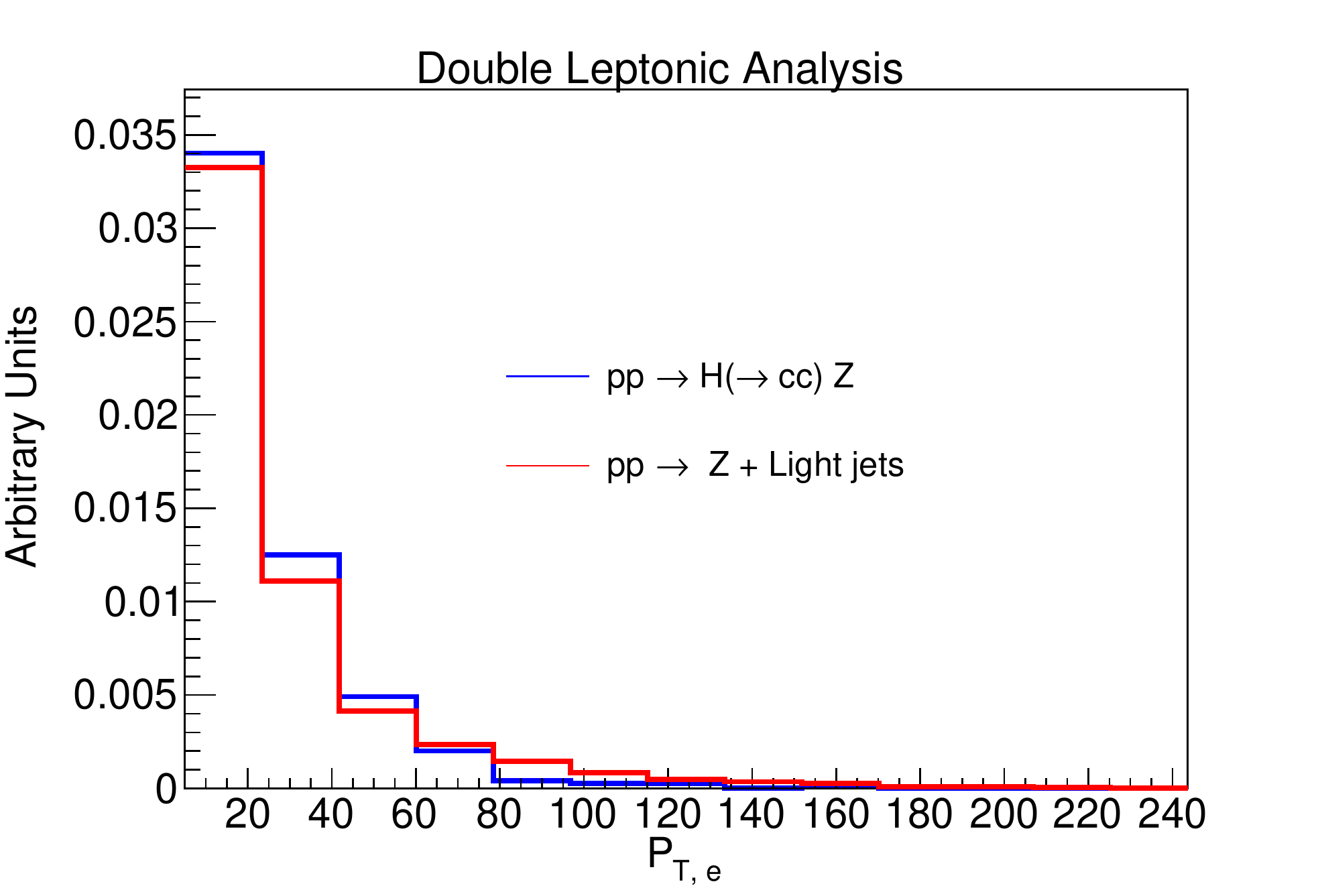}
\caption{Histograms for the main discrimination event shapes when selecting $pp\rightarrow H (\rightarrow c\bar{c})Z$ against
$pp\rightarrow  Z + \hbox{Light jets}$.}
\label{fig:HistoObsLightJets}
\end{figure*}

\appendix*
\section{\label{sec:EvS}Event Shapes}

This section summarizes most of the observables considered
during our analysis. For a more extensive discussion see   
\cite{Banfi:2003je, Banfi:2004nk, Banfi:2004yd} and the references cited therein.\\

We begin by introducing the definition of Thrust \cite{Brandt:1964sa, Farhi:1977sg}

\begin{eqnarray}
T=1-\max\Bigl(\frac{\sum_i |\vec{p}_i\cdot \vec{n}_T| }{\sum_i|\vec{p}_i | }\Bigl),
\label{eq:T}
\end{eqnarray}

where $\vec{n}_T$ is the direction that maximizes the numerator. To avoid confusion, 
in the subsequent discussion the symbol
``$\perp$'' will be used to denote the transverse contribution of different kinematical variables.  Then, the  {\bf Thrust major} is determined according to
\cite{Barber:1980uq}

\begin{eqnarray}
T_M=\max\limits_{\vec{n}\cdot \vec{n}_T=0}\Bigl(\frac{\sum_i |\vec{p}_i\cdot \vec{n}| }{\sum_i|\vec{p}_i | }\Bigl),
\end{eqnarray}

where it should be understood that $\vec{n}$ is perpendicular to $\vec{n}_T$.\\

We use the  {\bf Thrust of $e^{-\eta}$ momenta} \cite{Banfi:2004yd} calculated according to Eq. (\ref{eq:T})
but with the three-momenta of each one of the subjets in the event modified according to

\begin{eqnarray}
\vec{p}_i\rightarrow \vec{p}_i~ e^{-|\eta_i|}.
\end{eqnarray}

We include the Fox Wolfram moment inspired observable \cite{Fox:1978vu}

\begin{eqnarray}
H_n=\sum_{i,j}\frac{|\vec{p}_i| |\vec{p}_j|}{E^2_{Tot}}\sin^{2n}\theta_{ij},
\label{eq:Hn}
\end{eqnarray}

with $E_{Tot}$ being the total energy of the jet constituents, thus $E_{Tot}=\sum_i E_i$. The sum in the numerator of Eq. (\ref{eq:Hn})
considers only pairs of particles within the same hemisphere, i.e. those particles 
satisfying $\vec{p}_i\cdot \vec{p}_j>0$, and $n$ is a rational number. In what follows we will refer to $H_n$ as the {\bf  Fox Wolfram-like $n$ moment}, and
we will consider the value $n=1/4$.\\

The {\bf Transverse spherocity} \cite{Banfi:2010xy} is given by

\begin{eqnarray}
S_{ph}=\hbox{min}\Bigl(\frac{2}{\pi}\Bigl)^2\frac{\Bigl(\underset{i}{\sum} |\vec{p}_{\perp}\times \hat{n}''|\Bigl)^2}
{\Bigl(\underset{i}{\sum}|\vec{p}_{\perp,i}|\Bigl)^2},
\end{eqnarray}

where $\hat{n}''$ is the direction that minimizes the sum in the numerator.\\

The {\bf 3-jet resolution $y_3$}, defines the lower bound for the jet recombination parameter $y_{ij}$ in order to have a 3-jet event. 
Before presenting the determination algorithm for $y_3$, according to different schemes, let us first introduce the possible definitions
for the parameter $y_{ij}$

\begin{eqnarray}
y_{ij} &=&
\begin{cases}
   2~\hbox{min}(E^2_i, E^2_j)(1-\hbox{cos}\theta_{ij})/E^2_{vis}&\hbox{Durham}\\
   &\\
   \frac{8 E_i E_j (1-\hbox{cos}\theta_{ij})}{9 (E_i + E_j)^2}&\hbox{Geneva}\nonumber\\
   &\\
   (p_i + p_j)^2/E^2_{vis}&\hbox{Jade}.\\
\end{cases}
\label{eq:effWC}
\end{eqnarray}

with $E_{vis}$ is the sum of the energies for the different final state subjets before the recombinations.

In addition, the recombination schemes between the $i$-th and $j$-th subjets are

 \begin{eqnarray}
\hbox{ Schemes} &=&
\begin{cases}
   \hbox{P: }   \vec{p}=\vec{p}_i + \vec{p}_j, E_p=|\vec{p}|\\
   &\\
   \hbox{E: } p=p_i + p_j \nonumber\\
    &\\
   \hbox{E0: } \vec{p}=\frac{E_i + E_j}{|\vec{p}_i + \vec{p}_j|}(\vec{p}_i + \vec{p}_j), 
   E_p=E_{p,i} + E_{p,j}.
\end{cases}
\label{eq:schemes}
\end{eqnarray}

Then for example, in order to calculate the {\bf Resolution $y_3$ Durham (P-scheme)} \cite{Catani:1991hj, Banfi:2001bz}, we start by assigning an arbitrary high value to 
$y_3$. Next, we calculate the parameter $y_{ij}$ between all the subjets inside a given fat jet using the Durham recombination
rule shown before. We then determine the pair of elements whose $y_{ij}$ is minimum, $y^{min}_{ij}$ , and recombine them
applying the P-scheme presented above. Finally, if $y_3<y^{min}_{ij}$, we do the substitution $y_3=y^{min}_{ij}$ and
repeat the entire process, starting with the re-calculation of the values $y_{ij}$ over the set of subjets
determined in the last iteration. The algorithm stops when the total number of 
subjets left after all the recombinations is equal to 3. The value of  $y_3$ obtained in the final iteration is the number we are aiming for.
The determination of the {\bf Resolution $y_3$ Jade (E-scheme)} and  the {\bf Resolution $y_3$ Jade (E0-scheme)}   proceed in an analogous way; however the Durham parameter $y_{ij}$ should be substituted by the
Jade distance parameter; and the P recombination scheme should be replaced by the E-scheme (E0-scheme).\\

The {\bf Directly global $y_3$} \cite{Banfi:2010xy} is constructed using the $k_t$ jet algorithm. To begin with, for all $n$ 
final state particles we define the beam-distance measure 

\begin{eqnarray}
d_{k,B}&=&p^2_{\perp k},
\end{eqnarray}

and for constituent pairs we calculate 

\begin{eqnarray}
d_{kl}&=&\hbox{min}\{p^2_{\perp k}, p^2_{\perp l}\}\frac{(y_k - y_l)^2 + (\phi_k - \phi_l)^2}{R^2},
\end{eqnarray}
in terms of their corresponding pseudo rapidity $y$ and azimuthal angle $\phi$.

In our analysis we use $R=0.7$. Let $d^{(n)}=\hbox{min}\{d_{k B} , d_{kl}\}$, where the entire set of distances 
calculated at a given stage is considered. If $d^{(n)}$
is one of the values $d_{ij}$, then the pseudojets $i$ and $j$ are recombined using the E-scheme defined above. If  $d^{(n)}$ is one of the individual coefficients $d_{k B}$ then the pseudojet is removed and included in the beam. These steps are repeated until only 3-pseudo jets are left. At this stage we determine

\begin{eqnarray}
y_{23}&=&\frac{1}{P^2_{\perp}}\underset{n\geq 3}{\hbox{max}}\{d^{(n)}\},
\end{eqnarray}

with

\begin{eqnarray}
P_{\perp}= p_{\perp, 1} +  p_{\perp, 2},
\end{eqnarray}

being $p_{\perp, 1}$ and  $p_{\perp, 2}$  the transverse momenta of the jets obtained by continuing reclustering the event up to 2-pseudojets.\\

The observable $\tau_x$ can be modified to give the {\bf Fractional energy correlation} \cite{Banfi:2004yd}

\begin{eqnarray}
FC_x &=& \sum_{i\neq j}\frac{E_i E_j |\sin\theta_{ij}|\Bigl(1-|\cos\theta_{ij}| \Bigl)^{1-x}} 
{\Bigl(\underset{i}{\sum} E_i\Bigl)^2}\times\nonumber \\
&&\Theta[(\vec{p}_i\cdot \vec{n}_T)(\vec{p}_j\cdot \vec{n}_T)].
\end{eqnarray}

Here $x$ is a continuous parameter. During the analysis we
use $x=1.5$ that makes the observable particularly sensitive to collinear emissions for fixed transverse momentum.\\

To define the {\bf Transverse sphericity} let us first introduce the transverse momentum tensor

\begin{equation}
M_{xy} = \sum_i\left( \begin{array}{ccc}
p^2_{x, i}  & p_{x,i}  p_{y,i}   \\
p_{x,i}  p_{y,i} & p^2_{y,i}  \end{array} \right).
\end{equation}
\\
Then the transverse sphericity can be determined in terms of the eigenvalues $\lambda_1$ and $\lambda_2$ of $M_{xy}$
(for $\lambda_1 \geq \lambda_2$) as \cite{Bjorken:1969wi}

\begin{equation}
S^{pheri}_{\perp, g}\equiv \frac{2\lambda_2}{\lambda_1 + \lambda_2},
\end{equation}

for circular events in the transverse plane we have $S^{pheri}_{\perp, g}\rightarrow 1$, whereas for pencil like events
$S^{pheri}_{\perp, g}\rightarrow 0$.\\

To describe the cone jet mass let us start by introducing some definitions. The components of the highest $p_T$ fat jet
selected in our studies are first reclustered using the $k_t$ algorithm. Then, the region  $\mathcal{C}$
results from the union of the cones around the two new highest transverse momentum subjets (with coordinates
 $\eta_{J,j}, \phi_{J,j}$,  for $j=1,2$) according to

\begin{eqnarray}
\sqrt{(\eta_i-\eta_{J,j})^2 + (\phi_i-\phi_{J,j})^2}\leq R.
\end{eqnarray}

Where  the subindex $i$ runs over the rest of the newly generated subjets. During our implementation we considered $R=1$.
 The central transverse thrust axis $\vec{n}_{T, C}$ is then defined as the vector that 
maximizes 

\begin{eqnarray}
\frac{\underset{i\in \mathcal{C}}{\sum} |\vec{p}_{\perp i}\cdot \vec{n}_{T, C}| }{Q_{\perp, \mathcal{C}}},
\end{eqnarray}

where 

\begin{eqnarray}
Q_{\perp, \mathcal{C}}&=& \sum_{i\in \mathcal{C}} |\vec{p}_{\perp i}|.
\end{eqnarray}

The vector $\vec{n}_{T, C}$ allow us to divide the region $\mathcal{C}$ into the subregions $\mathcal{C}_U$ and $\mathcal{C}_D$, 
defined in terms of the conditions
$0<\vec{p}_{\perp}\cdot \vec{n}_{T, C}$ and $\vec{p}_{\perp}\cdot \vec{n}_{T, C}<0$ respectively. The partial masses
in each one of these regions are

\begin{eqnarray}
\rho_{U, \mathcal{C}}=\frac{\Biggl(\underset{i\in \mathcal{C}_U}{\sum}p_i \Biggl)^2}{Q^2_{\perp, \mathcal{C}}},&&
\rho_{D, \mathcal{C}}=\frac{\Biggl(\underset{i\in \mathcal{C}_D}{\sum}p_i \Biggl)^2}{Q^2_{\perp, \mathcal{C}}}.
\end{eqnarray}

Then, the {\bf Cone total jet mass} is

\begin{eqnarray}
\rho_{S, \mathcal{C}}&=&\rho_{U, \mathcal{C}} + \rho_{D, \mathcal{C}},
\end{eqnarray}

and the heavy jet mass is defined as

\begin{eqnarray}
\rho_{H,\mathcal{C}}&=&\hbox{max}\{\rho_{U, \mathcal{C}}, \rho_{D, \mathcal{C}}\}.
\end{eqnarray}

We can further add the exponentially suppressed term $\mathcal{E}_{\bar{\mathcal{C}}}$ given by

\begin{eqnarray}
\mathcal{E}_{\mathcal{\bar{C}}} &\equiv&\frac{1}{Q_{\perp, \mathcal{C}}}
\sum_{i \notin \mathcal{C}}|\vec{p}_{\perp, i}| e^{-|\eta_i - \eta_{\mathcal{C}}|},
\end{eqnarray}

where 

\begin{eqnarray}
\eta_{\mathcal{C}}\equiv \frac{1}{Q_{\perp, \mathcal{C}} }\sum_{i\in \mathcal{C}}\eta_i |\vec{p}_{\perp i}|.
\end{eqnarray}

The {\bf Central heavy jet mass with exponentially suppressed forward term} \cite{Banfi:2004yd}  is calculated according to

\begin{eqnarray}
\rho_{H, \mathcal{E}}&=&\rho_{H, \mathcal{C}} + \mathcal{E}_{\mathcal{\bar{C}}}.
\end{eqnarray}

Finally, we consider the following  {\bf $C$ parameter}-like observable  \cite{Parisi:1978eg, Donoghue:1979vi, Catani:1998sf}

\begin{eqnarray}
C&=&3 -\frac{3}{E^2_{Tot}} \sum_{i<j} \frac{(p_i\cdot p_j)^2}{E_i E_j},
\end{eqnarray}

with $E_{Tot}=\underset{i}{\sum} E_i$.

\bibliography{main}

\begin{thebibliography}{55}
\expandafter\ifx\csname natexlab\endcsname\relax\def\natexlab#1{#1}\fi
\expandafter\ifx\csname bibnamefont\endcsname\relax
  \def\bibnamefont#1{#1}\fi
\expandafter\ifx\csname bibfnamefont\endcsname\relax
  \def\bibfnamefont#1{#1}\fi
\expandafter\ifx\csname citenamefont\endcsname\relax
  \def\citenamefont#1{#1}\fi
\expandafter\ifx\csname url\endcsname\relax
  \def\url#1{\texttt{#1}}\fi
\expandafter\ifx\csname urlprefix\endcsname\relax\def\urlprefix{URL }\fi
\providecommand{\bibinfo}[2]{#2}
\providecommand{\eprint}[2][]{\url{#2}}

\bibitem[{\citenamefont{Aad et~al.}(2012)}]{Aad:2012tfa}
\bibinfo{author}{\bibfnamefont{G.}~\bibnamefont{Aad}} \bibnamefont{et~al.}
  (\bibinfo{collaboration}{ATLAS}), \bibinfo{journal}{Phys. Lett.}
  \textbf{\bibinfo{volume}{B716}}, \bibinfo{pages}{1} (\bibinfo{year}{2012}),
  \eprint{1207.7214}.

\bibitem[{\citenamefont{Chatrchyan et~al.}(2012)}]{Chatrchyan:2012xdj}
\bibinfo{author}{\bibfnamefont{S.}~\bibnamefont{Chatrchyan}}
  \bibnamefont{et~al.} (\bibinfo{collaboration}{CMS}), \bibinfo{journal}{Phys.
  Lett.} \textbf{\bibinfo{volume}{B716}}, \bibinfo{pages}{30}
  (\bibinfo{year}{2012}), \eprint{1207.7235}.

\bibitem[{\citenamefont{Khachatryan
  et~al.}(2015{\natexlab{a}})}]{Khachatryan:2014jba}
\bibinfo{author}{\bibfnamefont{V.}~\bibnamefont{Khachatryan}}
  \bibnamefont{et~al.} (\bibinfo{collaboration}{CMS}), \bibinfo{journal}{Eur.
  Phys. J.} \textbf{\bibinfo{volume}{C75}}, \bibinfo{pages}{212}
  (\bibinfo{year}{2015}{\natexlab{a}}), \eprint{1412.8662}.

\bibitem[{\citenamefont{Aad et~al.}(2016{\natexlab{a}})}]{Aad:2015gba}
\bibinfo{author}{\bibfnamefont{G.}~\bibnamefont{Aad}} \bibnamefont{et~al.}
  (\bibinfo{collaboration}{ATLAS}), \bibinfo{journal}{Eur. Phys. J.}
  \textbf{\bibinfo{volume}{C76}}, \bibinfo{pages}{6}
  (\bibinfo{year}{2016}{\natexlab{a}}), \eprint{1507.04548}.

\bibitem[{\citenamefont{Aad et~al.}(2016{\natexlab{b}})}]{Khachatryan:2016vau}
\bibinfo{author}{\bibfnamefont{G.}~\bibnamefont{Aad}} \bibnamefont{et~al.}
  (\bibinfo{collaboration}{ATLAS, CMS}), \bibinfo{journal}{JHEP}
  \textbf{\bibinfo{volume}{08}}, \bibinfo{pages}{045}
  (\bibinfo{year}{2016}{\natexlab{b}}), \eprint{1606.02266}.

\bibitem[{\citenamefont{Ghosh et~al.}(2016)\citenamefont{Ghosh, Gupta, and
  Perez}}]{Ghosh:2015gpa}
\bibinfo{author}{\bibfnamefont{D.}~\bibnamefont{Ghosh}},
  \bibinfo{author}{\bibfnamefont{R.~S.} \bibnamefont{Gupta}}, \bibnamefont{and}
  \bibinfo{author}{\bibfnamefont{G.}~\bibnamefont{Perez}},
  \bibinfo{journal}{Phys. Lett.} \textbf{\bibinfo{volume}{B755}},
  \bibinfo{pages}{504} (\bibinfo{year}{2016}), \eprint{1508.01501}.

\bibitem[{\citenamefont{Perez et~al.}(2015)\citenamefont{Perez, Soreq, Stamou,
  and Tobioka}}]{Perez:2015aoa}
\bibinfo{author}{\bibfnamefont{G.}~\bibnamefont{Perez}},
  \bibinfo{author}{\bibfnamefont{Y.}~\bibnamefont{Soreq}},
  \bibinfo{author}{\bibfnamefont{E.}~\bibnamefont{Stamou}}, \bibnamefont{and}
  \bibinfo{author}{\bibfnamefont{K.}~\bibnamefont{Tobioka}},
  \bibinfo{journal}{Phys. Rev.} \textbf{\bibinfo{volume}{D92}},
  \bibinfo{pages}{033016} (\bibinfo{year}{2015}), \eprint{1503.00290}.

\bibitem[{\citenamefont{collaboration}(2017)}]{ATLAS:2017bic}
\bibinfo{author}{\bibfnamefont{T.~A.} \bibnamefont{collaboration}}
  (\bibinfo{collaboration}{ATLAS}) (\bibinfo{year}{2017}).

\bibitem[{\citenamefont{Sirunyan et~al.}(2017)}]{Sirunyan:2017khh}
\bibinfo{author}{\bibfnamefont{A.~M.} \bibnamefont{Sirunyan}}
  \bibnamefont{et~al.} (\bibinfo{collaboration}{CMS}) (\bibinfo{year}{2017}),
  \eprint{1708.00373}.

\bibitem[{\citenamefont{Aad et~al.}(2014)}]{Aad:2014xva}
\bibinfo{author}{\bibfnamefont{G.}~\bibnamefont{Aad}} \bibnamefont{et~al.}
  (\bibinfo{collaboration}{ATLAS}), \bibinfo{journal}{Phys. Lett.}
  \textbf{\bibinfo{volume}{B738}}, \bibinfo{pages}{68} (\bibinfo{year}{2014}),
  \eprint{1406.7663}.

\bibitem[{\citenamefont{Khachatryan
  et~al.}(2015{\natexlab{b}})}]{Khachatryan:2014aep}
\bibinfo{author}{\bibfnamefont{V.}~\bibnamefont{Khachatryan}}
  \bibnamefont{et~al.} (\bibinfo{collaboration}{CMS}), \bibinfo{journal}{Phys.
  Lett.} \textbf{\bibinfo{volume}{B744}}, \bibinfo{pages}{184}
  (\bibinfo{year}{2015}{\natexlab{b}}), \eprint{1410.6679}.

\bibitem[{CMS(2013)}]{CMS:2013xfa}
in \emph{\bibinfo{booktitle}{{Proceedings, 2013 Community Summer Study on the
  Future of U.S. Particle Physics: Snowmass on the Mississippi (CSS2013):
  Minneapolis, MN, USA, July 29-August 6, 2013}}} (\bibinfo{year}{2013}),
  \eprint{1307.7135},
  \urlprefix\url{https://inspirehep.net/record/1244669/files/arXiv:1307.7135.p%
df}.

\bibitem[{\citenamefont{Altmannshofer et~al.}(2015)\citenamefont{Altmannshofer,
  Brod, and Schmaltz}}]{Altmannshofer:2015qra}
\bibinfo{author}{\bibfnamefont{W.}~\bibnamefont{Altmannshofer}},
  \bibinfo{author}{\bibfnamefont{J.}~\bibnamefont{Brod}}, \bibnamefont{and}
  \bibinfo{author}{\bibfnamefont{M.}~\bibnamefont{Schmaltz}},
  \bibinfo{journal}{JHEP} \textbf{\bibinfo{volume}{05}}, \bibinfo{pages}{125}
  (\bibinfo{year}{2015}), \eprint{1503.04830}.

\bibitem[{\citenamefont{d'Enterria}(2017)}]{dEnterria:2016sca}
\bibinfo{author}{\bibfnamefont{D.}~\bibnamefont{d'Enterria}}, in
  \emph{\bibinfo{booktitle}{{Proceedings, 17th Lomonosov Conference on
  Elementary Particle Physics: Moscow, Russia, August 20-26, 2015}}}
  (\bibinfo{year}{2017}), pp. \bibinfo{pages}{182--191}, \eprint{1602.05043},
  \urlprefix\url{https://inspirehep.net/record/1421932/files/arXiv:1602.05043.%
pdf}.

\bibitem[{\citenamefont{Aad et~al.}(2015)}]{Aad:2015sda}
\bibinfo{author}{\bibfnamefont{G.}~\bibnamefont{Aad}} \bibnamefont{et~al.}
  (\bibinfo{collaboration}{ATLAS}), \bibinfo{journal}{Phys. Rev. Lett.}
  \textbf{\bibinfo{volume}{114}}, \bibinfo{pages}{121801}
  (\bibinfo{year}{2015}), \eprint{1501.03276}.

\bibitem[{\citenamefont{Bodwin et~al.}(2013)\citenamefont{Bodwin, Petriello,
  Stoynev, and Velasco}}]{Bodwin:2013gca}
\bibinfo{author}{\bibfnamefont{G.~T.} \bibnamefont{Bodwin}},
  \bibinfo{author}{\bibfnamefont{F.}~\bibnamefont{Petriello}},
  \bibinfo{author}{\bibfnamefont{S.}~\bibnamefont{Stoynev}}, \bibnamefont{and}
  \bibinfo{author}{\bibfnamefont{M.}~\bibnamefont{Velasco}},
  \bibinfo{journal}{Phys. Rev.} \textbf{\bibinfo{volume}{D88}},
  \bibinfo{pages}{053003} (\bibinfo{year}{2013}), \eprint{1306.5770}.

\bibitem[{\citenamefont{König and Neubert}(2015)}]{Koenig:2015pha}
\bibinfo{author}{\bibfnamefont{M.}~\bibnamefont{König}} \bibnamefont{and}
  \bibinfo{author}{\bibfnamefont{M.}~\bibnamefont{Neubert}},
  \bibinfo{journal}{JHEP} \textbf{\bibinfo{volume}{08}}, \bibinfo{pages}{012}
  (\bibinfo{year}{2015}), \eprint{1505.03870}.

\bibitem[{\citenamefont{Perez et~al.}(2016)\citenamefont{Perez, Soreq, Stamou,
  and Tobioka}}]{Perez:2015lra}
\bibinfo{author}{\bibfnamefont{G.}~\bibnamefont{Perez}},
  \bibinfo{author}{\bibfnamefont{Y.}~\bibnamefont{Soreq}},
  \bibinfo{author}{\bibfnamefont{E.}~\bibnamefont{Stamou}}, \bibnamefont{and}
  \bibinfo{author}{\bibfnamefont{K.}~\bibnamefont{Tobioka}},
  \bibinfo{journal}{Phys. Rev.} \textbf{\bibinfo{volume}{D93}},
  \bibinfo{pages}{013001} (\bibinfo{year}{2016}), \eprint{1505.06689}.

\bibitem[{\citenamefont{Chisholm et~al.}(2016)\citenamefont{Chisholm,
  Kuttimalai, Nikolopoulos, and Spannowsky}}]{Chisholm:2016fzg}
\bibinfo{author}{\bibfnamefont{A.~S.} \bibnamefont{Chisholm}},
  \bibinfo{author}{\bibfnamefont{S.}~\bibnamefont{Kuttimalai}},
  \bibinfo{author}{\bibfnamefont{K.}~\bibnamefont{Nikolopoulos}},
  \bibnamefont{and}
  \bibinfo{author}{\bibfnamefont{M.}~\bibnamefont{Spannowsky}},
  \bibinfo{journal}{Eur. Phys. J.} \textbf{\bibinfo{volume}{C76}},
  \bibinfo{pages}{501} (\bibinfo{year}{2016}), \eprint{1606.09177}.

\bibitem[{\citenamefont{Delaunay et~al.}(2014)\citenamefont{Delaunay, Golling,
  Perez, and Soreq}}]{Delaunay:2013pja}
\bibinfo{author}{\bibfnamefont{C.}~\bibnamefont{Delaunay}},
  \bibinfo{author}{\bibfnamefont{T.}~\bibnamefont{Golling}},
  \bibinfo{author}{\bibfnamefont{G.}~\bibnamefont{Perez}}, \bibnamefont{and}
  \bibinfo{author}{\bibfnamefont{Y.}~\bibnamefont{Soreq}},
  \bibinfo{journal}{Phys. Rev.} \textbf{\bibinfo{volume}{D89}},
  \bibinfo{pages}{033014} (\bibinfo{year}{2014}), \eprint{1310.7029}.

\bibitem[{\citenamefont{Brivio et~al.}(2015)\citenamefont{Brivio, Goertz, and
  Isidori}}]{Brivio:2015fxa}
\bibinfo{author}{\bibfnamefont{I.}~\bibnamefont{Brivio}},
  \bibinfo{author}{\bibfnamefont{F.}~\bibnamefont{Goertz}}, \bibnamefont{and}
  \bibinfo{author}{\bibfnamefont{G.}~\bibnamefont{Isidori}},
  \bibinfo{journal}{Phys. Rev. Lett.} \textbf{\bibinfo{volume}{115}},
  \bibinfo{pages}{211801} (\bibinfo{year}{2015}), \eprint{1507.02916}.

\bibitem[{\citenamefont{Carpenter et~al.}(2017)\citenamefont{Carpenter, Han,
  Hendricks, Qian, and Zhou}}]{Carpenter:2016mwd}
\bibinfo{author}{\bibfnamefont{L.~M.} \bibnamefont{Carpenter}},
  \bibinfo{author}{\bibfnamefont{T.}~\bibnamefont{Han}},
  \bibinfo{author}{\bibfnamefont{K.}~\bibnamefont{Hendricks}},
  \bibinfo{author}{\bibfnamefont{Z.}~\bibnamefont{Qian}}, \bibnamefont{and}
  \bibinfo{author}{\bibfnamefont{N.}~\bibnamefont{Zhou}},
  \bibinfo{journal}{Phys. Rev.} \textbf{\bibinfo{volume}{D95}},
  \bibinfo{pages}{053003} (\bibinfo{year}{2017}), \eprint{1611.05463}.

\bibitem[{\citenamefont{Giudice and Lebedev}(2008)}]{Giudice:2008uua}
\bibinfo{author}{\bibfnamefont{G.~F.} \bibnamefont{Giudice}} \bibnamefont{and}
  \bibinfo{author}{\bibfnamefont{O.}~\bibnamefont{Lebedev}},
  \bibinfo{journal}{Phys. Lett.} \textbf{\bibinfo{volume}{B665}},
  \bibinfo{pages}{79} (\bibinfo{year}{2008}), \eprint{0804.1753}.

\bibitem[{\citenamefont{Botella et~al.}(2016)\citenamefont{Botella, Branco,
  Rebelo, and Silva-Marcos}}]{Botella:2016krk}
\bibinfo{author}{\bibfnamefont{F.~J.} \bibnamefont{Botella}},
  \bibinfo{author}{\bibfnamefont{G.~C.} \bibnamefont{Branco}},
  \bibinfo{author}{\bibfnamefont{M.~N.} \bibnamefont{Rebelo}},
  \bibnamefont{and} \bibinfo{author}{\bibfnamefont{J.~I.}
  \bibnamefont{Silva-Marcos}}, \bibinfo{journal}{Phys. Rev.}
  \textbf{\bibinfo{volume}{D94}}, \bibinfo{pages}{115031}
  (\bibinfo{year}{2016}), \eprint{1602.08011}.

\bibitem[{\citenamefont{Harnik et~al.}(2013)\citenamefont{Harnik, Kopp, and
  Zupan}}]{Harnik:2012pb}
\bibinfo{author}{\bibfnamefont{R.}~\bibnamefont{Harnik}},
  \bibinfo{author}{\bibfnamefont{J.}~\bibnamefont{Kopp}}, \bibnamefont{and}
  \bibinfo{author}{\bibfnamefont{J.}~\bibnamefont{Zupan}},
  \bibinfo{journal}{JHEP} \textbf{\bibinfo{volume}{03}}, \bibinfo{pages}{026}
  (\bibinfo{year}{2013}), \eprint{1209.1397}.

\bibitem[{\citenamefont{Bauer et~al.}(2016)\citenamefont{Bauer, Carena, and
  Gemmler}}]{Bauer:2015kzy}
\bibinfo{author}{\bibfnamefont{M.}~\bibnamefont{Bauer}},
  \bibinfo{author}{\bibfnamefont{M.}~\bibnamefont{Carena}}, \bibnamefont{and}
  \bibinfo{author}{\bibfnamefont{K.}~\bibnamefont{Gemmler}},
  \bibinfo{journal}{Phys. Rev.} \textbf{\bibinfo{volume}{D94}},
  \bibinfo{pages}{115030} (\bibinfo{year}{2016}), \eprint{1512.03458}.

\bibitem[{\citenamefont{Altmannshofer et~al.}(2016)\citenamefont{Altmannshofer,
  Eby, Gori, Lotito, Martone, and Tuckler}}]{Altmannshofer:2016zrn}
\bibinfo{author}{\bibfnamefont{W.}~\bibnamefont{Altmannshofer}},
  \bibinfo{author}{\bibfnamefont{J.}~\bibnamefont{Eby}},
  \bibinfo{author}{\bibfnamefont{S.}~\bibnamefont{Gori}},
  \bibinfo{author}{\bibfnamefont{M.}~\bibnamefont{Lotito}},
  \bibinfo{author}{\bibfnamefont{M.}~\bibnamefont{Martone}}, \bibnamefont{and}
  \bibinfo{author}{\bibfnamefont{D.}~\bibnamefont{Tuckler}},
  \bibinfo{journal}{Phys. Rev.} \textbf{\bibinfo{volume}{D94}},
  \bibinfo{pages}{115032} (\bibinfo{year}{2016}), \eprint{1610.02398}.

\bibitem[{\citenamefont{Bishara et~al.}(2016)\citenamefont{Bishara, Brod,
  Uttayarat, and Zupan}}]{Bishara:2015cha}
\bibinfo{author}{\bibfnamefont{F.}~\bibnamefont{Bishara}},
  \bibinfo{author}{\bibfnamefont{J.}~\bibnamefont{Brod}},
  \bibinfo{author}{\bibfnamefont{P.}~\bibnamefont{Uttayarat}},
  \bibnamefont{and} \bibinfo{author}{\bibfnamefont{J.}~\bibnamefont{Zupan}},
  \bibinfo{journal}{JHEP} \textbf{\bibinfo{volume}{01}}, \bibinfo{pages}{010}
  (\bibinfo{year}{2016}), \eprint{1504.04022}.

\bibitem[{\citenamefont{Collaboration}(2015)}]{Atlas:2015}
\bibinfo{author}{\bibfnamefont{A.}~\bibnamefont{Collaboration}}
  (\bibinfo{collaboration}{ATLAS Collaboration}) (\bibinfo{year}{2015}).

\bibitem[{\citenamefont{Abe et~al.}(1995)}]{Abe:1995hr}
\bibinfo{author}{\bibfnamefont{F.}~\bibnamefont{Abe}} \bibnamefont{et~al.}
  (\bibinfo{collaboration}{CDF}), \bibinfo{journal}{Phys. Rev. Lett.}
  \textbf{\bibinfo{volume}{74}}, \bibinfo{pages}{2626} (\bibinfo{year}{1995}),
  \eprint{hep-ex/9503002}.

\bibitem[{\citenamefont{Abachi et~al.}(1995)}]{Abachi:1994td}
\bibinfo{author}{\bibfnamefont{S.}~\bibnamefont{Abachi}} \bibnamefont{et~al.}
  (\bibinfo{collaboration}{D0}), \bibinfo{journal}{Phys. Rev. Lett.}
  \textbf{\bibinfo{volume}{74}}, \bibinfo{pages}{2422} (\bibinfo{year}{1995}),
  \eprint{hep-ex/9411001}.

\bibitem[{\citenamefont{Marchesini and Webber}(1984)}]{Marchesini:1983bm}
\bibinfo{author}{\bibfnamefont{G.}~\bibnamefont{Marchesini}} \bibnamefont{and}
  \bibinfo{author}{\bibfnamefont{B.~R.} \bibnamefont{Webber}},
  \bibinfo{journal}{Nucl. Phys.} \textbf{\bibinfo{volume}{B238}},
  \bibinfo{pages}{1} (\bibinfo{year}{1984}).

\bibitem[{\citenamefont{Banfi et~al.}(2004{\natexlab{a}})\citenamefont{Banfi,
  Salam, and Zanderighi}}]{Banfi:2004nk}
\bibinfo{author}{\bibfnamefont{A.}~\bibnamefont{Banfi}},
  \bibinfo{author}{\bibfnamefont{G.~P.} \bibnamefont{Salam}}, \bibnamefont{and}
  \bibinfo{author}{\bibfnamefont{G.}~\bibnamefont{Zanderighi}},
  \bibinfo{journal}{JHEP} \textbf{\bibinfo{volume}{08}}, \bibinfo{pages}{062}
  (\bibinfo{year}{2004}{\natexlab{a}}), \eprint{hep-ph/0407287}.

\bibitem[{\citenamefont{Banfi et~al.}(2010)\citenamefont{Banfi, Salam, and
  Zanderighi}}]{Banfi:2010xy}
\bibinfo{author}{\bibfnamefont{A.}~\bibnamefont{Banfi}},
  \bibinfo{author}{\bibfnamefont{G.~P.} \bibnamefont{Salam}}, \bibnamefont{and}
  \bibinfo{author}{\bibfnamefont{G.}~\bibnamefont{Zanderighi}},
  \bibinfo{journal}{JHEP} \textbf{\bibinfo{volume}{06}}, \bibinfo{pages}{038}
  (\bibinfo{year}{2010}), \eprint{1001.4082}.

\bibitem[{\citenamefont{Englert et~al.}(2012)\citenamefont{Englert, Spannowsky,
  and Takeuchi}}]{Englert:2012ct}
\bibinfo{author}{\bibfnamefont{C.}~\bibnamefont{Englert}},
  \bibinfo{author}{\bibfnamefont{M.}~\bibnamefont{Spannowsky}},
  \bibnamefont{and} \bibinfo{author}{\bibfnamefont{M.}~\bibnamefont{Takeuchi}},
  \bibinfo{journal}{JHEP} \textbf{\bibinfo{volume}{06}}, \bibinfo{pages}{108}
  (\bibinfo{year}{2012}), \eprint{1203.5788}.

\bibitem[{\citenamefont{Bernaciak et~al.}(2013)\citenamefont{Bernaciak,
  Buschmann, Butter, and Plehn}}]{Bernaciak:2012nh}
\bibinfo{author}{\bibfnamefont{C.}~\bibnamefont{Bernaciak}},
  \bibinfo{author}{\bibfnamefont{M.~S.~A.} \bibnamefont{Buschmann}},
  \bibinfo{author}{\bibfnamefont{A.}~\bibnamefont{Butter}}, \bibnamefont{and}
  \bibinfo{author}{\bibfnamefont{T.}~\bibnamefont{Plehn}},
  \bibinfo{journal}{Phys. Rev.} \textbf{\bibinfo{volume}{D87}},
  \bibinfo{pages}{073014} (\bibinfo{year}{2013}), \eprint{1212.4436}.

\bibitem[{\citenamefont{Gleisberg et~al.}(2009)\citenamefont{Gleisberg, Hoeche,
  Krauss, Schonherr, Schumann, Siegert, and Winter}}]{Gleisberg:2008ta}
\bibinfo{author}{\bibfnamefont{T.}~\bibnamefont{Gleisberg}},
  \bibinfo{author}{\bibfnamefont{S.}~\bibnamefont{Hoeche}},
  \bibinfo{author}{\bibfnamefont{F.}~\bibnamefont{Krauss}},
  \bibinfo{author}{\bibfnamefont{M.}~\bibnamefont{Schonherr}},
  \bibinfo{author}{\bibfnamefont{S.}~\bibnamefont{Schumann}},
  \bibinfo{author}{\bibfnamefont{F.}~\bibnamefont{Siegert}}, \bibnamefont{and}
  \bibinfo{author}{\bibfnamefont{J.}~\bibnamefont{Winter}},
  \bibinfo{journal}{JHEP} \textbf{\bibinfo{volume}{02}}, \bibinfo{pages}{007}
  (\bibinfo{year}{2009}), \eprint{0811.4622}.

\bibitem[{\citenamefont{Cacciari et~al.}(2012)\citenamefont{Cacciari, Salam,
  and Soyez}}]{Cacciari:2011ma}
\bibinfo{author}{\bibfnamefont{M.}~\bibnamefont{Cacciari}},
  \bibinfo{author}{\bibfnamefont{G.~P.} \bibnamefont{Salam}}, \bibnamefont{and}
  \bibinfo{author}{\bibfnamefont{G.}~\bibnamefont{Soyez}},
  \bibinfo{journal}{Eur. Phys. J.} \textbf{\bibinfo{volume}{C72}},
  \bibinfo{pages}{1896} (\bibinfo{year}{2012}), \eprint{1111.6097}.

\bibitem[{\citenamefont{Buckley et~al.}(2013)\citenamefont{Buckley,
  Butterworth, Lonnblad, Grellscheid, Hoeth, Monk, Schulz, and
  Siegert}}]{Buckley:2010ar}
\bibinfo{author}{\bibfnamefont{A.}~\bibnamefont{Buckley}},
  \bibinfo{author}{\bibfnamefont{J.}~\bibnamefont{Butterworth}},
  \bibinfo{author}{\bibfnamefont{L.}~\bibnamefont{Lonnblad}},
  \bibinfo{author}{\bibfnamefont{D.}~\bibnamefont{Grellscheid}},
  \bibinfo{author}{\bibfnamefont{H.}~\bibnamefont{Hoeth}},
  \bibinfo{author}{\bibfnamefont{J.}~\bibnamefont{Monk}},
  \bibinfo{author}{\bibfnamefont{H.}~\bibnamefont{Schulz}}, \bibnamefont{and}
  \bibinfo{author}{\bibfnamefont{F.}~\bibnamefont{Siegert}},
  \bibinfo{journal}{Comput. Phys. Commun.} \textbf{\bibinfo{volume}{184}},
  \bibinfo{pages}{2803} (\bibinfo{year}{2013}), \eprint{1003.0694}.

\bibitem[{\citenamefont{Barnard et~al.}(2017)\citenamefont{Barnard, Dawe,
  Dolan, and Rajcic}}]{Barnard:2016qma}
\bibinfo{author}{\bibfnamefont{J.}~\bibnamefont{Barnard}},
  \bibinfo{author}{\bibfnamefont{E.~N.} \bibnamefont{Dawe}},
  \bibinfo{author}{\bibfnamefont{M.~J.} \bibnamefont{Dolan}}, \bibnamefont{and}
  \bibinfo{author}{\bibfnamefont{N.}~\bibnamefont{Rajcic}},
  \bibinfo{journal}{Phys. Rev.} \textbf{\bibinfo{volume}{D95}},
  \bibinfo{pages}{014018} (\bibinfo{year}{2017}), \eprint{1609.00607}.

\bibitem[{\citenamefont{Krinner et~al.}(2013)\citenamefont{Krinner, Lenz, and
  Rauh}}]{Lenz:2013}
\bibinfo{author}{\bibfnamefont{F.}~\bibnamefont{Krinner}},
  \bibinfo{author}{\bibfnamefont{A.}~\bibnamefont{Lenz}}, \bibnamefont{and}
  \bibinfo{author}{\bibfnamefont{T.}~\bibnamefont{Rauh}},
  \bibinfo{journal}{Nucl. Phys.} \textbf{\bibinfo{volume}{B876}},
  \bibinfo{pages}{31} (\bibinfo{year}{2013}), \eprint{1305.5390}.

\bibitem[{\citenamefont{Hocker et~al.}(2007)}]{Hocker:2007ht}
\bibinfo{author}{\bibfnamefont{A.}~\bibnamefont{Hocker}} \bibnamefont{et~al.},
  \bibinfo{journal}{PoS} \textbf{\bibinfo{volume}{ACAT}}, \bibinfo{pages}{040}
  (\bibinfo{year}{2007}), \eprint{physics/0703039}.

\bibitem[{\citenamefont{Dermisek and Gunion}(2010)}]{Dermisek:2010mg}
\bibinfo{author}{\bibfnamefont{R.}~\bibnamefont{Dermisek}} \bibnamefont{and}
  \bibinfo{author}{\bibfnamefont{J.~F.} \bibnamefont{Gunion}},
  \bibinfo{journal}{Phys. Rev.} \textbf{\bibinfo{volume}{D81}},
  \bibinfo{pages}{075003} (\bibinfo{year}{2010}), \eprint{1002.1971}.

\bibitem[{\citenamefont{Banfi et~al.}(2004{\natexlab{b}})\citenamefont{Banfi,
  Salam, and Zanderighi}}]{Banfi:2003je}
\bibinfo{author}{\bibfnamefont{A.}~\bibnamefont{Banfi}},
  \bibinfo{author}{\bibfnamefont{G.~P.} \bibnamefont{Salam}}, \bibnamefont{and}
  \bibinfo{author}{\bibfnamefont{G.}~\bibnamefont{Zanderighi}},
  \bibinfo{journal}{Phys. Lett.} \textbf{\bibinfo{volume}{B584}},
  \bibinfo{pages}{298} (\bibinfo{year}{2004}{\natexlab{b}}),
  \eprint{hep-ph/0304148}.

\bibitem[{\citenamefont{Banfi et~al.}(2005)\citenamefont{Banfi, Salam, and
  Zanderighi}}]{Banfi:2004yd}
\bibinfo{author}{\bibfnamefont{A.}~\bibnamefont{Banfi}},
  \bibinfo{author}{\bibfnamefont{G.~P.} \bibnamefont{Salam}}, \bibnamefont{and}
  \bibinfo{author}{\bibfnamefont{G.}~\bibnamefont{Zanderighi}},
  \bibinfo{journal}{JHEP} \textbf{\bibinfo{volume}{03}}, \bibinfo{pages}{073}
  (\bibinfo{year}{2005}), \eprint{hep-ph/0407286},
  \urlprefix\url{http://home.fnal.gov/~zanderi/Caesar/caesar.html}.

\bibitem[{\citenamefont{Brandt et~al.}(1964)\citenamefont{Brandt, Peyrou,
  Sosnowski, and Wroblewski}}]{Brandt:1964sa}
\bibinfo{author}{\bibfnamefont{S.}~\bibnamefont{Brandt}},
  \bibinfo{author}{\bibfnamefont{C.}~\bibnamefont{Peyrou}},
  \bibinfo{author}{\bibfnamefont{R.}~\bibnamefont{Sosnowski}},
  \bibnamefont{and}
  \bibinfo{author}{\bibfnamefont{A.}~\bibnamefont{Wroblewski}},
  \bibinfo{journal}{Phys. Lett.} \textbf{\bibinfo{volume}{12}},
  \bibinfo{pages}{57} (\bibinfo{year}{1964}).

\bibitem[{\citenamefont{Farhi}(1977)}]{Farhi:1977sg}
\bibinfo{author}{\bibfnamefont{E.}~\bibnamefont{Farhi}},
  \bibinfo{journal}{Phys. Rev. Lett.} \textbf{\bibinfo{volume}{39}},
  \bibinfo{pages}{1587} (\bibinfo{year}{1977}).

\bibitem[{\citenamefont{Barber et~al.}(1980)}]{Barber:1980uq}
\bibinfo{author}{\bibfnamefont{D.~P.} \bibnamefont{Barber}}
  \bibnamefont{et~al.} (\bibinfo{collaboration}{MARK-J,
  AACHEN-DESY-MIT-NIKHEF-BEIJING}), \bibinfo{journal}{Phys. Rept.}
  \textbf{\bibinfo{volume}{63}}, \bibinfo{pages}{337} (\bibinfo{year}{1980}).

\bibitem[{\citenamefont{Fox and Wolfram}(1978)}]{Fox:1978vu}
\bibinfo{author}{\bibfnamefont{G.~C.} \bibnamefont{Fox}} \bibnamefont{and}
  \bibinfo{author}{\bibfnamefont{S.}~\bibnamefont{Wolfram}},
  \bibinfo{journal}{Phys. Rev. Lett.} \textbf{\bibinfo{volume}{41}},
  \bibinfo{pages}{1581} (\bibinfo{year}{1978}).

\bibitem[{\citenamefont{Catani et~al.}(1991)\citenamefont{Catani, Dokshitzer,
  Olsson, Turnock, and Webber}}]{Catani:1991hj}
\bibinfo{author}{\bibfnamefont{S.}~\bibnamefont{Catani}},
  \bibinfo{author}{\bibfnamefont{Y.~L.} \bibnamefont{Dokshitzer}},
  \bibinfo{author}{\bibfnamefont{M.}~\bibnamefont{Olsson}},
  \bibinfo{author}{\bibfnamefont{G.}~\bibnamefont{Turnock}}, \bibnamefont{and}
  \bibinfo{author}{\bibfnamefont{B.~R.} \bibnamefont{Webber}},
  \bibinfo{journal}{Phys. Lett.} \textbf{\bibinfo{volume}{B269}},
  \bibinfo{pages}{432} (\bibinfo{year}{1991}).

\bibitem[{\citenamefont{Banfi et~al.}(2002)\citenamefont{Banfi, Salam, and
  Zanderighi}}]{Banfi:2001bz}
\bibinfo{author}{\bibfnamefont{A.}~\bibnamefont{Banfi}},
  \bibinfo{author}{\bibfnamefont{G.~P.} \bibnamefont{Salam}}, \bibnamefont{and}
  \bibinfo{author}{\bibfnamefont{G.}~\bibnamefont{Zanderighi}},
  \bibinfo{journal}{JHEP} \textbf{\bibinfo{volume}{01}}, \bibinfo{pages}{018}
  (\bibinfo{year}{2002}), \eprint{hep-ph/0112156}.

\bibitem[{\citenamefont{Bjorken and Brodsky}(1970)}]{Bjorken:1969wi}
\bibinfo{author}{\bibfnamefont{J.~D.} \bibnamefont{Bjorken}} \bibnamefont{and}
  \bibinfo{author}{\bibfnamefont{S.~J.} \bibnamefont{Brodsky}},
  \bibinfo{journal}{Phys. Rev.} \textbf{\bibinfo{volume}{D1}},
  \bibinfo{pages}{1416} (\bibinfo{year}{1970}).

\bibitem[{\citenamefont{Parisi}(1978)}]{Parisi:1978eg}
\bibinfo{author}{\bibfnamefont{G.}~\bibnamefont{Parisi}},
  \bibinfo{journal}{Phys. Lett.} \textbf{\bibinfo{volume}{74B}},
  \bibinfo{pages}{65} (\bibinfo{year}{1978}).

\bibitem[{\citenamefont{Donoghue et~al.}(1979)\citenamefont{Donoghue, Low, and
  Pi}}]{Donoghue:1979vi}
\bibinfo{author}{\bibfnamefont{J.~F.} \bibnamefont{Donoghue}},
  \bibinfo{author}{\bibfnamefont{F.~E.} \bibnamefont{Low}}, \bibnamefont{and}
  \bibinfo{author}{\bibfnamefont{S.-Y.} \bibnamefont{Pi}},
  \bibinfo{journal}{Phys. Rev.} \textbf{\bibinfo{volume}{D20}},
  \bibinfo{pages}{2759} (\bibinfo{year}{1979}).

\bibitem[{\citenamefont{Catani and Webber}(1998)}]{Catani:1998sf}
\bibinfo{author}{\bibfnamefont{S.}~\bibnamefont{Catani}} \bibnamefont{and}
  \bibinfo{author}{\bibfnamefont{B.~R.} \bibnamefont{Webber}},
  \bibinfo{journal}{Phys. Lett.} \textbf{\bibinfo{volume}{B427}},
  \bibinfo{pages}{377} (\bibinfo{year}{1998}), \eprint{hep-ph/9801350}.

\end{thebibliography}

\end{document}